\documentclass[12pt]{article}

\usepackage{amssymb,amsmath,amsthm,bbm,enumitem,xcolor,mathtools}
\usepackage[a4paper, total={6.3in, 9.2in}]{geometry}
\usepackage{setspace}
\usepackage{hyperref}
\usepackage{graphicx,caption,subcaption}
\usepackage[title]{appendix}
\usepackage{makecell}
\usepackage{upgreek}
\usepackage{comment}

\newtheorem{theorem}{Theorem}[section]
\theoremstyle{definition}
\newtheorem{definition}{Definition}[section]
\theoremstyle{definition}

\theoremstyle{definition}
\newtheorem{assumption}{Assumption}[section]
\theoremstyle{definition}
\newtheorem{proposition}{Proposition}[section]
\theoremstyle{definition}

\theoremstyle{remark}
\newtheorem{remark}{Remark}[section]
\theoremstyle{definition}
\newtheorem{lemma}{Lemma}[section]

\newcommand{\bunderline}[1]{\underline{#1\mkern-4mu}\mkern4mu }

\allowdisplaybreaks
\usepackage{comment}

\setlist[itemize]{leftmargin=0.4cm,labelindent=\parindent}

\title{Forward Performance Processes under Multiple Default Risks}

\author{\small Wing Fung Chong\thanks{Department of Statistics and Actuarial Science, School of Computing and Data Science, The University of Hong Kong, Hong Kong, China. Email: \href{mailto:chongwf@hku.hk}{chongwf@hku.hk}.}\;, Roxana Dumitrescu\thanks{CREST, ENSAE, Institut Polytechnique de Paris, Palaiseau, France. Email: \href{mailto:roxana.dumitrescu@ensae.fr}{roxana.dumitrescu@ensae.fr}.}\;, Gechun Liang\thanks{Department of Statistics, University of Warwick, Coventry, United Kingdom. Email: \href{mailto:g.liang@warwick.ac.uk}{g.liang@warwick.ac.uk}.}\;, and Kenneth Tsz Hin Ng\thanks{Department of Mathematics, The Ohio State University, Columbus, United States. Email: \href{mailto:ng.499@osu.edu}{ng.499@osu.edu}.}}
\date{\small \today}

\begin{document}
\maketitle
\abstract{

This article constructs a forward exponential utility in a market with multiple defaultable risks. Using the Jacod-Pham decomposition for random fields, we first characterize forward performance processes in a defaultable market under the default-free filtration. We then construct a forward utility via a system of recursively defined, indexed infinite-horizon backward stochastic differential equations (BSDEs) with discounting, and establish the existence, uniqueness, and boundedness of their solutions. To verify the required (super)martingale property of the performance process, we develop a rigorous characterization of this property with respect to the general filtration in terms of a set of (in)equalities relative to the default-free filtration. We further extend the analysis to a stochastic factor model with ergodic dynamics. In this setting, we derive uniform bounds for the Markovian solutions of the infinite-horizon BSDEs, overcoming technical challenges arising from the special structure of the system of BSDEs in the defaultable setting. Passing to the ergodic limit, we identify the limiting BSDE and relate its constant to the risk-sensitive long-run growth rate of the optimal wealth process.}


\begin{flushleft}

\end{flushleft}

\section{Introduction}


 A default-free market assumes that all financial institutions can fulfill their financial obligations, meaning there is no risk of financial failure that could trigger sudden price changes. In such a market, asset prices evolve smoothly and are often modeled mathematically as diffusion processes. However, especially following the 2008 financial crisis, there has been growing interest in defaultable assets \textcolor{black}{(subject to obligor default)}. In particular, the contagion effect, whereby the default of \textcolor{black}{one obligor} can significantly impact others, may create a chain reaction that profoundly influences market dynamics, leading to multiple default events and, consequently, affecting the optimal investment strategies of investors. 

To incorporate defaultable assets, \cite{jiao2011optimal} formulated the optimal investment problem under a single default model, which was later extended to multiple default models in \cite{PHAM20101795,pham:2013:AOAP,kharroubi2014progressive}. In these works, the filtration $\mathbb{G}$
 of the defaultable market is constructed by enlarging a default-free filtration 
$\mathbb{F}$ to include knowledge of default times and the corresponding loss marks when they occur. Using the \textit{Jacod-Pham decomposition}, the optimal investment strategy is then characterized by a recursive sequence of indexed exponential-quadratic backward stochastic differential equations (BSDEs) driven by the Brownian motion in 
$\mathbb{F}$. Subsequent studies explored related problems: \cite{henderson:liang:2016:indifferent} addressed indifference pricing for non-traded assets subject to default risk; \cite{CREPEY20153023} studied BSDEs with random terminal times motivated by counterparty risk; \cite{bo:capponi:2016:contagion} examined optimal allocation between credit default swaps and a money market; and \cite{bo:liao:Yu:2019} analyzed portfolio allocation in a regime-switching market with default contagion.

The aforementioned works are based on classical utility theory, in which a terminal time $T$ and a utility function are specified at the outset of the planning horizon. The value function and optimal strategy are then determined backward in time using the dynamic programming principle. This pre-commitment is often considered a limitation of classical theory. Accordingly, the objective of this article is to complement this line of research on defaultable markets by considering a forward-looking utility preference that eliminates the need for pre-commitment.

The notion of forward preferences was first introduced in a series of papers \cite{Musiela2007,Musiela2008,Musiela2009,Musiela2010a,Musiela2010b,Musiela2011}. Unlike classical utility, a forward preference is a dynamically evolving process (more precisely, a random field) that, when evaluated at the controlled wealth process (or the optimally controlled wealth process), satisfies a supermartingale (or martingale) property. It provides a forward-looking mechanism by continuously incorporating evolving market conditions and investors’ preferences, thereby evaluating the performance of investment strategies in real time. This approach naturally accommodates forward features such as model updates and learning. In continuous-time models, forward preferences are often constructed or characterized via stochastic partial differential equations (SPDEs) \cite{karoui:2013,shkolnikov:2016,El:2018:FBSPDE}, dual formulations \cite{Zitkovic2009}, infinite-horizon BSDEs, and ergodic BSDEs \cite{liang:2017:bsde,chong2019ergodic}. Forward utilities have also been studied in general semimartingale models \cite{B02023,Choulli2016}, discrete-time binomial models \cite{angoshtari:2020,robo:2023,strub2021evolution,Angoshtari:2023:predictable}, and in rank-dependent preferences \cite{He:2021:rank:dependent,Angoshtari:2024}.

The flexibility of forward preferences has led to extensive applications in finance and related fields, including: investment and reinsurance \cite{colaneri:katia:reinsurance:2021,colaneri:katia:reinsurance:2022}, valuation of American options \cite{leung2012forward}, optimal consumption problems \cite{Kallblad2020,Hillairet2024}, model ambiguity \cite{Kallblad2018,Liang2021}, stochastic factor models \cite{Nadtochiy2014,Sircar2020}, indifference valuations \cite{Anthropelos2014,musiela2010indifference}, pension fund management \cite{dynamic:preference:pension,anthropelos:relative,hillairet:social,anthropelos2025time,NG2024192}, forward entropic risk measures \cite{chong2019ergodic}, equity-linked life insurance \cite{CHONG201993}, as well as robo-advising \cite{capponi2022personalized,robo:2023}.

This article consists of two main parts. In the first part, using a BSDE approach, we construct a forward exponential utility in a generic market consisting of $m$ defaultable assets and provide the corresponding optimal investment strategy. The construction is based on the Jacod-Pham decomposition, originally proposed in \cite{jacod2006grossissement} and later used to solve BSDEs with jumps in \cite{PHAM20101795} (see also \cite{pham:2013:AOAP,jiao2011optimal,kharroubi2014progressive}). The central idea is to decompose any $\mathbb{G}$-adapted (resp.~predictable) process into a sequence of $\mathbb{F}$-adapted (resp.~predictable) indexed processes, thereby allowing us to work under the default-free filtration, where the Brownian motion retains its martingale property -- a crucial feature that may fail in the augmented filtration $\mathbb{G}$.

In this generic defaultable market setting, our main contributions are as follows. First, we provide a rigorous characterization of the  $\mathbb{G}$-(super)martingale property of a process in terms of a set of integral (in)equalities based on the process’s Jacod-Pham decomposition. These (in)equalities incorporate the conditional density processes of the default times and losses given $\mathbb{F}$. Using this technical result, we then provide a characterization of the  $\mathbb{G}$-martingale and $\mathbb{G}$-supermartingale properties of the forward performance process under the optimal strategy and any admissible strategy, respectively (Theorem \ref{thm:U:G:F}). This characterization provides the key building blocks for constructing the forward performance process, particularly by leveraging the martingale property of the 
$\mathbb{F}$-Brownian motion.

Second, building on Theorem \ref{thm:U:G:F}, we propose an \textit{ansatz} for the forward preference 
$U$ based on a sequence of recursively defined, indexed BSDEs driven by an 
$\mathbb{F}$-Brownian motion. To establish the existence and uniqueness of the solutions to these BSDEs, we generalize the approach in \cite{pham:2013:AOAP} to the infinite-horizon setting, employing truncation arguments and the comparison principle to handle the exponential jump-intensity term in the driver. Indeed, unlike the backward finite-horizon BSDEs considered therein, a forward utility is constructed independently of any planning horizon and gives rise to infinite-horizon BSDEs. A positive discount rate is applied to both the \textit{ansatz} and the driver to ensure boundedness and well-posedness.

We then establish the existence and uniqueness of solutions recursively, forming the second main result of this article (Theorem \ref{thm:exist:Y}). Using these BSDE solutions together with Theorem \ref{thm:U:G:F}, we verify that the \textit{ansatz} indeed defines a forward preference and provide the associated optimal investment strategy in Theorem \ref{thm:forward:optimal}. Compared to the construction of forward performance processes in a general semimartingale setting by \cite{B02023,Choulli2016}, our work complements theirs by providing an explicit characterization of the conditional density processes and a BSDE-based framework that enables a concrete and systematic construction.


In the second part of the article, we extend our study to a factor model in which the market parameters are driven by a stochastic factor exhibiting ergodicity. Under this factor model, we first show that the system of infinite-horizon BSDEs admits a Markovian solution. We then establish bounds for the Markovian solution, similar to those in \cite{liang:2017:bsde}, as well as one-sided bounds for the deviations of consecutive solution components, analogous to those in \cite{hu2020systems}, which are uniform in the discount rate of the infinite-horizon BSDEs.

There are two main challenges in the current defaultable market. First, the index mismatch across default regimes causes solution components, even within the same regime, to differ when evaluated at different default times or loss levels. This complication also prevents the direct application of standard multi-dimensional comparison principles for BSDEs. Second, unlike the regime-switching setting in \cite{hu2020systems}, the jump sizes resulting from defaults are controlled by the investment strategy. This prevents us from exploiting the property of the quadratic distance function.

To address these challenges, we combine a truncation argument with a novel adaptation of the proof of the comparison principle for multi-dimensional BSDEs to establish bounds on the Markovian solution of the infinite-horizon BSDEs in a recursive manner, leading to Theorem \ref{thm:Z:markovian}. We then analyze the deviations between consecutive solution components, showing that these deviations are upper-bounded under Assumption \ref{ass:ergodic:kappa:g}, uniformly in the discount rate. This allows us to control the deviations recursively, in the spirit of \cite{hu2020systems}, leading to Proposition \ref{pp:Delta:Y}.

The bounds established in Theorem \ref{thm:Z:markovian} and Proposition \ref{pp:Delta:Y} guarantee uniform control of the BSDEs' 
$Z$-component, imply a uniform Lipschitz property of the Markovian solution, and prevent the exponential linkage across default intervals from approaching to infinity. These properties ensure that the solution remains stable as the discount rate vanishes. Utilizing these estimates, we adapt a perturbation argument from \cite{liang:2017:bsde} and \cite{hu2020systems} to study the limiting behavior of the Markovian solution as the discount rate tends to zero. We interpret the resulting ergodic constant, following the last default event, as the long-term risk-sensitive growth rate of the optimal wealth process in the zero-discount limit.

In the final part of the study, we discuss the challenges of constructing a forward utility via ergodic BSDEs due to the one-directional dependence within the system. Under an additional monotonicity assumption, we illustrate a possible construction and show that the system of infinite-horizon BSDEs converges to a system of ergodic BSDEs.



The remainder of the article is organized as follows. Section \ref{sec:model} introduces the market model. Section \ref{sec:forward} discusses the notion of forward preferences and their characterization under the default-free filtration, as presented in Theorem \ref{thm:U:G:F}. In Section \ref{sec:exp}, we propose an \textit{ansatz} for a forward exponential utility based on the infinite-horizon BSDEs \eqref{eq:Ym}-\eqref{eq:Yn}, establish the existence and uniqueness of the solution to the system, and verify that the \textit{ansatz} indeed defines a forward exponential utility by checking the relevant integrability conditions. Section \ref{sec:stochastic:factor} extends the analysis to an ergodic stochastic factor model and establishes the limiting behavior of the infinite-horizon BSDEs. Section \ref{sec:conclusion} concludes the article, and the appendices contain the proofs of the main results.

\section{Model Formulation}
\label{sec:model}

\subsection{Preliminaries}
 Let \((\Omega, \mathcal{F}, \mathbb{P})\) be a complete probability space, and \(m \geq 1\) be a fixed positive integer representing the number of defaults. Let \((T_1, \dots, T_m)\) be a sequence of \(\mathcal{F}\)-measurable random times, where each \(T_i : \Omega \to (0, \infty)\) satisfies \(0 = T_0 < T_1 < \cdots < T_m < \infty\) almost surely. This sequence represents the default times for the risky assets. We also let \((L_1, \dots, L_m)\) be a sequence of \(\mathcal{F}\)-measurable random variables taking values in a Polish space \(E\). For each \(i = 1, \dots, m\), \(L_i\) denotes the loss at the default time \(T_i\).

Let \(\mathbb{F} = (\mathcal{F}_t)_{t \geq 0}\) be the completed, right-continuous filtration generated by a \(d\)-dimensional Brownian motion \((W_t)_{t \geq 0}\), where \(d \geq 1\) is a fixed positive integer, representing the default-free market information. For each \(n = 1, \dots, m\) and \(t \geq 0\), define \(\mathcal{N}^n_t := \sigma(\mathbbm{1}_{\{T_n \leq s\}},L_n \mathbbm{1}_{\{T_n \leq s\}} : s \leq t)\), which captures the information about the \(n\)-th default time and loss up to time \(t\), and let \(\tilde{\mathcal{N}}^n_t := \mathcal{N}^n_{t^+} = \bigcap_{u > t} \mathcal{N}^n_u\), the right-continuous version of \(\mathcal{N}^n_t\).  We then define the default filtration $\tilde{\mathbb{N}}:=\bigvee_{n=1}^m\tilde{\mathbb{N}}^n$, where $\tilde{\mathbb{N}}^n:=(\tilde{\mathcal{N}}^n_t)_{t \geq 0},$ and the market filtration $\mathbb{G} := \mathbb{F}\vee\tilde{\mathbb{N}}$. 
By construction, each \(T_n\) is a \(\mathbb{G}\)-stopping time, and each \(L_n\) is \(\mathcal{G}_{T_n}\)-measurable.  

 For each \(n = 0, 1, \dots, m\), let \(\Delta_n := \{ (\theta_0, \theta_1, \dots, \theta_n) \in [0, \infty)^{n+1} : 0=\theta_0 < \theta_1 < \cdots < \theta_n<\infty \}\), with elements denoted by \(\theta_{(n)} := (\theta_0,\theta_1, \dots, \theta_n) \in \Delta_n\), representing strictly ordered sequences of times starting at zero. For each \(n = 1, \dots, m\), let \(E^n := E \times \cdots \times E\) ($n$-fold Cartesian product of the Polish space \(E\)), with elements denoted by \(l_{(n)} := (l_1, \dots, l_n) \in E^n\), where \(l_i \in E\) for \(i = 1, \dots, n\), representing the vector of the first $n$ losses.\footnote{With a slight abuse of notations, in this paper, when a mathematical object depends on $E^0$, $l_{\left(0\right)}$, or $L_{\left(0\right)}$, this is interpreted as that the object is independent of $E^0$, $l_{\left(0\right)}$, or $L_{\left(0\right)}$.}

We let \(\mathcal{P}(\mathbb{F})\) (resp.\ \(\mathcal{O}(\mathbb{F})\)) be the predictable (resp.\ optional) \(\sigma\)-algebra on \(\mathbb{R}_+ \times \Omega\) associated with the filtration \(\mathbb{F}\). For each \(n = 0, 1, \dots, m\), we denote by \(\mathcal{P}_\mathbb{F}(\Delta_n, E^n; \mathcal{D})\) the set of \(\mathbb{F}\)-predictable indexed processes \((\varphi^n_t(\cdot, \cdot))_{t \geq 0}\) taking values in a Borel set \(\mathcal{D}\) (e.g., \(\mathbb{R}\), \(\mathbb{R}_+\), \(\mathbb{R}^m\), \(\mathbb{R}^{m \times d}\)), such that the map \((t, \omega, \theta_{(n)}, l_{(n)}) \mapsto \varphi^n_t(\theta_{(n)}, l_{(n)}; \omega)\) is \(\mathcal{P}(\mathbb{F}) \otimes \mathcal{B}(\Delta_n) \otimes \mathcal{B}(E^n)\)-measurable. Likewise, we denote by \(\mathcal{O}_\mathbb{F}(\Delta_n, E^n; \mathcal{D})\) the set of \(\mathbb{F}\)-optional, \(\mathcal{D}\)-valued indexed processes \((\psi^n_t(\cdot, \cdot))_{t \geq 0}\) such that the map \((t, \omega, \theta_{(n)}, l_{(n)}) \mapsto \psi^n_t(\theta_{(n)}, l_{(n)}; \omega)\) is \(\mathcal{O}(\mathbb{F}) \otimes \mathcal{B}(\Delta_n) \otimes \mathcal{B}(E^n)\)-measurable.\footnote{For \(n = 0\), \(\mathcal{B}(\Delta_0)\) and \(\mathcal{B}(E^0)\) are trivial \(\sigma\)-algebras.}

For any $n=0,\dots,m$, we define the subspaces $\mathcal{S}(\Delta_n,E^n;\mathcal{D})\subseteq \mathcal{O}_\mathbb{F}(\Delta_n,E^n;\mathcal{D})$ and $\mathcal{S}_\mathcal{P}(\Delta_n,E^n;\mathcal{D})\subseteq \mathcal{P}_\mathbb{F}(\Delta_n,E^n;\mathcal{D})$ such that, $\varphi^n \in \mathcal{S}(\Delta_n,E^n;\mathcal{D})$ (resp.~$\varphi^n \in \mathcal{S}_\mathcal{P}(\Delta_n,E^n;\mathcal{D})$) if and only if $\varphi^n \in \mathcal{O}_\mathbb{F}(\Delta_n,E^n;\mathcal{D})$ (resp.~$\varphi^n \in \mathcal{P}_\mathbb{F}(\Delta_n,E^n;\mathcal{D})$), and
    \begin{equation*}
        \sup_{(\theta_{(n)},l_{(n)}) \in \Delta_n\times E^n} \mathop{\text{ess}\,\text{sup}}_{(t,\omega)\in [\theta_n,\infty)\times\Omega }\left|\varphi^n_t(\theta_{(n)},l_{(n)}) \right| < \infty,
    \end{equation*}%
where $|\cdot|$ denotes the usual Euclidean or matrix norm in $\mathcal{D}$. We also denote by $\mathcal{L}^2_{\text{loc}}(\Delta_n,E^n;\mathcal{D})$ a subspace of $\mathcal{P}_\mathbb{F}(\Delta_n,E^n;\mathcal{D})$, which is the set of all $\mathbb{F}$-predictable indexed process $Z(\cdot,\cdot)$ such that, for any $\theta_{(n)}\in\Delta_n, l_{(n)} \in E^n$, and $t\geq \theta_n$,
     \begin{equation*}
        \mathbb{E}\left[\int_{\theta_n}^t  |Z_s(\theta_{(n)},l_{(n)})|^2 ds  \right] < \infty.
    \end{equation*}
For any $\varepsilon>0$, we let $\mathcal{M}^{2,\varepsilon}(\Delta_n,E^n;\mathcal{D})$ be the set of all $\mathbb{F}$-predictable indexed process $Z(\cdot,\cdot)$ such that, for any $\theta_{(n)}\in\Delta_n, l_{(n)} \in E^n$,
  \begin{equation*}
        \mathbb{E}\left[\int_{\theta_n}^\infty e^{-2\varepsilon s} |Z_s(\theta_{(n)},l_{(n)})|^2 ds  \right] < \infty.
    \end{equation*}
We also define $\mathcal{M}^2(\Delta_n,E^n;\mathcal{D}) := \cap_{\varepsilon>0} \mathcal{M}^{2,\varepsilon}(\Delta_n,E^n;\mathcal{D})$.

Finally, 
for any $n\in\mathbb{N}$, we let ${\bf 1}_n\in\mathbb{R}^n$ to be the vector with all entries being 1. We also use the notation $v'$ to denote the transpose of the vector or matrix $v$.

\subsection{Market Model}
We consider an optimal portfolio selection problem where a portfolio of $m$ risky assets is subject to default risks,\footnote{For mathematical simplicity and notational convenience, we assume that the number of risky assets is the same as the number of defaults. In general, these two values can differ.} where $m\leq d$. Let $\mu=(\mu_t)_{t\geq 0}$ and $\sigma = (\sigma_t)_{t\geq 0}$ be $\mathbb{G}$-predictable $\mathbb{R}^m$ and $\mathbb{R}^{m\times d}$-valued processes, which represent the rate of return and volatility of the underlying risky assets, respectively.
Using the Jacod-Pham decomposition, the processes $\mu$ and $\sigma$ can be decomposed as
    \begin{equation}
    \label{eq:model:parameter:decompose}
        \begin{aligned}
            \mu_t &= \sum_{n=0}^{m-1} \mu^n_t(T_{(n)},L_{(n)}) \mathbbm{1}_{ \{T_n < t \leq T_{n+1} \} } + \mu^m_t(T_{(m)},L_{(m)}) \mathbbm{1}_{\{t > T_m\}}, \\
            \sigma_t &= \sum_{n=0}^{m-1} \sigma^n_t(T_{(n)},L_{(n)}) \mathbbm{1}_{ \{T_n < t \leq T_{n+1} \} } + \sigma^m_t(T_{(m)},L_{(m)}) \mathbbm{1}_{\{t > T_m\}},
        \end{aligned}
    \end{equation}
where, for any $n=0,1,\dots,m$, $T_{(n)}:=(T_0,T_1, \dots,T_n)$ and for any $n=1,\dots,m$, $L_{(n)}:=(L_1,\dots,L_n)$. Here, $\mu^n(\cdot,\cdot) \in \mathcal{P}_\mathbb{F}(\Delta_n,E^n;\mathbb{R}^m)$ and $\sigma^n(\cdot,\cdot) \in \mathcal{P}_\mathbb{F}(\Delta_n,E^n;\mathbb{R}^{m\times d})$.  We also let $\beta = (\beta_t(l))_{t\geq 0,\ l\in E}$ be a $\mathbb{G}$-predictable process  which admits the following decomposition: 
    \begin{equation}
        \label{eq:model:parameter:beta:decompose}
        \beta_t(l) = \sum_{n=0}^{m-1} \beta^n_t(T_{(n)},L_{(n)},l) \mathbbm{1}_{\{T_n < t \leq T_{n+1}\} },
    \end{equation}
where for each $n=0,\dots,m-1$, $\beta^n(\cdot,\cdot,\cdot) \in \mathcal{P}_\mathbb{F}(\Delta_n,E^n, E;\mathbb{R}^m)$, i.e., the map $(t,\omega,\theta_{(n)},l_{(n)},l)\mapsto \beta^n_t(\theta_{(n)},l_{(n)},l;\omega)$ is $\mathcal{P}(\mathbb{F})\otimes \mathcal{B}(\Delta_n)\otimes \mathcal{B}(E^n)\otimes \mathcal{B}(E)$-measurable. The process $\beta$ represents the  jump size factor of the underlying risky assets; see \eqref{eq:Sn} below. We assume that, for any $n=0,\dots,m-1$, $(\theta_{(n)},l_{(n)},l)\in \Delta_n \times E^n\times E$, and $t\geq \theta_n$, $\beta^{n,i}_t(\theta_{(n)},l_{(n)},l)>-1$ $\mathbb{P}$-a.s., where for $i=1,\dots,m$, $\beta^{n,i}_t(\theta_{(n)},l_{(n)},l)$ is the $i$-th entry of $\beta^{n}_t(\theta_{(n)},l_{(n)},l)$.  We also denote by $\mathcal{S}_\mathcal{P}(\Delta_n,E^n,E;\mathbb{R}^m)$ a subspace of $\mathcal{P}_\mathbb{F}(\Delta_n,E^n,E;\mathbb{R}^m)$ such that $\varphi(\cdot,\cdot,\cdot) \in \mathcal{S}_\mathcal{P}(\Delta_n,E^n,E;\mathbb{R}^m)$ if and only if
      \begin{equation*}
        \sup_{(\theta_{(n)},l_{(n)},l) \in \Delta_n\times E^n\times E} \ \mathop{\text{ess}\,\text{sup}}_{(t,\omega)\in [\theta_n,\infty)\times\Omega }\left|\varphi^n_t(\theta_{(n)},l_{(n)},l) \right| < \infty.
    \end{equation*}

 The $m$ risky assets' prices $S=(S_t)_{t\geq 0}$ is a $\mathbb{R}^m$-valued $\mathbb{G}$-optional process, which admits the following Jacod-Pham decomposition:
    \begin{equation}
    \label{eq:S:decompose}
        S_t = \sum_{n=0}^{m-1} S^n_t(T_{(n)},L_{(n)})\mathbbm{1}_{ \{ T_n \leq t < T_{n+1} \} } + S^m_t(T_{(m)},L_{(m)}) \mathbbm{1}_{\{t\geq T_m\}},
    \end{equation}
where for each $n=0,\dots,m$, $S^n(\cdot,\cdot) = (S^n_t(\cdot,\cdot))_{t\geq 0} \in \mathcal{O}_\mathbb{F}(\Delta_n,E^n;\mathbb{R}^m)$. The dynamics of the sequence of indexed processes are governed by the following: for $t\geq 0$,
    \begin{equation}
        \label{eq:S0}
   dS^0_t(0)  = S^0_t(0) *\bigg( \mu^{0}_t(0)  dt  + \sigma^{0}_t(0) dW_t \bigg),
    \end{equation}
and for any $n=1,\dots,m$, $\theta_{(n)}\in \Delta_{n}$, and $l_{(n)}\in E^{n}$,
    \begin{equation}
    \label{eq:Sn}
    \left\{ \begin{aligned}
        dS^{n}_t(\theta_{(n)},l_{(n)}) &= S^{n}_t(\theta_{(n)},l_{(n)})*\left( \mu^{n}_t(\theta_{(n)},l_{(n)}) dt   + \sigma^{n}_t(\theta_{(n)},l_{(n)})dW_t \right), \ t \geq \theta_{n}, \\
        S^{n}_{\theta_{n}}(\theta_{(n)},l_{(n)}) &= S^{n-1}_{\theta_{n}^-}(\theta_{(n-1)},l_{(n-1)})*( {\bf 1}_m +\beta^{n-1}_{\theta_n}(\theta_{(n-1)},l_{(n-1)},l_n)).
    \end{aligned} \right.
    \end{equation}
Here, for any $x=(x_1,\dots,x_m)'\in \mathbb{R}^m$ and $y = (y_1,\dots,y_m)' \in \mathbb{R}^{m}$, the product  $x*y $ is given by  $(x_1y_1,\dots,x_my_m)'\in \mathbb{R}^{m}$.

\begin{remark}
   By introducing the random measure $N(\cdot,\cdot)$ associated with the default times and loss values $(T_n,L_n)_{n=1}^m$ by, for any $t\geq 0$ and $B\in \mathcal{B}(E)$,
    \begin{equation}
    \label{eq:N}
       N([0,t]\times B ) = \sum_{n=1}^m \mathbbm{1}_{ \{ T_n \leq t \}}\mathbbm{1}_{\{L_n\in B\} }.
    \end{equation}
Using \eqref{eq:N}, we can also  represent \eqref{eq:S:decompose}-\eqref{eq:Sn} as follows: for $t\geq 0$,
\begin{equation}
    \label{eq:S}
        dS_t = S_{t}*(\mu_t dt + \sigma_t dW_t) +  \int_E S_{t^-}*\beta_t(l) N(dt,dl).
    \end{equation}
 Although \eqref{eq:S} is more succinct than \eqref{eq:S:decompose}–\eqref{eq:Sn}, it conceals the detailed information about the jump structure. Our aim here is to explore this concrete jump structure through more explicit results; therefore, we will work with \eqref{eq:S:decompose}–\eqref{eq:Sn}. \hfill $\square$     
\end{remark}

Throughout this article, we impose Assumption \ref{ass:bound} below on the market parameters. 
\begin{assumption}
\label{ass:bound}
For any $n=0,1,\dots,m$,
    \begin{enumerate}
        \item    $\mu^n \in \mathcal{S}_\mathcal{P}(\Delta_n,E^n;\mathbb{R}^m)$, $\sigma^n  \in \mathcal{S}_\mathcal{P}(\Delta_n,E^n;\mathbb{R}^{m\times d})$, and  $\beta^n \in \mathcal{S}_\mathcal{P}(\Delta_n,E^n,E;\mathbb{R}^m)$;
        \item for any $(\theta_{(n)},l_{(n)})\in \Delta_n\times E^n$ and $t\geq \theta_n$, $\sigma^n_t(\theta_{(n)},l_{(n)})$ is a full-rank matrix.
    \end{enumerate}
\begin{remark}
    The condition $\beta^{n,i}_t(\theta_{(n)},l_{(n)},l) > -1$ $\mathbb{P}$-a.s.~for any $n=0,\dots,m-1$, $i=1,\dots,m$,  $(\theta_{(n)},l_{(n)},l) \in \Delta_n \times E^n\times E$ and $t\geq \theta_n$, can be relaxed to a non-strict inequality. In that case, the price of the $i$-th risky asset can hit zero when $\beta^{n,i}_t(\theta_{(n)},l_{(n)},l)=-1$, after which the asset will have no value and cannot be traded in a meaningful way, i.e., investing in this asset will not affect the investor's wealth. To accommodate the reduced number of effectively tradable assets, we can modify the process $\mu$ by setting its $i$-th entry to zero. Likewise, the $i$-th row of the volatility process $\sigma$ can be set to zero, so that the remaining $(m-1)\times d$ block matrix of $\sigma$, which is formed by removing its $i$-th row, retains full rank; see Remark 2.2 in \cite{pham:2013:AOAP}. For mathematical simplicity and notational convenience, we assume that $\beta^{n,i}_t(\theta_{(n)},l_{(n)},l)> -1$ $\mathbb{P}$-a.s.\hfill $\square$
\end{remark}




\end{assumption}



Let $\pi:=(\pi_t)_{t\geq 0}$ be the vector of the amount invested into the $m$ risky assets, which admits the following decomposition:
    \begin{equation}
    \label{eq:pi:decompose}
        \pi_t = \sum_{n=0}^{m-1} \pi^n_t(T_{(n)},L_{(n)}) \mathbbm{1}_{\{T_n < t \leq T_{n+1}\} } +  \pi^m_t(T_{(m)},L_{(m)}) \mathbbm{1}_{\{t > T_m\}} ,
    \end{equation}
where for any $n=0,\dots,m$, $\pi^n(\cdot,\cdot) = (\pi^n_t(\cdot,\cdot))_{t\geq 0}\in \mathcal{P}_\mathbb{F}(\Delta_n,E^n;\mathbb{R}^m)$  takes values in $\Pi_n \subseteq \mathbb{R}^m$, where $\Pi_n$ is a closed convex set representing the set of investment constraints. For convenience in the subsequent calculations, we assume that ${\bf 0} \in \cap_{n=0}^m\Pi_n$ in the rest of the article. However, we remark that all results remain valid even without this assumption.

Using the strategy $\pi$, the wealth process $X^\pi:=(X_t^\pi)_{t\geq 0}$ admits the following decomposition: for $t\geq 0$,
    \begin{equation}
    \label{eq:X:decompose}
               X^\pi_t = \sum_{n=0}^{m-1}X^{\pi,n}_t(T_{(n)},L_{(n)}) \mathbbm{1}_{\{T_n \leq  t < T_{n+1}\} } + X^{\pi,m}_t(T_{(m)},L_{(m)}) \mathbbm{1}_{\{t \geq  T_m\}} ,
    \end{equation}
where for $n=0,\dots,m$, the indexed processes $X^{\pi,n}(\cdot,\cdot) \in \mathcal{O}_\mathbb{F}(\Delta_n,E^n;\mathbb{R})$ are defined as follows: for $t\geq 0$,
    \begin{equation}
    \label{eq:X0}
        dX^{\pi,0}_t(0) = (\pi^0_t(0))'\sigma^0_t(0)\left(\alpha^0_t(0) dt + dW_t \right),
    \end{equation}
and for any $n=1,\dots,m$, $X^{\pi,n} = (X^{\pi,n}_t(\cdot,\cdot))_{t\geq 0}$ is governed by the following SDE: for any $(\theta_{(n)},l_{(n)})\in \Delta_{n}\times E^{n}$,
    \begin{equation}
   \left\{ \begin{aligned}
            \label{eq:Xn}
        dX^{\pi,n}_t(\theta_{(n)},l_{(n)}) &= \left(\pi^{n}_t(\theta_{(n)},l_{(n)})\right)'\sigma_t^{n}(\theta_{(n)},l_{(n)})\left(\alpha_t^{n}(\theta_{(n)},l_{(n)})dt + dW_t\right), \ t \geq \theta_n, \\
        X^{\pi,n}_{\theta_{n}}(\theta_{(n)},l_{(n)}) &= X^{\pi,n-1}_{\theta_{n}^-}(\theta_{(n-1)},l_{(n-1)}) \\
        &\quad + \left(\pi_{\theta_n}^{n-1}(\theta_{(n-1)},l_{(n-1)})\right)'\beta_{\theta_n}^{n-1}(\theta_{(n-1)},l_{(n-1)},l_n) .
    \end{aligned} \right.
    \end{equation}
Here, $\alpha^n(\cdot,\cdot)   \in \mathcal{P}_\mathbb{F}(\Delta_n,E^n;\mathbb{R}^d)$ represents the \textit{market price of risk}; for any $(\theta_{(n)},l_{(n)})\in \Delta_n\times E^n$  and $t\geq \theta_n$,
    \begin{equation*}
    \sigma^n_t(\theta_{(n)},l_{(n)})\alpha^n_t(\theta_{(n)},l_{(n)}) = \mu^n_t(\theta_{(n)},l_{(n)}), 
    \end{equation*}
{and is given by 
    \begin{equation*}
        \alpha^n_t(\theta_{(n)},l_{(n)}) = \sigma^n(\theta_{(n)},l_{(n)})'\left(\sigma^n(\theta_{(n)},l_{(n)})\sigma^n(\theta_{(n)},l_{(n)})' \right)^{-1}\mu^n_t(\theta_{(n)},l_{(n)}). 
    \end{equation*}
}%
    Combining \eqref{eq:X:decompose}-\eqref{eq:Xn}, the process $X^\pi$ can be represented by the following dynamics: for any $t\geq 0$,
    \begin{equation}
    \label{eq:X}
        dX^{\pi}_t = (\pi_t)'\sigma_t\left( \alpha_t dt + dW_t\right) + \int_E (\pi_t)'\beta_t(l)    N(dt,dl),
    \end{equation}
where
     \begin{equation}
    \label{eq:model:parameter:decompose:2}
        \begin{aligned}
            \alpha_t &= \sum_{n=0}^{m-1} \alpha^n_t(T_{(n)},L_{(n)}) \mathbbm{1}_{ \{T_n < t \leq T_{n+1} \} } + \alpha^m_t(T_{(m)},L_{(m)}) \mathbbm{1}_{\{t > T_m\}}.
        \end{aligned}
    \end{equation}

We suppose that there exists a conditional density for $(T_{(m)},L_{(m)})$ with respect to the filtration $\mathbb{F}$:

\begin{assumption}
\label{ass:density}
    There exists  $\eta(\cdot,\cdot) \in \mathcal{O}_\mathbb{F}(\Delta_m,E^m;\mathbb{R}_+)$ such that, for any $t\geq 0$ and bounded measurable function $g$ on $\Delta_m\times E^m$,
        \begin{equation*}
            \mathbb{E}[g(T_{(m)},L_{(m)})|\mathcal{F}_t] = \int_{E^m\times \Delta_m} g(\theta_{(m)},l_{(m)}) \eta_t(\theta_{(m)},l_{(m)}) d\theta_{(m)}\boldsymbol{\lambda}(dl_{(m)} ),
        \end{equation*}
    where for $n=1,\dots,m$, $$d\theta_{(n)}:= d\theta_n\cdots d\theta_1 \text{ and } \boldsymbol{\lambda}(dl_{(n)}) :=\lambda_1(dl_1) \prod_{i=1}^{n-1}\lambda_{i+1}(l_{(i)},dl_{i+1});$$ here,  $\lambda_1(dl_1)$ is a non-negative Borel measure on $E$, and for $n=1,\dots,m-1$, $\lambda_{n+1}(l_{(n)},dl_{n+1})$ is a {probability} transition kernel on $E^n\times E$. For notational convenience, we may also write $\lambda_1(l_0,dl_1)$ to represent $\lambda_1(dl_1)$ in the sequel.
\end{assumption}
This density assumption, also known as the (\textbf{H}') hypothesis, ensures that any $\mathbb{F}$-semimartingale is also a $\mathbb{G}$-semimartingale; see, such as, \cite{PHAM20101795}. 
Finally, we  introduce the following notation: for $n=0,1,\dots,m-1$, we define $ \hat{\eta}^n(\cdot,\cdot) \in \mathcal{O}_\mathbb{F}(\Delta_n,E^n;\mathbb{R}_+)$ by, for any $(\theta_{(n)},l_{(n)})\in \Delta_n\times E^n$ and {$t\geq 0$},
    \begin{equation}
    \label{eq:hat:eta}
    \begin{aligned}
                &\ \hat{\eta}^n_t(\theta_{(n)},l_{(n)}) \\
                :=&\ \int_{E^{m-n}}\int_t^\infty \int_{\theta_{n+1}}^\infty \cdots \int_{\theta_{m-1}}^\infty \eta_t\left(\theta_{(m)},l_{(m)}\right) d\theta_m\cdots d\theta_{n+1} \prod_{j=n}^{m-1} \lambda_{j+1}(l_{(j)},dl_{j+1}).
    \end{aligned}
    \end{equation}
The function $\hat{\eta}^n(\cdot,\cdot)$, $n=0,\dots,m-1$, can be interpreted as the survival density of $T_{n+1}$ conditional on $\mathbb{F}$: for any $t\geq 0$,
\begin{equation*}
    \begin{aligned}
        \mathbb{P}(T_1>t|\mathcal{F}_t) &= \hat{\eta}^0_t(0), \\
        \mathbb{P}(T_{n+1}>t|\mathcal{F}_t) &= \int_{E^n}\int_{\Delta_n} \hat{\eta}_t^n(\theta_{(n)},l_{(n)}) d\theta_{(n)}\boldsymbol{\lambda}(dl_{(n)}).
    \end{aligned}
    \end{equation*}
    In addition, we take the convention $\hat{\eta}^m(\cdot,\cdot) := \eta(\cdot,\cdot)$. 

    

 We assume that the survival density functions are positive, 
 which later allows us to construct a forward utility preference by scaling the decomposition of the random field; see \eqref{eq:U:hat:U} below.
    \begin{assumption}
    \label{ass:eta:zero}
    For any $(\theta_{(m)},l_{(m)})\in \Delta_m\times E^m$, and $t\geq \theta_m$, $\eta_t(\theta_{(m)},l_{(m)})>0$.

        \end{assumption}
In particular, by the definition of the survival density functions,  Assumption \ref{ass:eta:zero} implies that $\hat{\eta}^n_t(\theta_{(n)},l_{(n)})>0$, for any $n=0,\dots,m$, $(\theta_{(n)},l_{(n)})\in \Delta_n\times E^n$, and $t\geq \theta_n$.



\section{Forward Utility Preference in Defaultable Markets}
\label{sec:forward}

In this section, we review the definition of forward utility preferences, which is formulated in terms of the (super)martingale property under the market filtration $\mathbb{G}$. As the Brownian motion $W$ is not necessarily a martingale under $\mathbb{G}$, we need to project the martingale condition onto the default-free filtration $\mathbb{F}$ via the Jacod-Pham decomposition. In Section~\ref{section2.6}, we provide a characterization of the $\mathbb{G}$-(super)martingale property under the filtration $\mathbb{F}$. Building on this, Section~\ref{sec:G-forward-F} introduces and characterizes $\mathbb{G}$-forward preferences under $\mathbb{F}$.

 \subsection{Characterization of $\mathbb{G}$-Martingales under the Filtration $\mathbb{F}$ }\label{section2.6}

The upcoming Lemma \ref{lem:M:G:F:martingale} characterizes the $\mathbb{G}$-martingale property in terms of $\mathbb{F}$, which is proved with the aid of the following Lemma \ref{lem:M:inductive}. Their proofs are deferred to Appendix~\ref{sec: proof Sec 2.6}.

\begin{lemma}
\label{lem:M:inductive}
    Let $M=(M_t)_{t\geq 0}$ be an $\mathbb{R}$-valued, $\mathbb{G}$-optional and integrable process (i.e., $\mathbb{E}[|M_t|]<\infty$ for any $t\geq 0$) with the following Jacod-Pham decomposition:
        \begin{equation*}
            M_t = \sum_{n=0}^{m-1} M^n_t\left(T_{(n)},L_{(n)}\right)\mathbbm{1}_{ \{ T_n \leq t < T_{n+1} \} } + M^m_t(T_{(m)},L_{(m)}) \mathbbm{1}_{\{ t\geq T_m\} },
        \end{equation*}
    where for any $n=0,\dots,m$, $M^n(\cdot,\cdot)\in \mathcal{O}_\mathbb{F}(\Delta_n,E^n;\mathbb{R})$. Suppose that 
    $(M_t)_{t\geq 0}$ satisfies the following recursive relation: for any $n=0,\dots,m-1$, $(\theta_{(n)},l_{(n)})\in \Delta_{n}\times E^{n}$  and any $s\geq t \geq \theta_n$,
        \begin{equation}
        \label{eq:M:G:F:martingale}
        \begin{aligned}
                     &\    M^n_t\left(\theta_{(n)},l_{(n)}\right)  \hat{\eta}^n_t(\theta_{(n)},l_{(n)})\\ \geq&\  \mathbb{E}\bigg[ M^n_s\left(\theta_{(n)},l_{(n)}\right) \hat{\eta}^n_s\left(\theta_{(n)},l_{(n)}\right) \\ &\quad + \int_E\int_t^s M^{n+1}_{\theta_{n+1}}\left((\theta_{(n)},\theta_{n+1}),(l_{(n)},l_{n+1})\right) \\
                        &\quad \cdot  \hat{\eta}^{n+1}_{\theta_{n+1}}((\theta_{(n)},\theta_{n+1}),(l_{(n)},l_{n+1}))d\theta_{n+1}\lambda_{n+1}(l_{(n)},dl_{n+1})   \big| \mathcal{F}_t  \bigg],
        \end{aligned}
        \end{equation}
    and for $n=m$, $(\theta_{(m)},l_{(m)})\in \Delta_m\times E^m$, and  $s\geq t\geq \theta_m$,
        \begin{equation}
        \label{eq:M:m:martingale}
            M^m_t\left(\theta_{(m)},l_{(m)}\right)   {\eta}_t(\theta_{(m)},l_{(m)}) \geq \mathbb{E}\left[M^m_s\left(\theta_{(m)},l_{(m)}\right)   {\eta}_s(\theta_{(m)},l_{(m)})\big|\mathcal{F}_t \right].
        \end{equation}
    Then, for any $n=1,2,\dots, m$, $(\theta_{(n-1)},l_{(n-1)})\in \Delta_{n-1}\times E^{n-1}$, and for any $s\geq t \geq \theta_{n-1}$,
        \begin{align}
            \label{eq:M:lemma}
           &\ \mathbb{E}\bigg[\int_E\int_t^s M^n_s\left((\theta_{(n-1)},\theta_n),(l_{(n-1)},l_n) \right)  \hat{\eta}^n_s\left((\theta_{(n-1)},\theta_n),(l_{(n-1)},l_n) \right)  \nonumber \\
           &\quad d\theta_n\lambda_{n}(l_{(n-1)},dl_{n}) \big| \mathcal{F}_t    \bigg]\nonumber \\
           &\ + \mathbb{E}\bigg[ \int_E\int_t^s \bigg\{ \sum_{j=n+1}^m \int_{E^{j-n}} \int_{\theta_n}^s\cdots \int_{\theta_{j-1}}^s  M^j_s\left((\theta_{(n-1)},\theta_{(n,j)}),(l_{(n-1)},l_{(n,j)}) \right) \nonumber \\
           &\quad \cdot  \hat{\eta}^{j}_s((\theta_{(n-1)},\theta_{(n,j)}),(l_{(n-1)},l_{(n,j)})) d\theta_j\cdots d\theta_{n+1} \prod_{i=n}^{j-1} \lambda_{i+1}(l_{(i)},dl_{i+1}) \bigg\} \nonumber\\
           &\quad d\theta_n \lambda_n(l_{(n-1)},dl_n) \big| \mathcal{F}_t \bigg]\nonumber \\
           \leq &\ \mathbb{E}\bigg[\int_E \int_t^s M^n_{\theta_n}\left((\theta_{(n-1)},\theta_n),(l_{(n-1)},l_n) \right)\hat{\eta}^n_{\theta_n}\left((\theta_{(n-1)},\theta_n),(l_{(n-1)},l_n) \right)   \nonumber\\
           &\quad d\theta_n \lambda_n(l_{(n-1)},dl_n) \big| \mathcal{F}_t \bigg],
        \end{align}
where, for any  $n=1,2,\dots, m-1$ and $j=n+1,n+2,\dots,m$, $\theta_{(n,j)}:=(\theta_n,\dots,\theta_j)$, $l_{(n,j)}:= (l_n,\dots,l_j)$. In addition, equality holds in \eqref{eq:M:lemma} if equalities hold in \eqref{eq:M:m:martingale}, and in \eqref{eq:M:G:F:martingale} for all $n=0,\dots,m-1$.
\end{lemma}

\begin{proof} See  Appendix \ref{sec:pf:lem:M:inductive}.
\end{proof}

The result indicates that, if the Jacod-Pham decomposition of $M$ exhibits the recursive properties \eqref{eq:M:G:F:martingale} and \eqref{eq:M:m:martingale}, its conditional expectation under $\mathbb{F}$ after the $(n-1)$-th default can be bounded by the conditional expectation of the single component $M^{n}(\cdot,\cdot)$.  This recursive characterization of the tail components of $M$ plays a key role in projecting the (super)martingale property in $\mathbb{G}$ on $\mathbb{F}$, as established in the next lemma.

\begin{lemma}
\label{lem:M:G:F:martingale}
    An $\mathbb{R}$-valued, $\mathbb{G}$-optional and integrable process $(M_t)_{t\geq 0}$ is a $\mathbb{G}$-supermartingale if its Jacod-Pham decomposition satisfies \eqref{eq:M:G:F:martingale} and \eqref{eq:M:m:martingale}. In addition, it is a $\mathbb{G}$-martingale  if equalities hold in \eqref{eq:M:G:F:martingale} and \eqref{eq:M:m:martingale}.
\end{lemma}

\begin{proof} See  Appendix \ref{sec:pf:lem:M:G:F:martingale}.
\end{proof}

\begin{remark}
   \label{rmk:M:necessary}
It can be proved that the equalities in \eqref{eq:M:G:F:martingale} and \eqref{eq:M:m:martingale} are not only sufficient but also necessary conditions for $(M_t)_{t \geq 0}$ to be a $\mathbb{G}$-martingale. In contrast, the inequality in \eqref{eq:M:G:F:martingale} may be slightly stronger than the minimal condition required for the supermartingale property.

Indeed, by following the proof of Lemma~\ref{lem:M:G:F:martingale}, one can show that if $(M_t)_{t \geq 0}$ is a $\mathbb{G}$-supermartingale, then \eqref{eq:M:m:martingale} necessarily holds, and for any $n = 0, \dots, m-1$, the following inequality holds for $\mathbb{P} \otimes d\theta_{(n)} \otimes \prod_{j=0}^{n-1} \lambda_{j+1}(l_{(j)}, dl_{j+1})$-almost all $(\omega, \theta_{(n)}, l_{(n)}) \in \Omega \times \Delta_n \times E^n$, and any $s \geq t \geq \theta_n$:
   \begin{align*}
       &\  M^n_t\left(\theta_{(n)},l_{(n)}\right) \hat{\eta}^n_t\left(\theta_{(n)},l_{(n)}\right) \\
       \geq &\   \mathbb{E}\Bigg[M^n_s\left(\theta_{(n)},l_{(n)} \right)\hat{\eta}^n_s\left(\theta_{(n)},l_{(n)} \right)   + \int_E \int_t^s \Bigg(  M^{n+1}_s\left(\theta_{(n+1)},l_{(n+1)} \right)\\
       &\quad \cdot \hat{\eta}^{n+1}_s\left(\theta_{(n+1)},l_{(n+1)} \right)  + \sum_{j=n+2}^m \int_{E^{j-n-1}} \int_{\theta_{n+1}}^s\cdots \int_{\theta_{j-1}}^s M^{j}_s\left(\theta_{(j)},l_{(j)} \right)\hat{\eta}^{j}_s\left(\theta_{(j)},l_{(j)} \right)   \nonumber \\
       &\quad  d\theta_j\cdots d\theta_{n+2}\prod_{i=n+2}^{j}\lambda_i(l_{(i-1)},dl_i) \Bigg)d\theta_{n+1}\lambda_{n+1}(l_{(n)},dl_{n+1})  \bigg|\mathcal{F}_t \Bigg].
    \end{align*} \hfill$\square$
\end{remark}

\subsection{$\mathbb{G}$-Forward Utility Preference and Characterizations under $\mathbb{F}$}
\label{sec:G-forward-F}
    We briefly review the definition of a forward utility preference. In the sequel, we let $\mathcal{A}$ be an admissible set of strategies, which is a subset of all $\mathbb{G}$-predictable processes,  subject only to an integrability condition on the resulting forward random field. The exact formulation of $\mathcal{A}$ shall be introduced in Section \ref{sec:verification}, {\color{black}and in Theorem \ref{thm:ergodic:pi:value} under the stochastic factor model.}

\begin{definition}
\label{def:forward}
    Let $u_0(\cdot)$ be a strictly increasing and strictly concave function.
    A random field $U=(U_t(x;\omega))_{x\in \mathbb{R},t\geq 0, \omega\in\Omega}$ is a forward utility preference with initial condition $u_0$ for the wealth process $X^\pi=(X^\pi_t)_{t\geq 0}$ with $\pi\in \mathcal{A}$, if it satisfies the following conditions:
    \begin{enumerate}
        \item $U_0(\cdot)\equiv u_0(\cdot)$;
        \item for any $x\in\mathbb{R}$, $U_\cdot(x)$ is $\mathbb{G}$-progressively measurable;
        \item for any $t\geq 0$, $x\mapsto U_t(x)$ is strictly increasing and strictly concave $\mathbb{P}$-almost surely;
        \item for any $\pi\in \mathcal{A}$, and for any $0\leq t\leq s$,
            \begin{equation*}
                U_t(X^\pi_t) \geq \mathbb{E}\left[U_s(X^\pi_s) | \mathcal{G}_t \right];
            \end{equation*}
        \item there exists a $\pi^*\in \mathcal{A}$ such that, for any $0\leq t\leq s$,
         \begin{equation*}
                U_t(X^{\pi^*}_t) = \mathbb{E}\left[U_s(X^{\pi^*}_s) | \mathcal{G}_t \right].
            \end{equation*}
    \end{enumerate}

\end{definition}

In the sequel, we shall construct a forward utility preference for $X^\pi$ based on the Jacod-Pham decomposition, where the random field $(U_t(x))_{x\in \mathbb{R},t\geq 0}$ can be represented as
    \begin{equation}
    \label{eq:U:decompose}
        U_t(x) = \sum_{n=0}^{m-1} U^n_t\left(x,T_{(n)},L_{(n)}\right)\mathbbm{1}_{ \{ T_n \leq t < T_{n+1} \} } + U^m_t\left(x,T_{(m)},L_{(m)}\right)\mathbbm{1}_{ \{ t\geq T_m \} }.
    \end{equation}
Here, for any $n=0,\dots,m$ and $x\in \mathbb{R}$,  $U^n(x,\cdot,\cdot) \in \mathcal{O}_\mathbb{F}(\Delta_n,E^n;\mathbb{R})$.

Using Lemma \ref{lem:M:G:F:martingale}, a forward utility preference in $\mathbb{G}$ can be constructed by its Jacod-Pham decomposition in $\mathbb{F}$:
\begin{theorem}
\label{thm:U:G:F}
    A random field $U=(U_t(x;\omega))_{x\in \mathbb{R},t\geq 0, \omega\in \Omega}$ is a forward utility preference for the process $X^\pi$ defined in \eqref{eq:X0}-\eqref{eq:Xn} with $\pi \in \mathcal{A}$ if it satisfies Properties 1-3 in Definition \ref{def:forward}, along with the following:
        \begin{enumerate}
            \item[4'.] for any $\pi \in \mathcal{A}$, $(\theta_{(m)},l_{(m)})\in \Delta_m\times E^m$ and $s\geq t\geq \theta_m$,
                    \begin{equation}
                    \begin{aligned}
                         &\ U^m_t\left(X^{\pi,m}_t (\theta_{(m)},l_{(m)}),\theta_{(m)},l_{(m)}\right)\eta_t(\theta_{(m)},l_{(m)})   \\
                         \geq &\ \mathbb{E}\left[U^m_s\left(X^{\pi,m}_s (\theta_{(m)},l_{(m)}),\theta_{(m)},l_{(m)}\right)\eta_s(\theta_{(m)},l_{(m)}) \big| \mathcal{F}_t\right],
                    \end{aligned}
                    \label{eq:U^m_condition}
                    \end{equation}
             and  for any $n=0,\dots,m-1$, $(\theta_{(n)},l_{(n)})\in \Delta_n\times E^n$, and $s\geq t\geq \theta_n$,
              \begin{equation}
                    \begin{aligned}
                         &\ U^n_t\left(X^{\pi,n}_t (\theta_{(n)},l_{(n)}),\theta_{(n)},l_{(n)}\right)\hat{\eta}^n_t(\theta_{(n)},l_{(n)})   \\
                         \geq &\ \mathbb{E}\bigg[U^n_s\left(X^{\pi,n}_s (\theta_{(n)},l_{(n)}),\theta_{(n)},l_{(n)}\right)\hat{\eta}^n_s(\theta_{(n)},l_{(n)}) \\
                         &\quad + \int_E\int_t^s U^{n+1}_{\theta_{n+1}}\left( X^{\pi,n+1}_{\theta_{n+1}}\left(\theta_{(n+1)},l_{(n+1)}\right)
 ,\theta_{(n+1)} ,l_{(n+1)} \right) \\
 &\quad  \hat{\eta}^{n+1}_{\theta_{n+1}}\left((\theta_{(n)},\theta_{n+1}),(l_{(n)},l_{n+1}) \right)     d\theta_{n+1}\lambda_{n+1}(l_{(n)},dl_{n+1}) \big| \mathcal{F}_t\bigg];
                    \end{aligned}
                    \label{eq:U^n_condition}
                    \end{equation}

            \item[5'] there exists $\pi^* \in \mathcal{A}$ such that the equalities in \eqref{eq:U^m_condition}, and \eqref{eq:U^n_condition} for all $n=0,\dots,m-1$, hold by such $\pi^*$.

        \end{enumerate}
 In addition, $U$ is a forward utility preference implies that Property 5' holds.
\end{theorem}

Motivated by Theorem \ref{thm:U:G:F}, we let $\hat{U}=(\hat{U}_t(x;\omega))_{x\in\mathbb{R}, t\geq 0, \omega\in\Omega}$ be a random field in $\mathbb{G}$ with the following decomposition:
    \begin{equation}
    \label{eq:hat:U}
        \hat{U}_t(x) = \sum_{n=0}^{m-1} \hat{U}^n_t\left(x,T_{(n)},L_{(n)}\right)\mathbbm{1}_{ \{ T_n \leq t < T_{n+1} \} } + \hat{U}^m_t\left(x,T_{(m)},L_{(m)}\right)\mathbbm{1}_{ \{ t\geq T_m \} },
    \end{equation}
where for each $n=0,\dots,m$ and $x\in \mathbb{R}$, $\hat{U}^n(x,\cdot,\cdot)\in \mathcal{O}_{\mathbb{F}}(\Delta_n,E^n;\mathbb{R})$ is defined by, for any  $(\theta_{(n)},l_{(n)})\in \Delta_n\times E^n$, and $t\geq \theta_n$,
    \begin{equation}
    \label{eq:hat:U:n}
        \hat{U}^n_t\left(x,\theta_{(n)},l_{(n)}\right) := U^n_t\left(x,\theta_{(n)},l_{(n)}\right)\hat{\eta}^n_t\left(\theta_{(n)},l_{(n)}\right).
    \end{equation}
For any $\pi\in \mathcal{A}$, we also define, for $n=0,1,\dots,m$, $(\theta_{(n)},l_{(n)}) \in \Delta_n\times E^n$, and $t\geq \theta_n$,
   \begin{equation}
    \label{eq:hat:U:martingale}
        \begin{aligned}
V^{\pi,n}_t\left(\theta_{(n)},l_{(n)}\right) &:=  \hat{U}^n_t\left(X^{\pi,n}_t(\theta_{(n)},l_{(n)}),\theta_{(n)},l_{(n)}\right) \\
&\  + \int_E\int_{\theta_n}^t \hat{U}^{n+1}_{\theta_{n+1}}\left(X^{\pi,n+1}_{\theta_{n+1}}(\theta_{(n+1)},l_{(n+1)}),\theta_{(n+1)},l_{(n+1)} \right) \\
&\quad d\theta_{n+1}\lambda_{n+1}(l_{(n)},dl_{n+1}),\text{ for $n<m$}, \\
V^{\pi,m}_t\left(\theta_{(m)},l_{(m)}\right) &:=  \hat{U}^m_t\left(X^{\pi,m}_t(\theta_{(m)},l_{(m)}),\theta_{(m)},l_{(m)}\right),\text{ for $n=m$}.
        \end{aligned}
    \end{equation}
Then,  Properties 4' and 5' in Theorem \ref{thm:U:G:F} are equivalent to requiring that the indexed process $(V^{\pi,n}_t(\theta_{(n)},l_{(n)}))_{t\geq \theta_n}$  
is a $\mathbb{F}$-supermartingale for any $n=0,\dots,m$,  $(\theta_{(n)},l_{(n)})\in \Delta_n\times E^n$, $t\geq \theta_n$, and any admissible strategy $\pi\in \mathcal{A}$, which becomes a true $\mathbb{F}$-martingale under an optimal strategy $\pi^*\in \mathcal{A}$. In the sequel, we shall construct a forward utility preference $U$ in $\mathbb{G}$ through the decomposition of $\hat{U}$ in $\mathbb{F}$. {By Assumption \ref{ass:eta:zero}}, the preference $U$ can be retrieved by utilizing the relationship of the decompositions of $U$ and $\hat{U}$: for any $n=0,\dots,m$, $x\in \mathbb{R}$, $(\theta_{(n)},l_{(n)})\in \Delta_n\times E^n$ and $t\geq \theta_n$,
    \begin{equation}
    \label{eq:U:hat:U}
        U^n_t\left(x,\theta_{(n)},l_{(n)}\right) = \frac{  \hat{U}^n_t\left(x,\theta_{(n)},l_{(n)}\right)}{  \hat{\eta}^n_t\left(\theta_{(n)},l_{(n)}\right)}.
    \end{equation}

\section{Exponential Forward Utility}
\label{sec:exp}
In this section, we propose a forward utility for the wealth process \eqref{eq:X}, with the initial condition $u_0$ given by
    \begin{equation*}
        u_0(x) =-e^{-\gamma x}, \ x\in\mathbb{R},
    \end{equation*}
where $\gamma>0$.

\textcolor{black}{To do so, we first introduce recursive infinite horizon BSDEs, for which we provide their well-posedness and boundedness results, and then proceed with the construction of the forward utility and conclude with a verification result.}

\subsection{Recursive Infinite Horizon BSDEs}\label{sec:3.1}
The construction grounds on a $\mathbb{G}$-optional process $Y=(Y_t)_{t\geq 0}$ with the following Jacod-Pham decomposition:
    \begin{equation}
    \label{eq:Y:decompose}
        Y_t = \sum_{n=0}^{m-1}Y^n_t(T_{(n)},L_{(n)}) \mathbbm{1}_{ \{ T_n \leq t < T_{n+1} \} } + Y^m_t(T_{(m)},L_{(m)}) \mathbbm{1}_{  \{  t \geq T_m  \} },
    \end{equation}
where for any $n=0,\dots,m$, $Y^n(\cdot,\cdot) \in \mathcal{O}_\mathbb{F}(\Delta_n,E^n;\mathbb{R})$ satisfies an indexed BSDE, which is defined recursively as follows: for  $\left(\theta_{(m)},l_{(m)}\right) \in \Delta_m\times E^m$, and $t\geq \theta_m$,
    \begin{equation}
        \label{eq:Ym}
        \begin{aligned}
            dY^{m}_t(\theta_{(m)},l_{(m)}) &= \left( \rho Y^{m}_t(\theta_{(m)},l_{(m)}) - \min_{\pi\in\Pi_m} f^m\left(t,\pi,Z^{m}_t(\theta_{(m)},l_{(m)}),\theta_{(m)},l_{(m)}\right) \right)dt \\
            &\quad + Z^{m}_t(\theta_{(m)},l_{(m)})'dW_t,
        \end{aligned}
    \end{equation}
and for $n=m-1,\dots,0$, $(\theta_{(n)},l_{(n)})\in \Delta_n\times E^n$ and $t \geq \theta_n$,
    \begin{equation}
\label{eq:Yn}
\begin{aligned}
    dY^{n}_t(\theta_{(n)},l_{(n)}) &=\Bigg( \rho Y^{n}_t(\theta_{(n)},l_{(n)}) \\
    &\qquad - \min_{\pi\in\Pi_n} f^n\left(t,\pi,Y^{n}_t(\theta_{(n)},l_{(n)}),Z^{n}_t(\theta_{(n)},l_{(n)}),\theta_{(n)},l_{(n)}\right)  \Bigg) dt \\
    &\quad + Z^{n}_t(\theta_{(n)},l_{(n)})'dW_t.
\end{aligned}
\end{equation}
 Here, $\rho>0$ is a fixed constant, and the drivers of the BSDEs \eqref{eq:Ym}-\eqref{eq:Yn} are given by, for any $n=0,\dots,m$,  $(z,\pi,y)\in \mathbb{R}^d\times \mathbb{R}^m\times \mathbb{R}$, $(\theta_{(n)},l_{(n)})\in \Delta_n \times E^n$, and $t\geq \theta_n$,
    \begin{align*}
        f^m(t,\pi,z,\theta_{(m)},l_{(m)}) &= F^m_1(t,\pi,z,\theta_{(m)},l_{(m)}),\\
        f^n(t,\pi,y,z,\theta_{(n)},l_{(n)}) &= F^n_1(t,\pi,z,\theta_{(n)},l_{(n)}) +  F^n_2(t,\pi,y,\theta_{(n)},l_{(n)}),
    \end{align*}
for $n=m-1,\dots,0$, where
    \begin{align*}
                 F_1^n(t,\pi,z,\theta_{(n)},l_{(n)}) &:= -\pi' \mu_t^n(\theta_{(n)},l_{(n)}) + \frac{\gamma|z-\sigma_t^n(\theta_{(n)},l_{(n)} )'\pi|^2}{2} \\
                 &= \frac{\gamma}{2}\left|\sigma^n_t(\theta_{(n)},l_{(n)})' \pi -  \left(z + \frac{\alpha^n_t(\theta_{(n)},l_{(n)})}{\gamma} \right)  \right|^2  \\
                 &\quad - \alpha^n_t(\theta_{(n)},l_{(n)})'z - \frac{\left|\alpha^n_t(\theta_{(n)},l_{(n)})\right|^2}{2\gamma}, \\
              F^n_2(t,\pi,y,\theta_{(n)}, l_{(n)}) &:=  \frac{1}{\gamma} \int_E e^{\gamma (Y^{n+1}_t( (\theta_{(n)},t),(l_{(n)},l)) - y -\pi'\beta_t^n(\theta_{(n)},l_{(n)},l)  ) }   \lambda_{n+1}(l_{(n)},dl).
    \end{align*}

\begin{remark}
    \textcolor{black}{Using a single equation, we can express the process $Y$ by the solution of the following infinite-horizon BSDE}:
    \begin{equation*}
        dY_t = \left(\rho Y_t - f^*(t,Y_t,Z_t) \right)dt + Z_t'dW_t + \int_E G_t(l) N(dt,dl),
    \end{equation*}
where for any $t\geq 0$ and $l\in E$, 
    \begin{align*}
        Z_t &= \sum_{n=0}^{m-1}Z^n_t(T_{(n)},L_{(n)}) \mathbbm{1}_{ \{ T_n < t \leq T_{n+1} \} } + Z^m_t(T_{(m)},L_{(m)}) \mathbbm{1}_{  \{  t > T_m  \} },\\
        f^*(t,Y_t,Z_t) &= \sum_{n=0}^{m-1} \min_{\pi \in \Pi_n} f^n\left(t,\pi,Y_t^n(T_{(n)},L_{(n)}),Z_t^n(T_{(n)},L_{(n)}),T_{(n)},L_{(n)} \right) \mathbbm{1}_{\{ T_n < t \leq T_{n+1} \} } \\
        &\quad + \min_{\pi\in\Pi_m} f^m\left(t,\pi,Z_t^m(T_{(m)},L_{(m)}),T_{(m)},L_{(m)}\right) \mathbbm{1}_{\{ t > T_m \}}, \\
        G_t(l) &= \sum_{n=0}^{m-1} \left[Y^{n+1}_t\left( (T_{(n)},t),(L_{(n)},l) \right) - Y_t^n\left( T_{(n)},L_{(n)}\right) \right]  \mathbbm{1}_{\{ T_n < t \leq T_{n+1} \} } .
    \end{align*}
\hfill$\square$
\end{remark}


We establish the well-posedness of the indexed infinite-horizon BSDEs \eqref{eq:Ym}–\eqref{eq:Yn} in the following theorem.

\begin{theorem}
\label{thm:exist:Y}
   Under Assumptions \ref{ass:bound}-\ref{ass:density}, for every $n=0,\dots,m$, the indexed infinite horizon BSDE \eqref{eq:Ym}-\eqref{eq:Yn} admits a unique solution $(Y^n,Z^n)$, where $Y^n \in \mathcal{S}(\Delta_n,E^n;\mathbb{R})$ and $Z^n\in  \mathcal{M}^{2}(\Delta_n,E^n;\mathbb{R}^d)\cap \mathcal{L}^2_{\text{loc}}(\Delta_n,E^n;\mathbb{R}^d)$.
\end{theorem}

\begin{proof} See Appendix \ref{sec:pf:thm:exist:Y}. 
\end{proof}

We end this subsection by providing an explicit bound for $Y^n$, $n=0, 1,\dots,m$, which will be useful in Section \ref{sec:stochastic:factor}.  

\begin{proposition}
\label{pp:Yn:bdd:factor}
    Suppose that Assumptions \ref{ass:bound}-\ref{ass:density} hold. Then,   for any $n=0,\dots,m$, the unique bounded solution $(Y^n,Z^n)$  of the infinite-horizon BSDEs \eqref{eq:Ym} and \eqref{eq:Yn} satisfies $|Y^n_t(\theta_{(n)},l_{(n)})| \leq K_Y/\rho$ for all $(\theta_{(n)},l_{(n)})\in \Delta_n\times E^n$ and $t\geq \theta_n$, where 
        \begin{equation*}
            \textcolor{black}{K_{Y}} :=  \frac{1}{\gamma}
              \max\left\{ 1 ,\max_{0\leq n\leq m}  \frac{\|\alpha^n\|^2_{\mathcal{S}(\Delta_n,E^n;\mathbb{R}^m)}}{2} \right\}. 
        \end{equation*}
\end{proposition}

\begin{proof} See Appendix \ref{sec:pf:pp:Yn:bdd:factor}.
\end{proof}


By Theorem \ref{thm:U:G:F} and the discussion following it, we shall focus on the construction of the random field $\hat{U}$ defined in \eqref{eq:hat:U}-\eqref{eq:hat:U:n} such that the processes defined in \eqref{eq:hat:U:martingale} satisfy the relevant (super)martingale property under $\mathbb{F}$. To this end, we propose  the following \textit{ansatz} for the random field $\hat{U}$ by utilizing the solution of the BSDEs \eqref{eq:Ym}-\eqref{eq:Yn}: for any $n=0,\dots,m$, $x\in \mathbb{R}$,  $(\theta_{(n)},l_{(n)}) \in \Delta_n\times E^n$ and $t\geq \theta_{n}$,
    \begin{equation}
    \label{eq:Un}
    \begin{aligned}
        &\ \hat{U}^n_t\left(x,\theta_{(n)},l_{(n)}\right) \\
        =&\ -e^{-\gamma x }e^{\gamma\left( Y^n_t(\theta_{(n)},l_{(n)}) -   \sum_{j=0}^{n-1} \int_{\theta_j}^{\theta_{j+1}} \rho Y^{j}_s(\theta_{(j)},l_{(j)}) ds   - \int_{\theta_n}^t \rho Y^n_s(\theta_{(n)},l_{(n)})ds  - Y_0 \right)}.
    \end{aligned}
    \end{equation}
The \textit{ansatz} for the forward preference $U$ can then be retrieved by the relation \eqref{eq:U:hat:U}. Using the decompositions \eqref{eq:hat:U}, \eqref{eq:U:decompose} and  \eqref{eq:Y:decompose}, we can also write the \textit{ansatz} of $\hat{U}$ and $U$ collectively as follows: for $x\in \mathbb{R}$ and $t\geq 0$,
       \begin{equation}
    \label{eq:U:exp}
    \begin{aligned}
              \hat{U}_t(x) &= -e^{-\gamma x} e^{\gamma(Y_t - \int_0^t \rho Y_s ds - Y_0 )}, \\
        U_t(x) &= \frac{\hat{U}_t(x)}{\hat{\eta}_t},
    \end{aligned}
    \end{equation}
where $(\hat{\eta}_t)_{t\geq 0}$ is a $\mathbb{G}$-optional process with the following decomposition:
    \begin{equation}
        \hat{\eta}_t = \sum_{n=0}^{m-1} \hat{\eta}^n_t\left(T_{(n)},L_{(n)}\right) \mathbbm{1}_{\{T_n\leq t<T_{n+1}\}} + \eta_t\left(T_{(m)},L_{(m)}\right)\mathbbm{1}_{ \{t\geq T_m \} }, \ t\geq 0.
        \label{eq:eta_decomposition}
    \end{equation}
In particular, the initial condition of the forward utility is clearly met, since $\hat{\eta}^0_0(0) = 1$ and so $U_0(x)=\hat{U}_0(x) = u_0(x)$.

\subsection{Verification Theorem}
\label{sec:verification}
In Theorem \ref{thm:exist:Y}, we have established the unique existence of solutions of the indexed BSDEs \eqref{eq:Ym}-\eqref{eq:Yn}, and thus the random field \eqref{eq:U:exp} is well-defined. In this section, we verify that for any $n=0,\dots,m$ and $(\theta_{(n)},l_{(n)})\in \Delta_n\times E^n$, the process $(V^{\pi,n}_t(\theta_{(n)},l_{(n)}))_{t\geq \theta_n}$ defined in  \eqref{eq:hat:U:martingale} is a $\mathbb{F}$-supermartingale for any  $\pi\in \mathcal{A}$, and is a $\mathbb{F}$-martingale under an admissible  strategy $\pi^*\in\mathcal{A}$.    To be precise, using \eqref{eq:hat:U:martingale} and \eqref{eq:Un}, for $n=0,\dots,m$ and $\pi\in\mathcal{A}$, the process $(V^{\pi,n}_t(\theta_{(n)},l_{(n)}))_{t\geq \theta_n}$ is given by
 \begin{equation}
    \label{eq:hat:U:martingale:exp}
        \begin{aligned}
&\ V^{\pi,n}_t\left(\theta_{(n)},l_{(n)}\right)\\
=&\  -e^{-\gamma X^{\pi,n}_t(\theta_{(n)},l_{(n)})}  e^{\gamma\left( Y^n_t(\theta_{(n)},l_{(n)}) -   \sum_{j=0}^{n-1} \int_{\theta_j}^{\theta_{j+1}} \rho Y^{j}_s(\theta_{(j)},l_{(j)}) ds   - \int_{\theta_n}^t \rho Y^n_s(\theta_{(n)},l_{(n)})ds  - Y_0 \right)}  \\
&\  - \int_E\int_{\theta_n}^t e^{-\gamma X^{\pi,n+1}_{\theta_{n+1}}(\theta_{(n+1)},l_{(n+1)}) }e^{\gamma\left( Y^{n+1}_{\theta_{n+1}}(\theta_{(n+1)},l_{(n+1)}) -   \sum_{j=0}^{n} \int_{\theta_j}^{\theta_{j+1}} \rho Y^{j}_s(\theta_{(j)},l_{(j)}) ds    - Y_0 \right)}\\
&\qquad d\theta_{n+1}\lambda_{n+1}(l_{(n)},dl_{n+1}), \ n=m-1,\dots,0,  (\theta_{(n)},l_{(n)}) \in \Delta_n\times E^n, t\geq \theta_n, \\
&\ V^{\pi,m}_t\left(\theta_{(m)},l_{(m)}\right) \\
=&\  -e^{-\gamma X^{\pi,m}_t(\theta_{(m)},l_{(m)})}  e^{\gamma\left( Y^m_t(\theta_{(m)},l_{(m)}) -   \sum_{j=0}^{m-1} \int_{\theta_j}^{\theta_{j+1}} \rho Y^{j}_s(\theta_{(j)},l_{(j)}) ds   - \int_{\theta_m}^t \rho Y^m_s(\theta_{(m)},l_{(m)})ds  - Y_0 \right)}, \\ &\quad\quad  (\theta_{(m)},l_{(m)}) \in \Delta_m\times E^m, t\geq \theta_m,
        \end{aligned}
    \end{equation}
and the admissible set of strategies $\mathcal{A}$ will be defined  below.  Once the $\mathbb{F}$-(super)martingale property of $(V^{\pi,n}_t(\theta_{(n)},l_{(n)}))_{t\geq \theta_n}$ is established, it follows by Theorem \ref{thm:U:G:F} that   $(U_t(X^\pi_t))_{t\geq 0}$ is a $\mathbb{G}$-supermartingale for any $\pi\in \mathcal{A}$, and   $(U_t(X^{\pi^*}_t))_{t\geq 0}$ is a $\mathbb{G}$-martingale. This would thus prove that $(U_t(x))_{t\geq 0, x\in \mathbb{R}}$ is indeed a (exponential) forward utility.



 We define the set of admissible strategies as follows. For any $n=0,\dots,m$, let
    \begin{equation}
        \label{eq:An}
        \begin{aligned}
               \mathcal{A}_n &:= \bigg\{ \pi^n \in \mathcal{P}_\mathbb{F}(\Delta_n,E^n;\mathbb{R}^m): (V^{\pi,n}_{\tau}(\theta_{(n)},l_{(n)}))_{\tau\geq \theta_n} \text{ is $\mathbb{F}$-uniformly integrable} \\
        &\text{over $\mathbb{F}$-stopping times $\tau$},\   \pi^n_t(\theta_{(n)},l_{(n)}) \in \Pi_n \text{ for $(\theta_{(n)},l_{(n)})\in \Delta_n\times E^n$, $t\geq \theta_n$} \bigg\},
        \end{aligned}
    \end{equation}
where the process $V^{\pi,n}(\cdot,\cdot)$ is given by  \eqref{eq:hat:U:martingale:exp}. The uniform integrability condition is a specific requirement for constructing a forward exponential utility, which ensures that the associated Dol\'eans-Dade exponentials are  $\mathbb{F}$-martingales.
 Using \eqref{eq:An} and the Jacod-Pham decomposition \eqref{eq:pi:decompose},  we define the admissible set $\mathcal{A}$ by
    \begin{equation}
    \label{eq:A}
        \begin{aligned}
            \mathcal{A} := \bigg\{ \pi =(\pi_t)_{t\geq 0} &: \pi_t =  \sum_{n=0}^{m-1} \pi^n_t(T_{(n)},L_{(n)}) \mathbbm{1}_{\{T_n < t \leq T_{n+1}\} } + \pi^m_t(T_{(m)},L_{(m)})\mathbbm{1}_{\{t>T_m\} }, \\
            &\quad  \pi^n \in \mathcal{A}_n, \ n=0,1,\dots,m  \bigg\}.
        \end{aligned}
    \end{equation}

In the following, we deduce equations satisfied by \eqref{eq:hat:U:martingale:exp} in order to verify the $\mathbb{F}$-(super)martingale property.  For any $n = 0,\dots,m-1$, $(\theta_{(n)},l_{(n)})\in \Delta_n\times E^n$, by applying It\^o's lemma on the indexed process \eqref{eq:hat:U:martingale:exp},  we have, for any $t\geq \theta_n$,
    \begin{align*}
    &\
    d\Bigg(\hat{U}^n_t\left(X^{\pi,n}_t(\theta_{(n)},l_{(n)}),\theta_{(n)},l_{(n)}\right)\nonumber  \\
    &\quad + \int_E\int_{\theta_n}^t \hat{U}^{n+1}_{\theta_{n+1}}\Bigg( X^{\pi,n}_{\theta_{n+1}^-}(\theta_{(n)},l_{(n)}) + \pi^n_{\theta_{n+1}}(\theta_{(n)},l_{(n)})'\beta^n_{\theta_{n+1}}(\theta_{(n)},l_{(n)},l_{n+1}) ,\nonumber\\&\quad\quad\quad\quad\quad\quad\quad\quad\quad\quad\quad\quad\quad\quad\quad\quad\quad\quad\theta_{(n+1)},l_{(n+1)}\Bigg)d\theta_{n+1}\lambda_{n+1}(l_{(n)},dl_{n+1}) \Bigg)\nonumber\\
    =&\ \hat{U}^n_t\left(X^{\pi,n}_t(\theta_{(n)},l_{(n)}),\theta_{(n)},l_{(n)}\right)\Bigg( -\gamma \rho Y^n_t(\theta_{(n)},l_{(n)}) dt - \gamma dX^{\pi,n}_t(\theta_{(n)},l_{(n)}) \nonumber \\
    &\quad+ \gamma dY^n_t(\theta_{(n)},l_{(n)})  -  \gamma^2d\left\langle X^{\pi,n}(\theta_{(n)},l_{(n)}),Y^n(\theta_{(n)},l_{(n)}) \right\rangle_t \nonumber \\
    &\quad+ \frac{\gamma^2}{2}\left(d\langle X^{\pi,n}(\theta_{(n)},l_{(n)})\rangle_t +d\langle Y^n(\theta_{(n)},l_{(n)})\rangle_t \right)  \Bigg)\nonumber \\
    &\ + \int_E \hat{U}^{n+1}_t\left(X^{\pi,n}_{t^-}(\theta_{(n)},l_{(n)}) + \pi^n_t(\theta_{(n)},l_{(n)})'\beta^n_t(\theta_{(n)},l_{(n)},l),(\theta_{(n)},t),(l_{(n)},l)  \right)dt\lambda_{n+1}(l_{(n)},dl)  \nonumber\\
    =&\ \hat{U}^n_t\left(X^{\pi,n}_t(\theta_{(n)},l_{(n)}),\theta_{(n)},l_{(n)}\right) \Bigg[   \Bigg( -\gamma \pi^n_t(\theta_{(n)},l_{(n)})' \mu^n_t(\theta_{(n)}, l_{(n)})\nonumber \\
    &\quad - \gamma \min_{\pi\in\Pi_n} f^n\left(t,\pi,Y^{n}_t(\theta_{(n)},l_{(n)}),Z^{n}_t(\theta_{(n)},l_{(n)}),\theta_{(n)},l_{(n)}\right)\nonumber \\
        &\quad  -\gamma^2 \pi^n_t(\theta_{(n)},l_{(n)})'
        \sigma^n_t(\theta_{(n)},l_{(n)})Z^n_t(\theta_{(n)},l_{(n)})\nonumber \\
        &\quad + \frac{\gamma^2}{2}\left( |Z^n_t(\theta_{(n)},l_{(n)})|^2 + \left|\sigma^n_t(\theta_{(n)},l_{(n)})'\pi^n_t(\theta_{(n)},l_{(n)})\right|^2\right) \nonumber \\
        &\quad + \int_E  e^{\gamma\left(Y^{n+1}_t\left( (\theta_{(n)},t),(l_{(n)},l)\right) - Y^n_t(\theta_{(n)},l_{(n)}) - \pi^n_t(\theta_{(n)},l_{(n)})'\beta^n_t\left(\theta_{(n)},l_{(n)},l\right) \right) } \lambda_{n+1}(l_{(n)},dl)  \Bigg)dt \nonumber\\
    &\quad  +\gamma \left( Z^n_t(\theta_{(n)},l_{(n)})-\sigma^n_t(\theta_{(n)},l_{(n)})'\pi^n_t(\theta_{(n)},l_{(n)})\right)' dW_t\Bigg].
    \end{align*}
Hence, using the definition of $V^{\pi,n}$ (see \eqref{eq:hat:U:martingale}), for any $s\geq t\geq \theta_n$, 
    \begin{align}
            \label{eq:U:cont}
       &\ V^{\pi,n}_s\left( \theta_{(n)},l_{(n)}\right)\nonumber\\ 
    =&\ V^{\pi,n}_t\left( \theta_{(n)},l_{(n)}\right) + \gamma \int_t^s  \hat{U}^n_\tau\left(X^{\pi,n}_\tau(\theta_{(n)},l_{(n)}),\theta_{(n)},l_{(n)}\right) \nonumber \\
    &\quad \cdot\bigg( - \pi^n_\tau(\theta_{(n)},l_{(n)})' \mu^n_\tau(\theta_{(n)}, l_{(n)})\nonumber \\
    &\quad -  \min_{\pi\in\Pi_n} f^n\left(\tau,\pi,Y^{n}_\tau(\theta_{(n)},l_{(n)}),Z^{n}_\tau(\theta_{(n)},l_{(n)}),\theta_{(n)},l_{(n)}\right)\nonumber \\
        &\quad  -\gamma \pi^n_\tau(\theta_{(n)},l_{(n)})'
        \sigma^n_\tau(\theta_{(n)},l_{(n)})Z^n_\tau(\theta_{(n)},l_{(n)})\nonumber \\
        &\quad + \frac{\gamma}{2}\left( |Z^n_\tau(\theta_{(n)},l_{(n)})|^2 + \left|\sigma^n_\tau(\theta_{(n)},l_{(n)})'\pi^n_\tau(\theta_{(n)},l_{(n)})\right|^2\right) \nonumber \\
        &\quad + \frac{1}{\gamma}\int_E  e^{\gamma\left(Y^{n+1}_\tau\left( (\theta_{(n)},\tau),(l_{(n)},l)\right) - Y^n_\tau(\theta_{(n)},l_{(n)}) - \pi^n_\tau(\theta_{(n)},l_{(n)})'\beta^n_\tau\left(\theta_{(n)},l_{(n)},l\right) \right) } \lambda_{n+1}(l_{(n)},dl)  \bigg)d\tau\nonumber\\
    &\  +\gamma \int_t^s \hat{U}^n_\tau\left(X^{\pi,n}_\tau(\theta_{(n)},l_{(n)}),\theta_{(n)},l_{(n)}\right) \nonumber\\
    &\quad\cdot \left( Z^n_\tau(\theta_{(n)},l_{(n)})-\sigma^n_\tau(\theta_{(n)},l_{(n)})'\pi^n_\tau(\theta_{(n)},l_{(n)})\right)' dW_\tau.
    \end{align}
Likewise, when $n=m$, for any $(\theta_{(m)},l_{(m)})\in \Delta_m\times E^m$ and $s\geq t\geq \theta_m$, we have
 \begin{align}
            \label{eq:U:cont:m}
       &\ V^{\pi,m}_s\left(\theta_{(m)},l_{(m)}\right)\nonumber\\ 
       =&\ V^{\pi,m}_t\left(\theta_{(m)},l_{(m)}\right) + \gamma \int_t^s  \hat{U}^m_\tau\left(X^{\pi,m}_\tau(\theta_{(m)},l_{(m)}),\theta_{(m)},l_{(m)}\right) \nonumber \\
    &\quad \cdot\bigg( - \pi^m_\tau(\theta_{(m)},l_{(m)})' \mu^m_\tau(\theta_{(m)}, l_{(m)})  -  \min_{\pi\in\Pi_m} f^m\left(\tau,\pi,Z^{m}_\tau(\theta_{(m)},l_{(m)}),\theta_{(m)},l_{(m)}\right)\nonumber \\
        &\quad  -\gamma \pi^m_\tau(\theta_{(m)},l_{(m)})'
        \sigma^m_\tau(\theta_{(m)},l_{(m)})Z^m_\tau(\theta_{(m)},l_{(m)})\nonumber \\
        &\quad + \frac{\gamma}{2}\left( |Z^m_\tau(\theta_{(m)},l_{(m)})|^2 + \left|\sigma^m_\tau(\theta_{(m)},l_{(m)})'\pi^m_\tau(\theta_{(m)},l_{(m)})\right|^2\right) \bigg) d\tau \nonumber \\
    &\  +\gamma \int_t^s \hat{U}^m_\tau\left(X^{\pi,m}_\tau(\theta_{(m)},l_{(m)}),\theta_{(m)},l_{(m)}\right) \nonumber\\
    &\quad \cdot \left( Z^m_\tau(\theta_{(m)},l_{(m)})-\sigma^m_\tau(\theta_{(m)},l_{(m)})'\pi^m_\tau(\theta_{(m)},l_{(m)})\right)' dW_\tau.
    \end{align}


 We first show that $(V^{\pi,n}_t(\theta_{(n)},l_{(n)}))_{t\geq \theta_n}$ is a $\mathbb{F}$-supermartingale for any $n=0,\dots,m$, $(\theta_{(n)},l_{(n)})\in \Delta_n\times E^n$, and $\pi \in \mathcal{A}$. By the definition of $f^n$ and the fact that $\hat{U}^n$ is non-positive, we see that the first integral on the right-hand side of \eqref{eq:U:cont} and \eqref{eq:U:cont:m} are non-positive. Hence, for any $\pi \in \mathcal{A}$, the process $(V^{\pi,n}_t(\theta_{(n)},l_{(n)}))_{t\geq \theta_n}$ is a $\mathbb{F}$-local supermartingale. By the admissibility of $\pi$, we know that $(V^{\pi,n}_t(\theta_{(n)},l_{(n)}))_{t\geq \theta_n}$ is $\mathbb{F}$-uniformly integrable, and thus it is a $\mathbb{F}$-supermartingale. By Theorem \ref{thm:U:G:F}, we infer that $(U_t(X^\pi_t))_{t\geq 0}$ is a $\mathbb{G}$-supermartingale for any $\pi\in \mathcal{A}$.



Next, we choose $\pi^*\in \mathcal{A}$ such that $(V^{\pi^*,n}_t(\theta_{(n)},l_{(n)}))_{t\geq \theta_n}$ is a $\mathbb{F}$-martingale for any $n=0,\dots,m$, and $(\theta_{(n)},l_{(n)})\in \Delta_n\times E^n$. Motivated by \eqref{eq:U:cont} and \eqref{eq:U:cont:m}, we let $\pi^*$ follow the following Jacod-Pham decomposition: for $t\geq 0$,
 \begin{equation}
        \label{eq:pi*}
        \pi^*_t = \sum_{n=0}^{m-1} \pi^{*n}_t(T_{(n)},L_{(n)})\mathbbm{1}_{ \{ T_n < t \leq T_{n+1} \} } + \pi^{*m}_t(T_{(m)},L_{(m)}) \mathbbm{1}_{ \{ t > T_m\} } ,
    \end{equation}
 where for each  $n = 0,\dots,m-1$, $\pi^{*n}(\cdot,\cdot) \in \mathcal{P}_\mathbb{F}(\Delta_n,E^n;\mathbb{R}^m)$ satisfies,  for any $(\theta_{(n)},l_{(n)})\in \Delta_n\times E^n$ and $t\geq \theta_n$,
    \begin{equation}
    \label{eq:pi*n}
    \begin{aligned}
                \pi^{*n}_t(\theta_{(n)},l_{(n)}) &=  \mathop{\arg\min}_{\pi \in \Pi_n } f^n\left(t,\pi,Y^n_t(\theta_{(n)},l_{(n)}),Z^n_t(\theta_{(n)},l_{(n)}),\theta_{(n)},l_{(n)}\right) ,
    \end{aligned}
    \end{equation}
and for $n=m$, $\pi^{*m}(\cdot,\cdot) \in \mathcal{P}_\mathbb{F}(\Delta_m,E^m;\mathbb{R}^m)$ is given by, for any $(\theta_{(m)},l_{(m)})\in \Delta_m\times E^m$ and $t\geq \theta_m$,
 \begin{equation}
    \label{eq:pi*m}
                 \pi^{*m}_t(\theta_{(m)},l_{(m)}) =  \mathop{\arg\min}_{\pi \in \Pi_m }  f^m\left(t,\pi,Z^m_t(\theta_{(m)},l_{(m)}),\theta_{(m)},l_{(m)}\right).
     \end{equation}
 The predictability of $\pi^*$ is guaranteed by a measurable selection argument in \cite{Benes1970}.      

 We verify that $\pi^*\in \mathcal{A}$  and $(V^{\pi^*,n}_t(\theta_{(n)},l_{(n)}))_{t\geq \theta_n}$ is a uniformly integrable martingale in $\mathbb{F}$.
To this end, for any $n=0,\dots,m$, using \eqref{eq:U:cont}, \eqref{eq:U:cont:m} and the definition of $\pi^{*n}(\cdot,\cdot)$, we have, for any $(\theta_{(n)},l_{(n)}) \in \Delta_n\times E^n$ and $t\geq \theta_n$,
    \begin{equation*}
         V^{\pi^*,n}_t\left(\theta_{(n)},l_{(n)}\right) =   V^{\pi^*,n}_{\theta_n}\left(\theta_{(n)},l_{(n)}\right)  \mathcal{E}_{\theta_n,t}^n\left(\varphi^n\left(\theta_{(n)},l_{(n)}\right)\right),
    \end{equation*}
where $\varphi^n(\cdot,\cdot) \in \mathcal{O}_\mathbb{F}(\Delta_n,E^n)$ is defined by
    \begin{equation*}
        \begin{aligned}
           &\  \varphi^n_t\left(\theta_{(n)},l_{(n)}\right) \\
           :=&\  \frac{\gamma \hat{U}^n_t \left(X^{\pi^*,n}_t\left(\theta_{(n)},l_{(n)} \right),\theta_{(n)},l_{(n)}\right)
 (Z^n_t(\theta_{(n)},l_{(n)})-\sigma^n_t(\theta_{(n)},l_{(n)})'\pi^{n*}_t(\theta_{(n)},l_{(n)}))'}{ V^{\pi^*,n}_t\left(\theta_{(n)},l_{(n)} \right) }  \\
 &\ \cdot  \mathbbm{1}_{\{ V^{\pi^*,n}_t(\theta_{(n)},l_{(n)}) \neq 0 \} },
        \end{aligned}
    \end{equation*}
and $(\mathcal{E}_{\theta_n,t}^n(\varphi^n(\theta_{(n)},l_{(n)})))_{t\geq \theta_n}$ is the  Dol\'eans-Dade exponential of $\varphi^n$:
    \begin{equation*}
        \begin{aligned}          \mathcal{E}_{\theta_n,t}^n(\varphi^n(\theta_{(n)},l_{(n)}))&:=           \exp\left( \int_{\theta_n}^t \varphi^n_s(\theta_{(n)},l_{(n)}) dW_s  - \frac{1}{2}\int_{\theta_n}^t  \left| \varphi^n_s(\theta_{(n)},l_{(n)})\right|^2 ds \right).
        \end{aligned}
    \end{equation*}

By following the derivation of \eqref{eq:pi:bound:tilde}, we can infer the existence of $C>0$ such that
    \begin{equation}
    \label{eq:pi:bound}
        |(\sigma^n_t(\theta_{(n)},l_{(n)}))'\pi^{*n}_t(\theta_{(n)},l_{(n)})| \leq C\left(1 + |Z^n_t(\theta_{(n)},l_{(n)})| \right),
    \end{equation}
for all $(\theta_{(n)},l_{(n)}) \in \Delta_n\times E^n$ and $t\geq \theta_n$. By \eqref{eq:hat:U:martingale} and the fact that $\hat{U}$ is non-positive, it is clear that $ V^{\pi^*,n}_t(\theta_{(n)},l_{(n)}) \leq \hat{U}^n_t(X^{\pi^*,n}_t(\theta_{(n)},l_{(n)}),\theta_{(n)},l_{(n)})\leq 0$. Using this, \eqref{eq:pi:bound}, and Assumption \ref{ass:bound}, we have
    \begin{align*}
       &\  \left|\varphi^n_t(\theta_{(n)},l_{(n)}) \right|\\
       \leq&\  \gamma \left| \frac{\hat{U}^n_t \left(X^{\pi^*,n}_t\left(\theta_{(n)},l_{(n)} \right),\theta_{(n)},l_{(n)}\right)}{V^{\pi^*,n}_t(\theta_{(n)},l_{(n)})} \right|\left( |Z^n_t(\theta_{(n)},l_{(n)})| +  |(\sigma^n_t(\theta_{(n)},l_{(n)}))'\pi^{*n}_t(\theta_{(n)},l_{(n)})| \right)  \\
       \leq&\ \gamma(1+C)\left(1 + |Z^n_t(\theta_{(n)},l_{(n)})| \right).
    \end{align*}
Since $\left(\int_{\theta_n}^\cdot Z^n_s(\theta_{(n)},l_{(n)})\right)_{t\geq 0}$ is a BMO martingale under  $\mathbb{F}$, the Dol\'eans-Dade exponential $ (\mathcal{E}_{\theta_n,t}^n(\varphi^n(\theta_{(n)},l_{(n)})))_{t\geq \theta_n}$ is a $\mathbb{F}$-uniformly integrable martingale. This in turn implies that $\pi^*\in \mathcal{A}$ and $(V^{\pi^*,n}_t(\theta_{(n)},l_{(n)}))_{t\geq \theta_n}$ is a $\mathbb{F}$-uniformly integrable martingale. By Theorem \ref{thm:U:G:F}, we have the following conclusion:

\begin{theorem}
\label{thm:forward:optimal}
    Under Assumptions \ref{ass:bound}-\ref{ass:eta:zero}, the random field $(U_t(x))_{x\in \mathbb{R},t\geq 0}$ defined by \eqref{eq:Un}-\eqref{eq:U:exp} is a forward exponential utility for \eqref{eq:X} with the set of admissible strategies $\mathcal{A}$ given by \eqref{eq:A}, and the optimal investment strategy $\pi^*$ given by \eqref{eq:pi*}-\eqref{eq:pi*m}.
\end{theorem}

\section{Stochastic Factor Market Model}
\label{sec:stochastic:factor}
In this section, we consider a Markovian model in which the model parameters $(\mu, \sigma, \beta)$ are driven by a stochastic factor process $\Phi=(\Phi_t)_{t\geq 0}$. 
{\color{black} In Section \ref{sec:factor:markovian}, we establish the Markovian solutions of the associated BSDEs, and derive bounds and properties of the solutions and the associated optimal investment strategies that hold uniformly in the discount rate $\rho$. In Section \ref{sec:factor:ergodic:unbounded}, we examine the long-term risk-sensitive growth rate of the optimal wealth process as the discount rate vanishes. In Section \ref{sec:factor:ergodic:construction}, we discuss the fundamental challenges of constructing a forward utility preference using ergodic BSDEs and illustrate a construction with additional monotonicity conditions on the model parameters.   } 

In the sequel, we assume that $\Phi$ is a $\mathbb{G}$-optional $\mathbb{R}^d$-valued process with the Jacod-Pham decomposition $(\Phi^1,\dots,\Phi^m)$, governed by the following SDEs: for $t\geq 0$,
    \begin{equation*}
        d\Phi^0_t(0) =  g^0(\Phi^0_t(0))dt + \kappa^0(0)dW_t,
    \end{equation*}
and for each $n=1,\dots,m$, $(\theta_{(n)},l_{(n)})\in \Delta_n\times E^n$ and $t\geq \theta_n$, 
    \begin{equation}
    \label{eq:phi}
    \left\{ \begin{aligned}
        d\Phi^n_t(\theta_{(n)},l_{(n)}) &= g^n\left(\Phi^n_t(\theta_{(n)},l_{(n)}),\theta_{(n)},l_{(n)}\right) dt + \kappa^n(\theta_{(n)},l_{(n)})dW_t,\\
        \Phi^n_{\theta_n}(\theta_{(n)},l_{(n)})& = \Phi^{n-1}_{\theta_n^-}(\theta_{(n-1)},l_{(n-1)}) + \varphi^{n-1}\left( \Phi^{n-1}_{\theta_n^-}(\theta_{(n-1)},l_{(n-1)})\right). 
        \end{aligned}
        \right.
    \end{equation}
Here, $g^n(\cdot, \cdot,\cdot):\mathbb{R}^d\times \Delta_n\times E^n \to\mathbb{R}^d$, $\kappa^n(\cdot,\cdot): \Delta_n\times E^n \to \mathbb{R}^{d\times d}$, $\varphi^{n}: \mathbb{R}^d \to \mathbb{R}^d$ are measurable functions. We impose the following dissipative condition on the drift coefficients $g^n$. 
 

\begin{assumption}
    \label{ass:lip:diss:factor} 
            
         There exists $C_g>0$ such that, for any $n=0,\dots,m$, $(\theta_{(n)},l_{(n)})\in \Delta_n\times E^n$ and $\phi_1,\phi_2\in \mathbb{R}^d$,
                \begin{equation}
                \label{eq:dissipative:gn}
                   \left( g^n(\phi_1,\theta_{(n)},l_{(n)}) - g^n(\phi_2,\theta_{(n)},l_{(n)})  \right)'(\phi_1-\phi_2) \leq -C_g|\phi_1-\phi_2|^2. 
                \end{equation}
\end{assumption}

\begin{remark}
    The dissipative condition \eqref{eq:dissipative:gn} on $g^n$ implies the following exponential ergodicity property: for any $n=0,\dots,m$, $\phi_1,\phi_2\in \mathbb{R}^d$, $(\theta_{(n)},l_{(n)})\in \Delta_n\times E^n$ and $t\geq \theta_n$, 
        \begin{equation}
        \label{eq:ergodic:phi}
            \left|\Phi^{n,\phi_1}_t(\theta_{(n)},l_{(n)}) - \Phi^{n,\phi_2}_t(\theta_{(n)},l_{(n)}) \right|^2 \leq e^{-2C_g (t-\theta_n)}|\phi_1-\phi_2|^2,
        \end{equation}
    where $\Phi^{n,\phi}$ is the solution of \eqref{eq:phi} with initial condition $\Phi^{n,\phi}_{\theta_n}(\theta_{(n)},l_{(n)}) = \phi\in\mathbb{R}^d$. \hfill $\square$
\end{remark}

For any $t \geq T_n$, we assume that the market parameters admit the following Markovian representation:
    \begin{align}           \label{eq:model:parameter:decompose:factor}
 \mu^n_t(T_{(n)},L_{(n)}) &= \hat{\mu}^n\left( \Phi_t^n(T_{(n)},L_{(n)}) \right), \             \sigma^n_t(T_{(n)},L_{(n)}) = \hat{\sigma}^n(\Phi^n_t(T_{(n)},L_{(n)})), \nonumber \\
             \alpha^n_t(T_{(n)},L_{(n)}) &= \hat{\alpha}^n(\Phi^n_t(T_{(n)},L_{(n)})), \ 
           \beta^n_t(T_{(n)},L_{(n)},l)  =  {\color{black}\hat{\beta}^n\left(\Phi^n_{t^-}(T_{(n)},L_{(n)}),l\right)},
    \end{align}
    where for $n=0,\dots,m$, $\hat{\mu}^n : \mathbb{R}^d\to\mathbb{R}^m$, $\hat{\sigma}^n : \mathbb{R}^d\to \mathbb{R}^{m\times d}$ such that $\hat{\sigma}^n(\phi)$ is a full-rank matrix for any $\phi\in \mathbb{R}^d$;     $\hat{\alpha}^n : \mathbb{R}^d\to\mathbb{R}^d$ is given by $ \hat{\alpha}^n(\cdot) = \hat{\sigma}^n(\cdot)'\left(\hat{\sigma}^n(\cdot)\hat{\sigma}^n(\cdot)' \right)^{-1}\hat{\mu}^n(\cdot)$, which satisfies $\hat{\sigma}^n(\cdot)\hat{\alpha}^n(\cdot) = \hat{\mu}^n(\cdot)$; and for $n=0,\dots,m-1$, {\color{black}$\hat{\beta}^n:\mathbb{R}^d\times E \to \mathbb{R}^m$}. All these measurable functions are deterministic.  Note that in this setting, the model parameters {\color{black}(except for $\beta^n$)} depend on the indexes $(\theta_{(n)},l_{(n)})$ only through the stochastic factor $\Phi$.

Throughout this section,  we assume that Assumption \ref{ass:density} holds. We also introduce the following boundedness and Lipschitz assumptions:
\begin{assumption}
    \label{ass:factor}
        For any $n=0,1,\dots,m$, 
        \begin{enumerate}
             \item $\hat{\sigma}^n(\phi)$ is a full-rank matrix, and there exists $\sigma_{\min}>0$ such that $|\hat{\sigma}^n(\phi)'x| \geq \sigma_{\min}|x|$ for all $\phi\in \mathbb{R}^d$ and $x\in \mathbb{R}^m$; 
            \item $\hat{\mu}^n(\cdot)$, $\hat{\sigma}^n(\cdot)$, $\hat{\alpha}^n(\cdot)$ are  bounded; 
            \item $\hat{\sigma}^n(\cdot)$, $\hat{\alpha}^n(\cdot)$ are Lipschitz continuous  with Lipschitz constants $C_\sigma,C_\alpha$, respectively;
            \item $|\kappa^n(\cdot,\cdot)|=1$; 
          
            \item For $n=0,\dots,m-1$, {\color{black}$\hat{\beta}^n(\cdot,\cdot)$} and $\varphi^n(\cdot)$ are  bounded and uniformly Lipschitz continuous, with Lipschitz constants $C_\beta$ and $C_\varphi$: {\color{black} for any $\phi_1,\phi_2\in \mathbb{R}^d$ and $l\in E$,
                \begin{align*}
                    & |\hat{\beta}^n(\phi_1,l) - \hat{\beta}^n(\phi_2,l)|\leq C_\beta|\phi_1-\phi_2|, \  |\varphi^n(\phi_1) - \varphi^n(\phi_2)|\leq C_\varphi|\phi_1-\phi_2|.
                \end{align*}} %
        \end{enumerate}
    \end{assumption}

\begin{remark}
For ease of exposition, we impose the assumption $|\kappa^n(\cdot,\cdot)|=1$ that simplifies the expressions of bounds that follow. This condition can be relaxed in a straightforward manner. \hfill$\square$
\end{remark}


\subsection{Infinite Horizon BSDEs and Markovian Solution}
\label{sec:factor:markovian}
Under the stochastic factor model, the drivers of the infinite-horizon BSDEs \eqref{eq:Ym}-\eqref{eq:Yn} can be written as follows:  for any $(\pi,y,z)\in \mathbb{R}^m\times \mathbb{R}\times  \mathbb{R}^d$, $(\theta_{(m)},l_{(m)})\in \Delta_m \times E^m$ and $t\geq \theta_m$,
\begin{equation}
f^m(t,\pi,z,\theta_{(m)},l_{(m)}) = \hat{f}^m(\pi,z,\Phi^m_t(\theta_{(m)},l_{(m)})) := \hat{F}^m_1(\pi,z,\Phi^m_t(\theta_{(m)},l_{(m)})),
\label{eq:factor:drivers_1}
\end{equation}
and for any $n=0,\dots,m-1$, $(\theta_{(n)},l_{(n)})\in \Delta_n \times E^n$, and $t\geq \theta_n$,
\begin{equation}
\begin{aligned}
f^n(t,\pi,y,z,\theta_{(n)},l_{(n)}) &= \hat{f}^n(t,\pi,y,z,\Phi^n_t(\theta_{(n)},l_{(n)}))\\
     &:= \hat{F}^n_1(\pi,z,\Phi^n_t(\theta_{(n)},l_{(n)})) + \hat{F}^n_2(t,\pi,y,\Phi^n_t(\theta_{(n)},l_{(n)}),\theta_{(n)},l_{(n)}),
\end{aligned}
\label{eq:factor:drivers_2}
\end{equation}
where for $\phi\in \mathbb{R}^d$, 
  \begin{align*}
                 \hat{F}_1^n(\pi,z,\phi) &:= 
                  \frac{\gamma}{2}\left|\hat{\sigma}^n(\phi)' \pi -  \left(z + \frac{\hat{\alpha}^n(\phi)}{\gamma} \right)  \right|^2   - \hat{\alpha}^n(\phi)'z - \frac{\left|\hat{\alpha}^n(\phi)\right|^2}{2\gamma},\nonumber \\
              \hat{F}^n_2(t,\pi,y,\phi, \theta_{(n)}, l_{(n)}) &:=  \frac{1}{\gamma} \int_E e^{\gamma (Y^{n+1}_t( (\theta_{(n)},t),(l_{(n)},l)) - y -\pi'\hat{\beta}^n(\phi,l)  ) }   \lambda_{n+1}(l_{(n)},dl). 
    \end{align*}

By Assumptions \ref{ass:density}, \ref{ass:factor}, and Theorem \ref{thm:exist:Y}, it is clear that the BSDEs \eqref{eq:Ym}-\eqref{eq:Yn}, with drivers \eqref{eq:factor:drivers_1}-\eqref{eq:factor:drivers_2}, admit a unique solution $(Y^n,Z^n)_{n=0}^m$ such that 
$$(Y^n,Z^n) \in \mathcal{S}(\Delta_n,E^n;\mathbb{R})\times(\mathcal{M}^{2}(\Delta_n,E^n;\mathbb{R}^d)\cap \mathcal{L}^2_{\text{loc}}(\Delta_n,E^n;\mathbb{R}^d)),$$ for any $n=0,\dots,m$. In addition, the random field $U$ defined by 
    \begin{equation*}
        U_t(x) := \frac{-e^{-\gamma x}e^{\gamma(Y_t - \int_0^t \rho Y_s ds - Y_0)}}{\hat{\eta}_t}, \ x\in \mathbb{R}, t\geq 0, 
    \end{equation*}
where $Y$ and $\hat{\eta}$ are respectively given as in \eqref{eq:Y:decompose} and \eqref{eq:hat:eta} \& \eqref{eq:eta_decomposition}, is a forward exponential preference. The ergodicity of the stochastic factor model allows us to further represent the solution of the infinite-horizon BSDEs in a Markovian form with stronger boundedness properties, as discussed below.

We first provide a  bound for the solution $(Y^n,Z^n)$ in $\mathcal{S}(\Delta_n,E^n)$. Recall from Proposition \ref{pp:Yn:bdd:factor}, we have already shown that $|Y^n_t(\cdot,\cdot)|\leq K_Y/\rho$ for any $n=0,\dots,m$. In Theorem \ref{thm:Z:markovian} below, utilizing the dissipative condition of the drift $g^n$, we provide a bound of $Z^n$ in $\mathcal{S}(\Delta_n,E^n)$ that is uniform in $\rho>0$. To this end, we introduce the following assumptions, which will be used to deduce the local Lipschitz continuity of the driver for the BSDEs with respect to the stochastic factor.


\begin{assumption}
    \label{ass:pi:compact}
    For $n=0,\dots,m$, $\Pi_n \subseteq \mathbb{R}^m$ is compact. In particular, there exists $C_\Pi>0$ such that $|\pi|\leq C_\Pi$ for any $\pi \in  \cup_{n=0}^m \Pi_n$.  
\end{assumption}    

Under Assumptions \ref{ass:factor} and \ref{ass:pi:compact}, it is straightforward to verify that, there exists $C_\phi>0$ such that for any $n=0,\dots,m$, $\pi\in\Pi_n$ and $z,\phi_1,\phi_2 \in \mathbb{R}^d$,
    \begin{equation}
    \label{eq:F:lip:factor}
        \left|\hat{F}^n_1(\pi,z,\phi_1) -\hat{F}^n_1(\pi,z,\phi_2) \right| \leq C_\phi(1+|z|)|\phi_1-\phi_2|.
    \end{equation}
{Indeed, using the boundedness and the uniform Lipschitz property of $\hat{\alpha}^n,\hat{\sigma}^n$, along with the fact that $|\pi|\leq C_\Pi$, we have 
 \begin{align*}
&\left|\hat{F}^n_1(\pi,z,\phi_1) -\hat{F}^n_1(\pi,z,\phi_2) \right| \\
   &\leq \Bigg[
         \gamma \Big( C_\sigma C_\Pi + \tfrac{C_\alpha}{\gamma} \Big)
                \Big( C_\Pi K_\sigma + \tfrac{K_\alpha}{\gamma} \Big)
         + \tfrac{C_\alpha K_\alpha}{\gamma}  
         + \Big( C_\sigma C_\Pi + \tfrac{C_\alpha}{\gamma} + C_\alpha \Big)|z|
       \Bigg] |\phi_1-\phi_2|,
\end{align*}
where $K_\alpha,K_\sigma>0$ are the uniform bound  for $\hat{\alpha}^n$ and $\hat{\sigma}^n$, $n=0,\dots,m$, respectively.  Hence, we can pick $C_\phi= C_\phi(C_\Pi)$ as
    \begin{equation}
            \label{eq:F:lip:factor2}
        C_\phi := \max\left\{  \gamma  \left( C_\sigma C_\Pi + \frac{C_\alpha}{\gamma} \right)  \left( C_\Pi K_\sigma + \frac{K_\alpha}{\gamma}   \right) + \frac{C_\alpha K_\alpha}{\gamma},  C_\sigma C_\Pi + \frac{C_\alpha}{\gamma} + C_\alpha  \right\}.
    \end{equation}

    
}

We also impose the following assumption on the constants $C_\phi$ and $C_g$. 
 
\begin{assumption} 
    \label{ass:Cphi:Cg}
        The constants $C_g,C_\phi>0$ satisfy $C_g-C_\phi>0$, where $C_\phi$ is defined in \eqref{eq:F:lip:factor2}.  
             
    
\end{assumption}

\begin{theorem}
\label{thm:Z:markovian}
    Under Assumptions \ref{ass:lip:diss:factor}-\ref{ass:Cphi:Cg}, the system of infinite-horizon BSDEs \eqref{eq:Ym}-\eqref{eq:Yn} with drivers \eqref{eq:factor:drivers_1}-\eqref{eq:factor:drivers_2} admit a unique bounded Markovian solution, i.e.,  for any $n=0,\dots,m$, there exist measurable functions $y^n:\mathbb{R}^d \to \mathbb{R}$ and $z^n:\mathbb{R}^d\to \mathbb{R}^d$ such that for any $(\theta_{(n)},l_{(n)})\in \Delta_n\times E^n$ and $t\geq \theta_n$, 
        \begin{equation*}
            Y^n_t(\theta_{(n)},l_{(n)}) = y^n(\Phi^n_t(\theta_{(n)},l_{(n)})) \quad \text{and} \quad Z^n_t(\theta_{(n)},l_{(n)}) = z^n(\Phi^n_t(\theta_{(n)},l_{(n)})). 
        \end{equation*}
    In addition, for any $\phi,\phi_1,\phi_2\in \mathbb{R}^d$, it holds that 
        \begin{equation}
        \label{eq:y:z:markov:bound:lip} 
            |y^n(\phi)|\leq \frac{K_Y}{\rho}, \ |z^n(\phi)| \leq K_{Z^n}, \ |y^n(\phi_1)-y^n(\phi_2)|\leq K_{Z^n}|\phi_1-\phi_2|, 
        \end{equation}
   where $K_Y>0$ is given in Proposition \ref{pp:Yn:bdd:factor}, and 
        \begin{equation*}
            K_{Z^n} :=  \frac{C_\phi(1+C_\varphi)^{m-n}}{C_g-C_\phi}+ C_\Pi C_\beta\sum_{j=0}^{m-n-1} (1+C_\varphi)^j.  
        \end{equation*}
  
\end{theorem}

\begin{proof} See Appendix \ref{sec:pf:thm:Z:markovian}. 
\end{proof}

\begin{remark}
     The proof of Theorem~\ref{thm:Z:markovian} adapts the methodology used in the proof of the multi-dimensional comparison theorem for BSDEs under regime-switching models in \cite{hu2020systems}. However, that result is not directly applicable in our defaultable market setting for three main reasons. First, the BSDEs considered herein are linked in a one-directional manner:  $Y^n$ depends on $Y^{n+1}$ but not vice versa, whereas in the regime-switching model, the BSDEs interact mutually in a closed-loop structure. Second, the BSDEs in our framework are indexed differently, which introduces additional deviations when establishing uniform bounds between the solutions. Third, the BSDEs under default depend exponentially on the controlled jump size $\pi'\hat{\beta}^n(\phi,l)$. Hence, one cannot simply rely on the quadratic distance structure of the driver as in the regime-switching setting. \hfill $\square$ 
\end{remark}


Next, we establish a {\color{black}one-sided} bound, uniform in $\rho$, of the difference between the solutions of successive infinite-horizon BSDEs. To this end, we will need the following assumption, which enables us to bound the exponent appearing in $\hat{F}^n_2$, $n=0,\dots,m-1$:

\begin{assumption}
\label{ass:ergodic:kappa:g}
   There exists $D_g>0$ such that for all $n=1,\dots,m$,
        \begin{equation*}
            \sup_{\phi\in\mathbb{R}^d, (\theta_{(n)},l_{(n)})\in \Delta_n\times E^n}|g^n(\phi,\theta_{(n)},l_{(n)})-g^{n-1}(\phi,\theta_{(n-1)},l_{(n-1)})| < D_g <\infty. 
        \end{equation*}

\end{assumption}

\begin{remark}
    Comparing with the regime-switching framework, the processes $Y^{n}$ and $Y^{n-1}$ are defined on different index sets ($\Delta_n\times E^n$ for the former, and $\Delta_{n-1}\times E^{n-1}$ for the latter). Hence, these processes are given by the respective Markovian functions evaluated at \textit{different} stochastic factor ($\Phi^n_t(\theta_{(n)},l_{(n)})$ for the former, and    $\Phi^{n-1}_t(\theta_{(n-1)},l_{(n-1)})$ for the latter). Assumption \ref{ass:ergodic:kappa:g} is then used to control the discrepancy of $Y^{n}$ and $Y^{n-1}$ due to the difference in the stochastic factor. To name a class of examples, Assumption \ref{ass:ergodic:kappa:g} is satisfied if $g^n(\phi,\theta_{(n)},l_{(n)}) = g(\phi) + \bar{g}^n(\theta_{(n)},l_{(n)})$, where $g:\mathbb{R}^d\to \mathbb{R}^d$ satisfies \eqref{eq:dissipative:gn}, and $\bar{g}^n:\Delta_n\times E^n\to\mathbb{R}^d$ is bounded. \hfill $\square$
\end{remark}

\begin{proposition}
\label{pp:Delta:Y}
    Suppose that Assumptions \ref{ass:lip:diss:factor}-\ref{ass:ergodic:kappa:g} hold. Then, for any $n=1,\dots,m$,
    \begin{equation}
    \label{eq:Delta:Y:n:bound:stronger}
       Y^n_t\left((\theta_{(n-1)},t),(l_{(n-1)},l) \right) - Y^{n-1}_t\left(\theta_{(n-1)},l_{(n-1)}\right)  \leq K_{\Delta Y^n}
    \end{equation}
    for all  $(\theta_{(n-1)},l_{(n-1)},l)\in \Delta_{n-1}\times E^{n-1}\times E$,  $t\geq \theta_{n-1}$, and $\rho>0$, where  
        \begin{equation*}
            K_{\Delta Y^n} := \gamma e^{C_\Pi K_\beta} C_{n} -1 + K_{Z^n}\left[2K_\varphi + \frac{\sqrt{\pi}}{2} e^{\frac{C_\Pi K_\beta}{2}} \sqrt{ C_g^{-1}D_g^2+4 } \right], 
        \end{equation*}
   $K_\varphi:= \max_{n=0,\dots,m-1}\|\varphi^n\|$,  $K_\beta:= \max_{n=0,\dots,m-1}\|\hat{\beta}^n\|$, and
    \begin{align*}
        C_m &= 
            \frac{\gamma}{2}(K_{Z^m})^2, \ 
        C_n =  
             \frac{\gamma}{2}(K_{Z^n})^2  + \frac{1}{\gamma} e^{\gamma K_{\Delta Y^{n+1}} }, \ n=0,\dots,m-1. 
    \end{align*} 
    
\end{proposition}

\begin{proof} See Appendix \ref{sec:pf:pp:Delta:Y}.
\end{proof}

By Proposition \ref{pp:Delta:Y}, the exponent in the driver $\hat{F}^n_2$ of the infinite-horizon BSDEs, for $n=0,\dots,m-1$, is upper bounded by a constant independent of $\rho$, and thus its exponential term, and also the driver $\hat{F}^n_2$ itself, are also uniformly upper bounded in $\rho$. 
Using this estimate, uniform in $\rho$, one can show that the optimal investment strategy is always bounded, and thus automatically satisfies Assumption \ref{ass:pi:compact}, under the following condition:
\begin{assumption}
    \label{ass:pi:compact:remove}
    There exists $C_\Pi\in(0,C_g)$ such that, for $n=0,\dots,m$,
            \begin{equation}
            \label{eq:Cpi:solution}
            \begin{aligned}
                     \sigma_{\min} C_\Pi &\geq   \frac{\sqrt{2}}{\gamma}e^{\frac{\gamma}{2} K_{\Delta Y^{n+1}}}\mathbbm{1}_{\{n\neq m\}} + \frac{2\|\hat{\alpha}^n\|}{\gamma} + 2K_{Z^n},\    C_g \geq C_\phi,
            \end{aligned}
        \end{equation}
    where the constants $K_{Z^n},K_{\Delta Y^n},$ are defined in Theorem \ref{thm:Z:markovian} and Proposition \ref{pp:Delta:Y}, respectively, which depend on $C_\Pi$ directly and via $C_\phi$ defined in \eqref{eq:F:lip:factor2}. 
\end{assumption}

\begin{remark}
    Note that $K_{\Delta Y^n}$ and $K_{Z^n}$ depend on $C_\Pi$ via $1/(C_g-C_\phi)$, $C_\Pi C_\beta$, $C_\Pi K_\beta$, and $C_\phi=C_\phi(C_\Pi)$ in \eqref{eq:F:lip:factor2} is non-decreasing in $C_\Pi$. Hence, Assumption \ref{ass:pi:compact:remove} can be met for large $C_g$ and small $C_\beta$, $K_\beta$.      \hfill $\square$
\end{remark}

\begin{theorem}
\label{thm:bound:pi}
    Under Assumptions \ref{ass:lip:diss:factor}, \ref{ass:factor}, \ref{ass:Cphi:Cg},   \ref{ass:ergodic:kappa:g}, and  \ref{ass:pi:compact:remove}, the Markovian solutions $(y^n(\cdot),z^n(\cdot))_{n=0}^m$ of the infinite-horizon BSDEs \eqref{eq:Ym}-\eqref{eq:Yn}  with drivers \eqref{eq:factor:drivers_1}-\eqref{eq:factor:drivers_2} satisfy the estimates    \eqref{eq:y:z:markov:bound:lip} and \eqref{eq:Delta:Y:n:bound:stronger}. In addition, the optimal investment strategy satisfies $|\pi^{*n}_t(\theta_{(n)},l_{(n)})|\leq C_\Pi$ for all $n=0,\dots,m$, $(\theta_{(n)},l_{(n)})\in \Delta_n\times E^n$, $t\geq\theta_{n}$, and $\rho>0$, where $C_\Pi$ satisfies \eqref{eq:Cpi:solution}. 

\end{theorem}
\begin{proof}
    See Appendix \ref{sec:pf:thm:bound:pi}. 
\end{proof}
\begin{remark}
    By the inductive argument used in the proof of Theorem \ref{thm:bound:pi}, the estimates 
    \eqref{eq:y:z:markov:bound:lip} and \eqref{eq:Delta:Y:n:bound:stronger} continue to hold for all $n = k,\dots,m$, where $k\leq m$, as long as \eqref{eq:Cpi:solution} holds for $n \ge k$ (instead of for every $n = 0,\dots,m$ as required in the theorem).
\end{remark}

\subsection{Risk-Sensitive Growth Rate}
\label{sec:factor:ergodic:unbounded}
In this subsection, we propose a risk-sensitive growth rate by examining a long-term behavior and when the discount rate $\rho\to 0$. To this end, we shall introduce an ergodic BSDE as the limit of the infinite-horizon BSDE \eqref{eq:Ym}, when $n=m$ with the driver \eqref{eq:factor:drivers_1}, making use of the solution $(y^m,z^m)$ from Theorem \ref{thm:Z:markovian} via $\rho\to 0$.
In the sequel, we use the superscript $\rho$ to emphasize the dependence of the relevant functions or processes on $\rho$. For instance, we will write $Y^{n,\rho}$ to represent $Y^n$ with discount rate $\rho$. 

Fix $\hat{\phi}_m\in\mathbb{R}^d$, and define $\bar{y}^{m,\rho}:\mathbb{R}^d\to \mathbb{R}$ by $\bar{y}^{m,\rho}(\phi):= y^{m,\rho}(\phi) - y^{m,\rho}(\hat{\phi}_m)$ for $\phi\in\mathbb{R}^d$. Using Theorem \ref{thm:Z:markovian}, particularly \eqref{eq:y:z:markov:bound:lip}, we see that $\bar{y}^{m,\rho}(\cdot)$ satisfies the following linear growth and Lipschitz conditions uniformly in $\rho$: for any $\rho>0$, and $\phi,\phi_1,\phi_2\in \mathbb{R}^d$, 
     \begin{equation}
        \label{eq:linear:growth:m}
            \begin{aligned}
                    &   | \bar{y}^{m,\rho}(\phi) | \leq K_{Z^m}|\phi - \hat{\phi}_m| \leq K_{Z^m}(|\phi|+|\hat{\phi}_m|), \\
                    & | \bar{y}^{m,\rho}(\phi_1) -  \bar{y}^{m,\rho}(\phi_2)| \leq K_{Z^m}|\phi_1-\phi_2|. 
            \end{aligned}
    \end{equation}

For any $(\theta_{(m)},l_{(m)})\in \Delta_m\times E^m$ and $t\geq \theta_m$, define the process $\bar{Y}^{m,\rho}_t(\theta_{(m)},l_{(m)}) := Y^{m,\rho}_t(\theta_{(m)},l_{(m)})-y^{m,\rho}(\hat{\phi}_m) = \bar{y}^{m,\rho}(\Phi^m_t(\theta_{(m)},l_{(m)}))$. From \eqref{eq:Ym}, it is clear that $\bar{Y}^{m,\rho}_t(\theta_{(m)},l_{(m)})$ satisfies the following infinite-horizon BSDE:  
        \begin{align}
\label{eq:Ym:peturb}
    & \ \ \ \ d\bar{Y}^{m,\rho}_t(\theta_{(m)},l_{(m)}) \nonumber \\
    &=\Bigg( \rho y^{m,\rho}(\hat{\phi}_m) + \rho \bar{Y}^{m,\rho}_t(\theta_{(m)},l_{(m)}) - \min_{\pi\in\Pi_m}  \hat{f}^m\left(\pi,Z^{m,\rho}_t(\theta_{(m)},l_{(m)}),\Phi^m_t(\theta_{(m)},l_{(m)})\right)\nonumber  \\
    &\quad + Z^{m,\rho}_t(\theta_{(m)},l_{(m)})'dW_t,
\end{align}
where the driver is given in \eqref{eq:factor:drivers_1}.

    Using \eqref{eq:y:z:markov:bound:lip},
    $\rho |y^{m,\rho}(\hat{\phi}_m) | \leq K_Y$, and a standard diagonal argument, by the uniform linear growth property \eqref{eq:linear:growth:m}, there exists a sequence $(\rho_i)_{i=1}^\infty$, with $\rho_i\downarrow 0$, a constant $\varrho_m \in \mathbb{R}$, and a function $\bar{y}^m(\cdot):\mathbb{R}^d\rightarrow\mathbb{R}$, such that, for $\phi$ lying in a dense subset of $\mathbb{R}^d$, 
        \begin{equation}
        \label{eq:ym:converge}
            \lim_{i\to \infty}\rho_i y^{m,\rho_i}(\hat{\phi}_m) = \varrho_m, \ \lim_{i\to\infty} \bar{y}^{m,\rho_i}(\phi) = \bar{y}^m({\phi}).
        \end{equation}
   By the uniform Lipschitz property \eqref{eq:linear:growth:m}, the convergence can be extended to the entire domain $\mathbb{R}^{d}$. 
This indicates the existence of a sequence $(\rho_i)_{i=1}^\infty$, $\rho_i\downarrow 0$, such that, for any $(\theta_{(m)},l_{(m)})\in \Delta_m\times E^m$, and $t\geq \theta_{m}$,
$$\lim_{i\to \infty} \bar{Y}^{m,\rho_i}_t(\theta_{(m)},l_{(m)})= \bar{y}^m\left(\Phi^m_t(\theta_{(m)},l_{(m)}) \right) =: \mathcal{Y}^m_t(\theta_{(m)},l_{(m)}).$$
Likewise, by \eqref{eq:y:z:markov:bound:lip} in Theorem \ref{thm:Z:markovian}, using the boundedness of $z^{m,\rho}$, uniformly in $\rho$, it is standard to show the existence of a function $\bar{z}^m(\cdot) :\mathbb{R}^{d}\to \mathbb{R}^{d}$ such that, for any $(\theta_{(m)},l_{(m)})\in \Delta_m\times E^m$, and $t\geq \theta_{m}$,
$$\lim_{i\to\infty} Z^{m,\rho_i}_t(\theta_{(m)},l_{(m)})= \bar{z}^m(\Phi^m_t(\theta_{(m)},l_{(m)}))  =: \mathcal{Z}^m_t(\theta_{(m)},l_{(m)}).$$
By \eqref{eq:Ym:peturb}, the tuple $\big( \mathcal{Y}^m,$ $\mathcal{Z}^m,\varrho_m)$ is then the solution of the following indexed ergodic BSDE, at $n=m$: for $(\theta_{(m)},l_{(m)})\in \Delta_m\times E^m$ and $t\geq \theta_m$,
    \begin{equation}
        \label{eq:ergodic:Ym:delta}
        \begin{aligned}
                 d\mathcal{Y}^{m}_t(\theta_{(m)},l_{(m)}) &=\Bigg(  \varrho_m   - \min_{\pi\in\Pi_m}  \hat{f}^m\left(\pi,\mathcal{Z}^{m}_t(\theta_{(m)},l_{(m)}),\Phi^m_t(\theta_{(m)},l_{(m)})\right)  \Bigg)  dt   \\
                 &\quad + \mathcal{Z}^{m}_t(\theta_{(m)},l_{(m)})'dW_t,
        \end{aligned}    
    \end{equation}
    where the driver is given in \eqref{eq:factor:drivers_1}. {\color{black}The uniqueness of the solution to the ergodic BSDE \eqref{eq:ergodic:Ym:delta} shall be discussed in the next subsection.}

The solution component $\varrho_m$ of the ergodic BSDE \eqref{eq:ergodic:Ym:delta} is the risk-sensitive growth rate. The following explains this economic interpretation. First, we have the following lemma, which holds not only for $n=m$, but also any $n=0,\dots,m$, using a Lyapunov-type argument.
\begin{lemma}
\label{lem:exp:estimate:factor}
    Suppose that Assumptions \ref{ass:lip:diss:factor}, \ref{ass:factor} and \ref{ass:ergodic:kappa:g} hold. Then, for any $c>0$, there exists $K_c>0$ such that, for any $n=0,\dots,m$, $(\theta_{(n)},l_{(n)})\in\Delta_n\times E^n$, and $t\geq \theta_n$, 
        \begin{equation*}
            \mathbb{E}\left[e^{c|\Phi^n_{t}(\theta_{(n)},l_{(n)})|} \right] \leq K_c, \  \mathbb{E}\left[e^{-c|\Phi^n_{t}(\theta_{(n)},l_{(n)})|} \right] \geq \frac{1}{K_c} > 0.  
        \end{equation*}
\end{lemma}
\begin{proof} See Appendix \ref{sec:pf:lem:exp:estimate:factor}.
\end{proof}
Next, we introduce the following assumption.


    \begin{assumption}
        \label{ass:density:growth}
        For any $t\geq 0$, there exist measurable functions $h_t(\cdot,\cdot), H_t(\cdot,\cdot): \Delta_m\times E^m \to \mathbb{R}_+$ (i.e., for any $t\geq0$, $(\theta_{(m)},l_{(m)}) \mapsto h_t(\theta_{(m)},l_{(m)})$ and  $(\theta_{(m)},l_{(m)}) \mapsto H_t(\theta_{(m)},l_{(m)})$ are $\mathcal{B}(\Delta_m)\otimes \mathcal{B}(E^m)$-measurable), such that for any $(\theta_{(m)},l_{(m)})\in\Delta_m\times E^m$,   $h_t(\theta_{(m)},l_{(m)}) \leq \eta_t(\theta_{(m)},l_{(m)}) \leq H_t(\theta_{(m)},l_{(m)})$. In addition,  the functions satisfy that
    \begin{equation}
                \label{eq:eta:condition:growth}
                \begin{aligned}
                    &             \sup_{T>0} \int_{\Delta_m\times E^m} H_T(\theta_{(m)},l_{(m)})\mathbbm{1}_{\{T\geq \theta_m\} } d\theta_{(m)}\boldsymbol{\lambda}(dl_{(m)}) < \infty, \\
                    &   \inf_{T>0} \int_{\Delta_m\times E^m} h_T(\theta_{(m)},l_{(m)})\mathbbm{1}_{\{T\geq \theta_m\} } d\theta_{(m)}\boldsymbol{\lambda}(dl_{(m)}) > 0 . 
                \end{aligned}
    \end{equation}
    \end{assumption}

{\color{black} 

Assumption \ref{ass:density:growth} allows us to estimate expectations that involve the density together with the exponential utility by handling their contributions separately. Specifically, the density is handled via its prescribed growth bounds \eqref{eq:eta:condition:growth}, whereas the exponential factor is estimated independently by Lemma \ref{lem:exp:estimate:factor}.}

\begin{proposition}
\label{pp:risk:sensitive:rho}
   Suppose that Assumptions \ref{ass:lip:diss:factor}, \ref{ass:factor}, \ref{ass:ergodic:kappa:g}, and \ref{ass:density:growth} hold, and that either \eqref{eq:Cpi:solution} holds for $n=m$, or both Assumptions \ref{ass:pi:compact}--\ref{ass:Cphi:Cg} are satisfied.
   Then, there exists a sequence $(\rho_i)_{i=1}^\infty$ with $\rho_i\to 0$, such that 
\begin{equation*}
        \varrho_m = \lim_{T\to\infty} \lim_{i\to\infty} \frac{1}{T}\log \mathbb{E}\left[\left. e^{-\gamma\left( X^{\pi^*,\rho_i}_T - X^{\pi^*,\rho_i}_{T_m} \right)} \right|\mathbbm{1}_{ \{T\geq T_m\}} \right],
    \end{equation*} 
    where $X^{\pi^*,\rho_i}$ represents the optimal  wealth process adapting the optimal investment strategy $\pi^*$ that depends on the discount rate $\rho_i$ via the solution $\left(Y^{m,\rho_i},Z^{m,\rho_i}\right)$ of the infinite-horizon BSDE \eqref{eq:Ym} with the driver \eqref{eq:factor:drivers_1}.
\end{proposition}
\begin{proof} See Appendix \ref{sec:pf:pp:risk:sensitive:rho}.
\end{proof}

The term 
$e^{-\gamma\left(X^{\pi^*,\rho_i}_T - X^{\pi^*,\rho_i}_{T_m}\right)}$ represents the risk-sensitive exponential transformation of the wealth increment between the $m$-th default time $T_m$ and the horizon $T$. When the horizon $T$ approaches to infinity, the limit $\varrho_m$ measures the asymptotic risk-sensitive (certainty equivalent) growth rate of the investor’s wealth following the $m$-th default as $\rho_i\to 0$. It captures the long-term impact of default events on risk-adjusted performance.

\subsection{Construction of Forward Performance Processes via Ergodic BSDEs} 
\label{sec:factor:ergodic:construction}
In the last subsection, we have shown the existence of the solution to a corresponding ergodic BSDE \eqref{eq:ergodic:Ym:delta}, with the driver given in \eqref{eq:factor:drivers_1}, from the infinite-horizon BSDE \eqref{eq:Ym}, when $n=m$. This section discusses the cases when $n=0,1,\dots,m-1$, and thus constructing a forward exponential utility with multiple defaults via a system of ergodic BSDEs, including $n=0,1,\dots,m$.


When $n=m$, the driver \eqref{eq:factor:drivers_1} of the BSDE \eqref{eq:Ym} does not depend on the solution of a BSDE after the next default event, since all default events happened in this case. Hence, by \eqref{eq:y:z:markov:bound:lip} and \eqref{eq:linear:growth:m}, diagonal arguments can be utilized to construct the sequence $(\rho_i)_{i=1}^\infty$, with $\rho_i\to 0$, to establish the convergence of $\bar{y}^{m,\rho_i}$ with respect to a fixed reference point. Therein, $\hat{\phi}_m\in\mathbb{R}^d$ serves as the fixed reference point only for the equation \eqref{eq:Ym} when $n=m$, but not for other equations \eqref{eq:Yn} when $n=0,1,\dots,m-1$.

To discuss a system of ergodic BSDEs, including $n=0,1,\dots,m$, one can fix $\hat{\boldsymbol{\phi}} = (\hat{\phi}_n)_{n=0}^m\in \mathbb{R}^{d\times (m+1)}$, and consider the perturbation $\bar{y}^{n,\rho}(\phi):= y^{n,\rho}(\phi)-y^{n,\rho}(\hat{\phi}_n)$, for $n=0,1,\dots,m$, $\rho>0$, and $\phi\in\mathbb{R}^{d}$. The uniform-in-$\rho$ Lipschitz continuity and linear growth of $y^{n,\rho}$ are inherited by $\bar{y}^{n,\rho}$, so the diagonal arguments can be applied to obtain subsequential convergence, for each $n=0,1,\dots,m$. Yet, for $n=0,1,\dots,m-1$, the associated process $\bar{Y}^{n,\rho}_t(\theta_{(n)},l_{(n)}):= \bar{y}^{n,\rho}(\Phi^n_t(\theta_{(n)},l_{(n)}))$, for $(\theta_{(n)},l_{(n)})\in \Delta_n\times E^n$ and $t\geq \theta_n$, would satisfy:
        \begin{align}
\label{eq:Yn:peturb}
    & \ \ \ \ d\bar{Y}^{n,\rho}_t(\theta_{(n)},l_{(n)}) \nonumber \\
    &=\Bigg( \rho y^{n,\rho}(\hat{\phi}_n) + \rho \bar{Y}^{n,\rho}_t(\theta_{(n)},l_{(n)}) - \min_{\pi\in\Pi_n} \bigg\{ \hat{F}^n_1\left(\pi,Z^{n,\rho}_t(\theta_{(n)},l_{(n)}),\Phi^n_t(\theta_{(n)},l_{(n)})\right)\nonumber  \\
    &\quad + \frac{e^{\gamma(y^{n+1,\rho}(\hat{\phi}_{n+1}) - y^{n,\rho}(\hat{\phi}_n))} }{\gamma}  \int_E e^{\gamma\left( \bar{Y}^{n+1,\rho}_t((\theta_{(n)},t),(l_{(n)},l)) - \bar{Y}^{n,\rho}_t(\theta_{(n)},l_{(n)}) - \pi'\hat{\beta}^n(\Phi^{n}_{t^-}(\theta_{(n)},l_{(n)}),l) \right)} \nonumber\\
    &\qquad \lambda_{n+1}(l_{(n)},dl) \bigg\}   \Bigg)  dt   + Z^{n,\rho}_t(\theta_{(n)},l_{(n)})'dW_t;
\end{align}
herein, while $y^{n+1,\rho}(\hat{\phi}_{n+1}) - y^{n,\rho}(\hat{\phi}_n)$ admits an upper bound uniformly in $\rho$ by Proposition \ref{pp:Delta:Y}, the absence of its uniform-in-$\rho$ lower bound allows this term approaching to negative infinity, and thus the term $e^{\gamma\,(y^{n+1,\rho}(\hat{\phi}_{n+1}) - y^{n,\rho}(\hat{\phi}_n))}$ degenerating to zero, as $\rho\to 0$. In that case, the resulting system of ergodic BSDEs, for $n=0,1,\dots,m$, would not be coupled by their solutions between adjacent default-time intervals, and hence preventing the construction of a consistent forward performance process.

One can then attempt the solution construction by fixing a common reference point $\hat{\phi}_0\in \mathbb{R}^d$, independent of the index $n=0,1,\dots,m$. Then, consider the perturbation $\bar{y}^{n,\rho}(\phi):= y^{n,\rho}(\phi)-y^{0,\rho}(\hat{\phi}_0)$, for $n=0,1,\dots,m$, $\rho>0$, and $\phi\in\mathbb{R}^{d}$. While the uniform-in-$\rho$ Lipschitz continuity of $y^{n,\rho}$ is inherited by $\bar{y}^{n,\rho}$, it does not satisfy a uniform-in-$\rho$ linear growth property; indeed, Proposition \ref{pp:Delta:Y} only entails an upper bound for $y^{n,\rho}(\phi + \varphi^{n-1}(\phi)) - y^{n-1,\rho}(\phi)$, for $n=1,\dots,m$, $\rho>0$, and $\phi\in \mathbb{R}^d$, but uniform-in-$\rho$ two-sided bounds for the successive differences between $y^{n,\rho}$ and $y^{n-1,\rho}$ is missed. Consequently, the diagonal arguments actually cannot be invoked to establish the desired subsequential convergence.

These observations deem the necessity to have a uniform-in-$\rho$ lower bound for the difference $y^{n+1,\rho}-y^{n,\rho}$, for $n=0,1,\dots,m-1$. Unlike the regime-switching setting of \cite{hu2020systems}, the symmetry of the BSDEs therein is lost due to the unidirectional dependence of the BSDEs' system here, which prevents the use of the methods in \cite{hu2020systems} to establish such a lower bound; see also, again, Proposition \ref{pp:Delta:Y}.
Therefore, below, we propose a construction of forward performance processes with additional monotonic conditions on the model parameters across default intervals to remedy this structural issue.

\begin{assumption}
\label{ass:extra:ergodic:alpha}
    For any $n=0,\dots,m-1$ and $\phi \in \mathbb{R}^d$, we require that $\hat{\pi}^n(\phi)\in \Pi_n$, where $\hat{\pi}^n(\phi) = \hat{\sigma}^n(\phi)[\hat{\sigma}^n(\phi)'\hat{\sigma}^n(\phi)]^{-1}\hat{\alpha}^n(\phi)/\gamma$. In addition, for any $\phi_1,\phi_2\in \mathbb{R}^d$, 
        \begin{equation}
        \label{eq:alpha:ergodic:ass}
        \begin{aligned}
                        \frac{|\hat{\alpha}^{n}(\phi_1)|^2}{2\gamma} &\geq \frac{|\hat{\alpha}^{n+1}(\phi_2)|^2}{2\gamma} + \frac{\gamma (K_{Z^n})^2 }{2} + |\hat{\alpha}^n(\phi_1)-\hat{\alpha}^{n+1}(\phi_2)| K_{Z^{n}} \\
                        &\quad + (K_{Z^n}+K_{Z^{n+1}})|\hat{\alpha}^{n+1}(\phi_2)|. 
        \end{aligned}
        \end{equation}
 \end{assumption}
\begin{remark}
    By \eqref{eq:drift:bound:Z:factor}, the constants $K_{Z^n}$ and $K_{Z^{n+1}}$ can be arbitrarily small when $C_g$ and $C_\beta$ are sufficiently large and small, respectively. In that case,  \eqref{eq:alpha:ergodic:ass} can be seen as a monotonic relationship on the risk premium. \hfill $\square$ 
\end{remark}

\begin{proposition}
\label{pp:lower:bound}
  Suppose that Assumptions \ref{ass:lip:diss:factor}, \ref{ass:factor}, \ref{ass:ergodic:kappa:g}, and \ref{ass:extra:ergodic:alpha} hold, and that either Assumption \ref{ass:pi:compact:remove} or both Assumptions \ref{ass:pi:compact}–\ref{ass:Cphi:Cg} are satisfied. Then, there exists a non-negative sequence $(\rho_i)_{i=1}^\infty$, $\rho_i\to 0$, such that  for the reference points $\hat{\boldsymbol{\phi}} =(\hat{\phi}_n)_{n=0}^m$,
        \begin{equation*}
           \lim_{i\to \infty} e^{\gamma( y^{n+1,\rho_i}(\hat{\phi}_{n+1}) - y^{n,\rho_i}(\hat{\phi}_n) )}  >0,
        \end{equation*}
    for all $n=0,\dots,m-1$.
\end{proposition}

\begin{proof}
  See Appendix \ref{sec:pf:pp:lower:bound}. 
  \end{proof}
 
With a uniform lower bound established in Proposition \ref{pp:lower:bound}, we construct a system of ergodic BSDEs with a common reference point via the perturbation approach as follows. 
Given $\hat{\phi}_0\in\mathbb{R}^d$, we define $\bar{y}^{n,\rho}(\phi):= y^{n,\rho}(\phi) - y^{0,\rho}(\hat{\phi}_0)$, for $n=0,1,\dots,m$, $\rho>0$, and $\phi\in\mathbb{R}^{d}$. It is clear that $\bar{y}^{n,\rho}$ is globally Lipschitz uniformly in $\rho$: for any $n=0,1,\dots,m$, $\phi_1,\phi_2\in \mathbb{R}^d$ and $\rho>0$, 
    \begin{equation*}
        |\bar{y}^{n,\rho}(\phi_1) - \bar{y}^{n,\rho}(\phi_2)|\leq K_{Z^n}|\phi_1-\phi_2|. 
    \end{equation*}
In addition, there exist $C>0$ and a sequence $(\rho_i)_{i=1}^\infty$, $\rho_i\to 0$, such that, for any $n=0,\dots,m$, $i\in\mathbb{N}$ and $\phi\in\mathbb{R}^d$,
    \begin{equation}
    \label{eq:linear:growth:ergodic}
        |\bar{y}^{n,\rho_i}(\phi)|\leq C(1+|\phi|). 
    \end{equation}
To see this, fix $\hat{\boldsymbol{\phi}}:=(\hat{\phi}_n)_{n=0}^m$ and using Proposition \ref{pp:lower:bound}, for any $C>0$, there exists a sequence $(\rho_i)_{i=1}^\infty$, $\rho_i\to 0$, such that, for any $i\in\mathbb{N}$ and $n=0,\dots,m-1$, 
    \begin{equation*}
        y^{n+1,\rho_i}(\hat{\phi}_{n+1}) - y^{n,\rho_i}(\hat{\phi}_n) \geq -C. 
    \end{equation*}
In addition, using Proposition \ref{pp:Delta:Y}, 
 \begin{align*}
        y^{n+1,\rho_i}(\hat{\phi}_{n+1}) - y^{n,\rho_i}(\hat{\phi}_n) &=  y^{n+1,\rho_i}(\hat{\phi}_{n+1}) -  y^{n+1,\rho_i}(\hat{\phi}_{n} + \varphi^n(\hat{\phi}_n)) \\
        &\quad + y^{n+1,\rho_i}(\hat{\phi}_{n} + \varphi^n(\hat{\phi}_n)) - y^{n,\rho_i}(\hat{\phi}_n)  \\
        &\leq K_{Z^{n+1}}|\hat{\phi}_{n+1} -  (\hat{\phi}_{n} + \varphi^n(\hat{\phi}_n))| + K_{\Delta Y^{n+1}}.
    \end{align*}
Hence, with the given sequence $(\rho_i)_{i=1}^\infty$, for any $n=0,\dots,m$ and $\phi\in \mathbb{R}^d$, 
\begin{align*}
        |\bar{y}^{n,\phi}(\phi)| &= \left|y^{n,\rho_i}(\phi) - y^{n,\rho_i}(\hat{\phi}_n) +  \sum_{j=1}^n \left[ y^{j,\rho_i}(\hat{\phi}_j) - y^{j-1,\rho_i}(\hat{\phi}_{j-1})  \right]\right| \\
        &\leq K_{Z^n}|\phi - \hat{\phi}_n| + \sum_{j=1}^n\left|y^{j,\rho_i}(\hat{\phi}_j) - y^{j-1,\rho_i}(\hat{\phi}_{j-1}) \right| \\
        &\leq C(1+|\phi|),
    \end{align*}
for some $C>0$ independent of $i$.

Using the same argument as in the construction of the solution for the ergodic BSDE \eqref{eq:ergodic:Ym:delta} when $n=m$, as well as that for the ergodic BSDEs when $n=0,1,\dots,m-1$ in the proof of Proposition \ref{pp:lower:bound}, we conclude that  there exists a subsequence $(\rho_{i_k})_{k\in\mathbb{N}}$ of $(\rho_i)_{i\in\mathbb{N}}$, with $\rho_{i_k}\to 0$, a constant $\varrho \in \mathbb{R}$, functions $\bar{y}^n(\cdot):\mathbb{R}^d\rightarrow\mathbb{R}$, for $n=0,\dots,m$, and functions $\bar{z}^n(\cdot):\mathbb{R}^d\rightarrow\mathbb{R}^d$, for $n=0,\dots,m$, such that, for any $n=0,\dots,m$, $\phi\in\mathbb{R}^d$,     \begin{equation*}
        \lim_{k\to\infty}\rho_{i_k}y^{0,\rho_{i_k}}(\hat{\phi}_{0}) = \varrho\in \mathbb{R}, \ \lim_{k\to\infty} \bar{y}^{n,\rho_{i_k}}(\phi) = \bar{y}^n(\phi), \ \lim_{k\to\infty}z^{n,\rho_{i_k}}(\phi) = \bar{z}^n(\phi),
    \end{equation*}
and $\bar{y}^n$ is Lipschitz continuous and satisfies a linear growth condition. In addition, for $n=0,\dots,m$, $(\theta_{(n)},l_{(n)})\in \Delta_n\times E^n$, and $t\geq \theta_n$, denote $\mathcal{Y}^n_t(\theta_{(n)},l_{(n)}):=\bar{y}^n(\Phi^n_t(\theta_{(n)},l_{(n)}))$ and $\mathcal{Z}^n_t(\theta_{(n)},l_{(n)}):=\bar{z}^n(\Phi^n_t(\theta_{(n)},l_{(n)}))$, which is the solution of the following ergodic BSDEs: at $n=m$, for $(\theta_{(m)},l_{(m)})\in \Delta_m\times E^m$, and $t\geq \theta_m$,
    \begin{equation}
        \label{eq:ergodic:Ym}
        \begin{aligned}
                 d\mathcal{Y}^{m}_t(\theta_{(m)},l_{(m)}) &=\Bigg(  \varrho   - \min_{\pi\in\Pi_m}  \hat{f}^m\left(\pi,\mathcal{Z}^{m}_t(\theta_{(m)},l_{(m)}),\Phi^m_t(\theta_{(m)},l_{(m)})\right)  \Bigg)  dt   \\
                 &\quad + \mathcal{Z}^{m}_t(\theta_{(m)},l_{(m)})'dW_t,
        \end{aligned}    
    \end{equation}
where the driver is given in \eqref{eq:factor:drivers_1}; for $n=0,1,\dots,m-1$, $(\theta_{(n)},l_{(n)})\in \Delta_n\times E^n$, and $t\geq \theta_n$,
     \begin{equation}
\label{eq:ergodic:Yn}
\begin{aligned}
    d\mathcal{Y}^{n}_t(\theta_{(n)},l_{(n)}) &=\Bigg(  \varrho  - \min_{\pi\in\Pi_n} \bigg\{ \hat{F}^n_1\left(\pi,\mathcal{Z}^{n}_t(\theta_{(n)},l_{(n)}),\Phi^n_t(\theta_{(n)},l_{(n)})\right)  \\
    &\qquad + \frac{1}{\gamma}\int_E e^{\gamma\left( \mathcal{Y}^{n+1}_t((\theta_{(n)},t),(l_{(n)},l)) - \mathcal{Y}^{n}_t(\theta_{(n)},l_{(n)}) - \pi'\hat{\beta}^n(\Phi^{n}_{t^-}(\theta_{(n)},l_{(n)}),l) \right)} \\
    &\qquad \lambda_{n+1}(l_{(n)},dl) \bigg\}   \Bigg)  dt   + \mathcal{Z}^{n}_t(\theta_{(n)},l_{(n)})'dW_t,
\end{aligned}
\end{equation}
where the driver is given in \eqref{eq:factor:drivers_2}.

The system of ergodic BSDEs \eqref{eq:ergodic:Ym}-\eqref{eq:ergodic:Yn} admits a unique solution as depicted below. 

\begin{theorem}\label{th:EBSDE}
Suppose that Assumptions \ref{ass:lip:diss:factor}, \ref{ass:factor}, \ref{ass:ergodic:kappa:g}, and \ref{ass:extra:ergodic:alpha} hold, and that either Assumption \ref{ass:pi:compact:remove} or both Assumptions \ref{ass:pi:compact}–\ref{ass:Cphi:Cg} are satisfied. Then, the system of ergodic BSDEs \eqref{eq:ergodic:Ym}-\eqref{eq:ergodic:Yn} admits a unique Markovian solution 
    \begin{equation*}
        \left((\mathcal{Y}^n(\theta_{(n)},l_{(n)}), \mathcal{Z}^n(\theta_{(n)},l_{(n)}))_{n=0}^m,\varrho\right) = \left((\bar{y}^n(\Phi^n_t(\theta_{(n)},l_{(n)})) ,\bar{z}^n(\Phi^n_t(\theta_{(n)},l_{(n)})))_{n=0}^m  ,\varrho\right), 
    \end{equation*}
    such that each $n=0,\dots,m$ and $\phi_n\in \mathbb{R}^d$,  $|\bar{y}^n(\phi_n)| \leq C(1+|\phi_n|)$ and $|\bar{z}^n({\phi}_n)| \leq C$   for some $C>0$, and the functions $\bar{y}^n(\cdot)$ are unique up to an additive constant. 
\end{theorem}
\begin{proof} See Appendix \ref{sec:pf:th:EBSDE}.
\end{proof}
{\color{black} 
\begin{remark}
   Since the ergodic BSDE \eqref{eq:ergodic:Ym} is decoupled from the remainder of the system, uniqueness of its solution (up to an additive constant in the $\mathcal{Y}^m$-component) follows under Assumptions \ref{ass:lip:diss:factor}, \ref{ass:factor}, \ref{ass:ergodic:kappa:g}, and either condition \eqref{eq:Cpi:solution} for $n=m$, or Assumptions \ref{ass:pi:compact}–\ref{ass:Cphi:Cg}.\hfill $\square$
\end{remark}

\begin{remark}
    By the uniqueness of the solution, the ergodic constants in \eqref{eq:ergodic:Ym:delta} and \eqref{eq:ergodic:Ym}, coincide; i.e., $\varrho_m=\varrho$.  \hfill $\square$
\end{remark}}

Using the ergodic BSDEs \eqref{eq:ergodic:Ym}-\eqref{eq:ergodic:Yn}, we can construct a random field $\mathcal{U}=(\mathcal{U}_t(x))_{x\in \mathbb{R},t\geq 0}$ by 
    \begin{equation}
    \label{eq:U:ergodic}
        \mathcal{U}_t(x) = \sum_{n=0}^m \mathcal{U}^n_t\left(x,T_{(n)},L_{(n)} \right) \mathbbm{1}_{\{T_n\leq t<T_{n+1}\}} + \mathcal{U}^m_t\left(x,T_{(m)},L_{(m)} \right) \mathbbm{1}_{\{t\geq T_m \}},
    \end{equation}
where for $n=0,\dots,m$, $(\theta_{(n)},l_{(n)})\in \Delta_n\times E^n$ and $t\geq \theta_n$, 
    \begin{equation}
        \label{eq:Un:ergodic}
        \mathcal{U}^n_t\left(x,\theta_{(n)},l_{(n)}\right) = \frac{1}{\hat{\eta}^n_t(\theta_{(n)},l_{(n)})} e^{-\gamma x} e^{\gamma\left( \mathcal{Y}^n_t(\theta_{(n)},l_{(n)}) - \mathcal{Y}_0 - \varrho t   \right) }. 
    \end{equation}

Following the same arguments as in Section \ref{sec:verification}, it is straightforward to verify that the random field $\mathcal{U}$ defined above is a forward exponential utility:
\begin{theorem}
\label{thm:ergodic:pi:value}
    Suppose that Assumptions \ref{ass:lip:diss:factor}, \ref{ass:factor}, \ref{ass:ergodic:kappa:g}, and \ref{ass:extra:ergodic:alpha} hold, and that either Assumption \ref{ass:pi:compact:remove} or both Assumptions \ref{ass:pi:compact}–\ref{ass:Cphi:Cg} are satisfied. Then, the random field $(\mathcal{U}_t(x))_{x\in \mathbb{R},t\geq 0}$ defined by \eqref{eq:U:ergodic}-\eqref{eq:Un:ergodic} is a forward exponential utility for \eqref{eq:X} under the factor model \eqref{eq:phi}-\eqref{eq:model:parameter:decompose:factor}, with the set of admissible strategies $\mathcal{A}$ given by
        \begin{equation*}
        \begin{aligned}
          \mathcal{A} &= \bigg\{ \pi = (\pi^n)_{n=0}^m :  \pi^n \in   \mathcal{P}_\mathbb{F}(\Delta_n,E^n;\mathbb{R}^m) , \   \pi^n_t(\cdot,\cdot)\in\Pi_n, \\
          &\quad \int_{\theta_n}^\cdot 
\hat{\sigma}^n(\Phi^n_s(\theta_{(n)},l_{(n)}))'\pi^n_s(\theta_{(n)},l_{(n)})dW_s\    \text{is an $\mathbb{F}$-BMO martingale} \\
&\quad \text{for any $(\theta_{(n)},l_{(n)})\in \Delta_n\times E^n$, $n=0,\dots,m$}  \bigg\}. 
        \end{aligned}
        \end{equation*}
    The optimal investment strategy $\uppi^*=(\uppi^{n*}(\cdot,\cdot))$ is given by, at $n=m$, for $(\theta_{(m)},l_{(m)})\in \Delta_m\times E^m$, $t\geq \theta_m$,
        \begin{equation*}
            \uppi^{*m}_t(\theta_{(m)},l_{(m)}) = \mathop{\arg\min}_{\pi\in\Pi_m}  \hat{f}^m\left(\pi,\mathcal{Z}^{m}_t(\theta_{(m)},l_{(m)}),\Phi^m_t(\theta_{(m)},l_{(m)})\right);
        \end{equation*}
    for $n=0,1,\dots,m-1$, $(\theta_{(n)},l_{(n)})\in \Delta_n\times E^n$, $t\geq \theta_n$, 
      \begin{equation*}
      \begin{aligned}
                    \uppi^{*n}_t(\theta_{(n)},l_{(n)}) &= \mathop{\arg\min}_{\pi\in\Pi_n}  \bigg\{ \hat{F}^n_1\left(\pi,\mathcal{Z}^{n}_t(\theta_{(n)},l_{(n)}),\Phi^n_t(\theta_{(n)},l_{(n)})\right)  \\
    &\quad + \frac{1}{\gamma}\int_E e^{\gamma\left( \mathcal{Y}^{n+1}_t((\theta_{(n)},t),(l_{(n)},l)) - \mathcal{Y}^{n}_t(\theta_{(n)},l_{(n)}) - \pi'\hat{\beta}^n(\Phi^{n}_{t^-}(\theta_{(n)},l_{(n)}),l) \right)}\lambda_{n+1}(l_{(n)},dl)\bigg\}. 
          \end{aligned}
        \end{equation*}
        
\end{theorem}


{\color{black}Similarly, in Proposition \ref{pp:risk:sensitive:rho:2} below, the ergodic constant $\varrho$ can be interpreted as the risk-sensitive growth rate of the wealth process under the optimal investment strategy, where the forward performance process is constructed via the ergodic BSDEs \eqref{eq:ergodic:Ym}-\eqref{eq:ergodic:Yn}. The result follows by an argument parallel to that of Proposition \ref{pp:risk:sensitive:rho}, and therefore the proof is omitted.    } 

\begin{proposition}
\label{pp:risk:sensitive:rho:2}
   Suppose that Assumptions \ref{ass:lip:diss:factor}, \ref{ass:factor}, \ref{ass:ergodic:kappa:g}, \ref{ass:density:growth} and \ref{ass:extra:ergodic:alpha} hold, and that either Assumption \ref{ass:pi:compact:remove} or both Assumptions \ref{ass:pi:compact}–\ref{ass:Cphi:Cg} are satisfied. Then, 
\begin{equation*}
        \varrho = \lim_{T\to\infty} \frac{1}{T}\log \mathbb{E}\left[\left. e^{-\gamma\left( X^{\uppi^*}_T - X^{\uppi^*}_{T_m} \right)} \right|\mathbbm{1}_{ \{T\geq T_m\}} \right],
    \end{equation*} 
where $X^{\uppi^*}$ is the wealth process under the optimal investment strategy $\uppi^*$ with the forward exponential utility given by \eqref{eq:U:ergodic}. 
\end{proposition}

\section{Concluding Remarks}
\label{sec:conclusion}
In this article, we employ the Jacod-Pham decomposition to characterize an exponential forward preference via a sequence of indexed, $\mathbb{F}$-optional infinite-horizon BSDEs. By combining a truncation argument with the comparison principle for BSDEs, we establish the unique existence of the solutions of the BSDEs, their uniform boundedness, and verify the desired (super)martingale property of the forward performance process. This framework also yields the optimal investment strategy associated with the proposed preference.

We further extend the analysis to a stochastic factor model. Under the factor model, a number of uniform estimates for the solutions of the infinite-horizon BSDEs are established, including the one to upper-bound the deviations of solution components across different default intervals and indices. We identify the structural challenges in constructing a forward exponential utility by ergodic BSDEs when uniform lower bounds on the successive differences are absent, and propose a monotonicity condition that could remedy this issue. \textcolor{black}{Under this proposed condition, together with the derived estimates, we demonstrate that the system of infinite-horizon BSDEs converges to a system of ergodic BSDEs. As a direction for future research, it remains to explore whether alternative techniques can fully overcome this one-directional dependence challenge.}

\bibliographystyle{acm}

\bibliography{ref}

\begin{appendices}



\section{Proofs of Section \ref{sec:forward}}
\label{sec: proof Sec 2.6}
This appendix consists of proofs of statements in Section \ref{sec:forward}. 

\subsection{Proof of Lemma \ref{lem:M:inductive}}
\label{sec:pf:lem:M:inductive}
   The case \( n = m \) is clear by taking the conditional expectation with respect to \( \mathcal{F}_{\theta_m} \) on the left-hand side of inequality~\eqref{eq:M:lemma} and using \eqref{eq:M:m:martingale}.
   
In the following, we shall show \eqref{eq:M:lemma} for $n=1,\dots, m-1$ inductively. For $n=m-1$, given any $(\theta_{(m-2)},l_{(m-2)})\in \Delta_{m-2}\times E^{m-2}$ and $s\geq t\geq \theta_{m-2}$, using the tower property of conditional expectation and recall $\hat{\eta}^m(\cdot,\cdot) = \eta(\cdot,\cdot)$, we have
    \begin{align*}
         &\ \mathbb{E}\bigg[\int_E\int_t^s M^{m-1}_s\left((\theta_{(m-2)},\theta_{m-1}),(l_{(m-2)},l_{m-1}) \right) \\
         &\quad\cdot\hat{\eta}^{m-1}_s\left((\theta_{(m-2)},\theta_{m-1}),(l_{(m-2)},l_{m-1}) \right) d\theta_{m-1}\lambda_{m-1}(l_{(m-2)},dl_{m-1}) \big| \mathcal{F}_t    \bigg] \\
           &\ + \mathbb{E}\bigg[ \int_E\int_t^s \bigg\{ \int_E   \int_{\theta_{m-1}}^s   M^m_s\left((\theta_{(m-2)},\theta_{m-1},\theta_m),(l_{(m-2)},l_{m-1},l_m ) \right)  \\
       &\quad \cdot   {\eta}_s((\theta_{(m-2)},\theta_{m-1},\theta_m),(l_{(m-2)},l_{m-1},l_m))d\theta_m \lambda_{m}(l_{(m-1)},dl_m) \bigg\} \\
       &\quad d\theta_{m-1} \lambda_{m-1}(l_{(m-2)},dl_{m-1}) \big| \mathcal{F}_t \bigg] \\
       =&\ \mathbb{E}\Bigg[   \int_E\int_t^s \bigg\{  \mathbb{E}\bigg[  M^{m-1}_s\left((\theta_{(m-2)},\theta_{m-1}),(l_{(m-2)},l_{m-1}) \right) \\
       &\quad \cdot \hat{\eta}^{m-1}_s\left((\theta_{(m-2)},\theta_{m-1}),(l_{(m-2)},l_{m-1}) \right) \\
       &\quad + \int_E   \int_{\theta_{m-1}}^s \mathbb{E}\bigg[   M^m_s\left((\theta_{(m-2)},\theta_{m-1},\theta_m),(l_{(m-2)},l_{m-1},l_m ) \right)  \\
       &\quad \cdot  {\eta}_s((\theta_{(m-2)},\theta_{m-1},\theta_m),(l_{(m-2)},l_{m-1},l_m))\big| \mathcal{F}_{\theta_m} \bigg] d\theta_m \lambda_{m}(l_{(m-1)},dl_m)  \big| \mathcal{F}_{\theta_{m-1}}    \bigg] \\
       &\quad d\theta_{m-1}\lambda_{m-1}(l_{(m-2)},dl_{m-1}) \bigg\} \big| \mathcal{F}_t \Bigg] \\
       \leq &\ \mathbb{E}\Bigg[   \int_E\int_t^s \bigg\{  \mathbb{E}\bigg[  M^{m-1}_s\left((\theta_{(m-2)},\theta_{m-1}),(l_{(m-2)},l_{m-1}) \right) \\
       &\quad \cdot \hat{\eta}^{m-1}_s\left((\theta_{(m-2)},\theta_{m-1}),(l_{(m-2)},l_{m-1}) \right) \\
       &\quad + \int_E   \int_{\theta_{m-1}}^s     M^m_{\theta_m}\left((\theta_{(m-2)},\theta_{m-1},\theta_m),(l_{(m-2)},l_{m-1},l_m ) \right)  \\
       &\quad \cdot   {\eta}_{\theta_m}((\theta_{(m-2)},\theta_{m-1},\theta_m),(l_{(m-2)},l_{m-1},l_m))  d\theta_m \lambda_{m}(l_{(m-1)},dl_m)  \big| \mathcal{F}_{\theta_{m-1}}    \bigg] \\
       &\quad d\theta_{m-1}\lambda_{m-1}(l_{(m-2)},dl_{m-1}) \bigg\} \big| \mathcal{F}_t \Bigg] \\
       \leq &\  \mathbb{E}\bigg[\int_E \int_t^s M^{m-1}_{\theta_{m-1}}\left((\theta_{(m-2)},\theta_{m-1}),(l_{(m-2)},l_{m-1}) \right) \\
       &\quad\cdot \hat{\eta}^{m-1}_{\theta_{m-1}}\left((\theta_{(m-2)},\theta_{m-1}),(l_{(m-2)},l_{m-1}) \right)      d\theta_{m-1} \lambda_{m-1}(l_{(m-2)},dl_{m-1}) \big| \mathcal{F}_t \bigg],
    \end{align*}
    where the last two inequalities follow from \eqref{eq:M:m:martingale} and \eqref{eq:M:G:F:martingale} with $n=m-1$, respectively.

   In general, suppose  \eqref{eq:M:lemma} holds for $n+1$ with $n\leq m-2$. Then, using  the tower property of conditional expectations, we have,
    \begin{align*}
           &\  \mathbb{E}\bigg[ \int_E\int_t^s \bigg\{ \sum_{j=n+1}^m \int_{E^{j-n}} \int_{\theta_n}^s\cdots \int_{\theta_{j-1}}^s  M^j_s\left((\theta_{(n-1)},\theta_{(n,j)}),(l_{(n-1)},l_{(n,j)}) \right)  \\
           &\quad \cdot  \hat{\eta}^{j}_s((\theta_{(n-1)},\theta_{(n,j)}),(l_{(n-1)},l_{(n,j)}))d\theta_j\cdots d\theta_{n+1} \prod_{i=n}^{j-1} \lambda_{i+1}(l_{(i)},dl_{i+1}) \bigg\}  \\
           &\quad d\theta_n \lambda_n(l_{(n-1)},dl_n) \big| \mathcal{F}_t \bigg] \\
           =&\ \mathbb{E}\bigg[ \int_E\int_t^s \bigg\{  \int_E \int_{\theta_n}^s \bigg( M^{n+1}_s\left((\theta_{(n)},\theta_{n+1}),(l_{(n)},l_{n+1}) \right) \\
           &\quad\cdot  \hat{\eta}^{n+1}_s\left((\theta_{(n)},\theta_{n+1}),(l_{(n)},l_{n+1}) \right)\\
           &\quad +
                       \sum_{j=n+2}^m \int_{E^{j-(n+1)}} \int_{\theta_{n+1}}^s\cdots \int_{\theta_{j-1}}^s  M^j_s\left((\theta_{(n)},\theta_{(n+1,j)}),(l_{(n)},l_{(n+1,j)}) \right)  \\
           &\quad \cdot   \hat{\eta}^{j}_s((\theta_{(n)},\theta_{(n+1,j)}),(l_{(n)},l_{(n+1,j)}))d\theta_j\cdots d\theta_{n+2} \prod_{i=n+1}^{j-1} \lambda_{i+1}(l_{(i)},dl_{i+1}) \bigg) \\
           &\quad d\theta_{n+1} \lambda_{n+1}(l_{(n)},dl_{n+1})\bigg\} d\theta_n \lambda_n(l_{(n-1)},dl_n) \big| \mathcal{F}_t \bigg] \\
           =&\ \mathbb{E}\bigg[ \int_E\int_t^s \mathbb{E}\bigg[   \int_E \int_{\theta_n}^s \bigg( M^{n+1}_s\left((\theta_{(n)},\theta_{n+1}),(l_{(n)},l_{n+1}) \right) \\
           &\quad\cdot  \hat{\eta}^{n+1}_s\left((\theta_{(n)},\theta_{n+1}),(l_{(n)},l_{n+1}) \right)\\
           &\quad +
                       \sum_{j=n+2}^m \int_{E^{j-(n+1)}} \int_{\theta_{n+1}}^s\cdots \int_{\theta_{j-1}}^s  M^j_s\left((\theta_{(n)},\theta_{(n+1,j)}),(l_{(n)},l_{(n+1,j)}) \right)  \\
           &\quad \cdot   \hat{\eta}^{j}_s((\theta_{(n)},\theta_{(n+1,j)}),(l_{(n)},l_{(n+1,j)}))d\theta_j\cdots d\theta_{n+2} \prod_{i=n+1}^{j-1} \lambda_{i+1}(l_{(i)},dl_{i+1}) \bigg){\color{black} \big| \mathcal{F}_{\theta_{n+1}} \bigg]}  \\
           &\quad d\theta_{n+1} \lambda_{n+1}(l_{(n)},dl_{n+1})  d\theta_n \lambda_n(l_{(n-1)},dl_n) \big| \mathcal{F}_t \bigg] \\
           \leq &\ \mathbb{E}\bigg[ \int_E\int_t^s \bigg\{ \int_E\int_{\theta_n}^s M^{n+1}_{\theta_{n+1}}\left((\theta_{(n)},\theta_{n+1}),(l_{(n)},l_{(n+1)}) \right) \\
           &\quad \cdot \hat{\eta}^{n+1}_{\theta_{n+1}}\left((\theta_{(n)},\theta_{n+1}),(l_{(n)},l_{(n+1)}) \right)d\theta_{n+1}\lambda_{n+1}(l_{(n)},dl_{n+1}) \bigg\}   d\theta_n\lambda_n(l_{(n)},dl_{n+1}) \big| \mathcal{F}_t \bigg],
    \end{align*}
where the last inequality follows from the induction assumption  at step $n+1$. Using this,  for any $n\leq m-2$, $(\theta_{(n)},l_{(n)})\in \Delta_n\times E^n$, and any $s\geq t \geq \theta_{n-1}$,
        \begin{align*}
           &\ \mathbb{E}\bigg[\int_E\int_t^s M^n_s\left((\theta_{(n-1)},\theta_n),(l_{(n-1)},l_n) \right)   \hat{\eta}^n_s\left((\theta_{(n-1)},\theta_n),(l_{(n-1)},l_n) \right)   \\
           &\quad d\theta_n\lambda_{n}(l_{(n-1)},dl_{n}) \big| \mathcal{F}_t    \bigg] \\
           &\ + \mathbb{E}\bigg[ \int_E\int_t^s \bigg\{ \sum_{j=n+1}^m \int_{E^{j-n}} \int_{\theta_n}^s\cdots \int_{\theta_{j-1}}^s  M^j_s\left((\theta_{(n-1)},\theta_{(n,j)}),(l_{(n-1)},l_{(n,j)}) \right)  \\
           &\quad \cdot    \hat{\eta}^{j}_s((\theta_{(n-1)},\theta_{(n,j)}),(l_{(n-1)},l_{(n,j)})) d\theta_j\cdots d\theta_{n+1} \prod_{i=n}^{j-1} \lambda_{i+1}(l_{(i)},dl_{i+1}) \bigg\} \\
           &\quad d\theta_n \lambda_n(l_{(n-1)},dl_n) \big| \mathcal{F}_t \bigg] \\
           \leq &\ \mathbb{E}\bigg[\int_E\int_t^s M^n_s\left((\theta_{(n-1)},\theta_n),(l_{(n-1)},l_n) \right)  \hat{\eta}^n_s\left((\theta_{(n-1)},\theta_n),(l_{(n-1)},l_n) \right)  \\
           &\quad d\theta_n\lambda_{n}(l_{(n-1)},dl_{n}) \big| \mathcal{F}_t    \bigg] \\
           &\ + \mathbb{E}\bigg[ \int_E\int_t^s \bigg\{ \int_E\int_{\theta_n}^s M^{n+1}_{\theta_{n+1}}\left((\theta_{(n)},\theta_{n+1}),(l_{(n)},l_{(n+1)}) \right) \\
           &\quad \cdot  \hat{\eta}^{n+1}_{\theta_{n+1}}\left((\theta_{(n)},\theta_{n+1}),(l_{(n)},l_{(n+1)}) \right)d\theta_{n+1}\lambda_{n+1}(l_{(n)},dl_{n+1}) \bigg\} \\
           &\quad d\theta_n\lambda_n(l_{(n)},dl_{n+1}) \big| \mathcal{F}_t \bigg]\\
           \leq &\ \mathbb{E}\bigg[\int_E \int_t^s M^n_{\theta_n}\left((\theta_{(n-1)},\theta_n),(l_{(n-1)},l_n) \right) \hat{\eta}^n_{\theta_n}\left((\theta_{(n-1)},\theta_n),(l_{(n-1)},l_n) \right)   \\
           &\quad d\theta_n \lambda_n(l_{(n-1)},dl_n) \big| \mathcal{F}_t \bigg],
        \end{align*}
    where the last line follows from \eqref{eq:M:G:F:martingale}. Therefore, we arrive at \eqref{eq:M:lemma}. 
    
    Following the above calculations, it is clear that the inequalities become equalities when equalities hold in \eqref{eq:M:m:martingale}, and in \eqref{eq:M:G:F:martingale} for any $n=0,\dots,m-1$. The proof is thus complete. \hfill $\square$

\subsection{Proof of  Lemma \ref{lem:M:G:F:martingale}}
\label{sec:pf:lem:M:G:F:martingale}
 We first show that $M$ is a $\mathbb{G}$-supermartingale given the inequalities \eqref{eq:M:G:F:martingale} for all $n=0,\dots,m-1$, and  \eqref{eq:M:m:martingale}. The martingale property can be shown by replacing the inequalities with equalities in the subsequent calculations.

  For any $0\leq t\leq s$, we have
    \begin{equation}
    \label{eq:M:1}
        \mathbb{E}[M_s|\mathcal{G}_t] = \sum_{n=0}^{m-1}\mathbb{E}\left[M_s \mathbbm{1}_{ \{T_n \leq t< T_{n+1} \} } | \mathcal{G}_t \right] + \mathbb{E}[M_s \mathbbm{1}_{\{t\geq T_m \} }|\mathcal{G}_t].
    \end{equation}
 By noticing that $\mathbbm{1}_{\{t\geq T_m\}}T_{(m)}$ and $\mathbbm{1}_{\{t\geq T_m\}}L_{(m)}$ are $\mathcal{G}_t$-measurable, the last term of \eqref{eq:M:1} can be computed as follows:
        \begin{align}
            \mathbb{E}[M_s \mathbbm{1}_{\{t\geq T_m \} }|\mathcal{G}_t] &= \mathbb{E}[M^m_s(T_{(m)},L_{(m)}) \mathbbm{1}_{\{t\geq T_m \} }|\mathcal{G}_t] \nonumber \\
            &= \mathbb{E}\left[M^m_s(\theta_{(m)},l_{(m)}) \mathbbm{1}_{\{t\geq \theta_m \} }|\mathcal{G}_t \right]\bigg|_{\theta_{(m)}=T_{(m)}, l_{(m)}=L_{(m)} } \nonumber \\
            &=  \frac{\mathbbm{1}_{\{t\geq T_m \} } \mathbb{E}\left[M^m_s(\theta_{(m)},l_{(m)})  \eta_s(\theta_{(m)},l_{(m)}) |\mathcal{F}_t \right]\bigg|_{\theta_{(m)}=T_{(m)}, l_{(m)}=L_{(m)} }}{\eta_t(T_{(m)},L_{(m)})}  \label{eq:M:s:m:intermediate}   \\
            &\leq  \frac{\mathbbm{1}_{\{t\geq T_m \} }}{\eta_t(T_{(m)},L_{(m)})}\left(M^m_t(T_{(m)},L_{(m)}) \eta_t(T_{(m)},L_{(m)})  \right) \nonumber\\
            &= M^m_t(T_{(m)},L_{(m)})\mathbbm{1}_{\{t\geq T_m \} },
                    \label{eq:M:s:m}
        \end{align}
    where \eqref{eq:M:s:m:intermediate} follows from Lemma 5.24 of \cite{aksamit:filtration:2017}, and the second-to-last line follows from \eqref{eq:M:m:martingale}.

    To proceed, for any $n=0,\dots,m-1$, $(\theta_{(m)},l_{(m)})\in \Delta_m\times E^m$ and $t\geq 0$, define
        \begin{equation*}
            \hat{M}^n_t\left(\theta_{(m)},l_{(m)}\right) := \sum_{j=n}^{m-1} M^j_t\left(\theta_{(j)},l_{(j)} \right) \mathbbm{1}_{ \{ \theta_j \leq t < \theta_{j+1} \} } + M^m_t(\theta_{(m)},l_{(m)}) \mathbbm{1}_{ \{ t\geq \theta_m\} },
        \end{equation*}
 with $\hat{M}_t^0(\cdot,\cdot)=M_t$.        
Using the filtration switching formula from $\mathcal{G}_t$ to $\mathcal{F}_t$ (See Lemma 2.9 of \cite{aksamit:filtration:2017}), and the fact that $\mathbbm{1}_{ \{ T_n\leq t \} }T_{(n)}$ and $\mathbbm{1}_{ \{ T_n\leq t \} }L_{(n)}$ are $\mathcal{G}_t$-measurable, we have
    \begin{align}
    \label{eq:M:n:conditional}
        &\ \mathbb{E}\left[ M_s \mathbbm{1}_{\{ T_n\leq t < T_{n+1} \} }\big| \mathcal{G}_t\right] \nonumber \\
         =&\ \Bigg[  \mathbbm{1}_{ \{ \theta_{n}\leq t < T_{n+1}\} } \mathbb{E}\left[ \hat{M}^{n}_s\left( (\theta_{(n)},T_{(n+1,m)} ),(l_{(n)},L_{(n+1,m)}) \right) \mathbbm{1}_{ \{ T_{n+1}> t\} } \big| \mathcal{F}_t  \right]  \nonumber \\
         &\ \times \bigg(  \int_{E^{m-n}}\int_t^\infty \int_{\theta_{n+1}}^\infty \cdots \int_{\theta_{m-1}}^\infty \eta_t\left( (\theta_{(n)}, \theta_{(n+1,m)}),(l_{(n)},l_{(n+1,m)}) \right)\nonumber \\
         &\quad d\theta_m\cdots d\theta_{n+1} \prod_{j=n+1}^m \lambda_j(l_{(j-1)},dl_{j}) \bigg)^{-1} \Bigg]_{\theta_{(n)}=T_{(n)}, l_{(n)} = L_{(n)} }\nonumber \\
        =&\ \Bigg[  \frac{ \mathbbm{1}_{ \{ \theta_{n}\leq t < T_{n+1}\} } }{ \hat{\eta}^{n}_t(\theta_{(n)},l_{(n)}) } \nonumber \\
        &\ \times \mathbb{E}\left[ \hat{M}^{n}_s\left( (\theta_{(n)},T_{(n+1,m)} ),(l_{(n)},L_{(n+1,m)}) \right) \mathbbm{1}_{ \{ T_{n+1}> t\} } \big| \mathcal{F}_t  \right] \Bigg]_{\theta_{(n)}=T_{(n)},l_{(n)} = L_{(n)} },
\end{align}
where $T_{(n,m)}:= (T_n,\dots,T_m)$ and $L_{(n,m)} := (L_n,\dots,L_m)$.

Consider for $s\geq t\geq \theta_{n}$,
    \begin{align}
    \label{eq:M:n-1}
     &\   \mathbb{E}\left[ \hat{M}^{n}_s\left( (\theta_{(n)},T_{(n+1,m)} ),(l_{(n)},L_{(n+1,m)}) \right) \mathbbm{1}_{ \{ T_{n+1}> t\} } \big| \mathcal{F}_t  \right]  \nonumber \\
     =&\    \mathbb{E}\left[ \hat{M}^{n}_s\left( (\theta_{(n)},T_{(n+1,m)} ),(l_{(n)},L_{(n+1,m)}) \right)\mathbbm{1}_{ \{t\geq \theta_n \} }  \big| \mathcal{F}_t  \right] \nonumber  \\
      &\ -  \mathbb{E}\left[ \hat{M}^n_s\left( (\theta_{(n)},T_{(n+1,m)} ),(l_{(n)},L_{(n+1,m)}) \right) \mathbbm{1}_{ \{ t \geq T_{n+1} \} } \big| \mathcal{F}_t  \right]  .
    \end{align}
We shall express the conditional expectations \eqref{eq:M:n-1} in terms of integrals with respect to the survival density functions in \eqref{eq:hat:eta}. To this end, consider for any $n<j<m$, $(\theta_{(n)},l_{(n)})\in \Delta_n\times E^n$ and $s\geq t\geq \theta_n$,
    \begin{align}
    \label{eq:M:n-1:1:term}
     &\   \mathbb{E}\left[M^j_s\left((\theta_{(n)}, T_{(n+1,j)}),(l_{(n)}, L_{(n+1,j)})\right) \mathbbm{1}_{ \{T_j \leq s < T_{j+1} \} } \mathbbm{1}_{ \{t\geq T_{n+1} \} } \big| \mathcal{F}_s \right] \nonumber \\
     =&\ \int_{E^{m-n}\times \{ \theta_n\leq \theta_{n+1}\leq \cdots \leq \theta_m<\infty \}  } M^j_s\left((\theta_{(n)}, \theta_{(n+1,j)}),(l_{(n)}, l_{(n+1,j)})\right)  \mathbbm{1}_{\{ \theta_j\leq s < \theta_{j+1}  \}}  \nonumber  \\
     &\quad \cdot \mathbbm{1}_{\{ t\geq \theta_{n+1} \} }  \eta_s\left((\theta_{(n)}, \theta_{(n+1,m)}),(l_{(n)}, l_{(n+1,m)})\right) d\theta_m\cdots d\theta_{n+1} \prod_{i=n+1}^{m} \lambda_i(l_{(i-1)},dl_i) \nonumber \\
     =&\   \int_{E^{j-n}} \int_{\theta_n}^t \int_{\theta_{n+1}}^s \cdots \int_{\theta_{j-1}}^s M^j_s\left((\theta_{(n)}, \theta_{(n+1,j)}),(l_{(n)}, l_{(n+1,j)})\right) \nonumber \\
     &\quad \cdot \Bigg\{ \int_{E^{m-j}}  \int_s^\infty \int_{\theta_{j+1}}^\infty \cdots \int_{\theta_{m-1}}^\infty  \eta_s\left(\theta_{(m)},l_{(m)}\right) d\theta_m\cdots d\theta_{j+2} d\theta_{j+1} \prod_{i =j+1}^m \lambda_i(l_{i-1},dl_i) \Bigg\}      \nonumber \\
     &\quad \cdot  d\theta_j \cdots  d\theta_{n+1} \prod_{i=n+1}^j \lambda_i(l_{i-1},dl_i) \nonumber \\
     =&\ \int_{E^{j-n}} \int_{\theta_n}^t \int_{\theta_{n+1}}^s \cdots \int_{\theta_{j-1}}^s M^j_s\left((\theta_{(n)}, \theta_{(n+1,j)}),(l_{(n)}, l_{(n+1,j)})\right) \nonumber \\
     &\quad \cdot \hat{\eta}^j_s\left((\theta_{(n)}, \theta_{(n+1,j)}),(l_{(n)}, l_{(n+1,j)})\right)  d\theta_j \cdots  d\theta_{n+1} \prod_{i=n+1}^j \lambda_i(l_{i-1},dl_i),
     \end{align}
where the last equality follows from the definition of the survival density function in \eqref{eq:hat:eta}. Using \eqref{eq:M:n-1:1:term} and the tower property of conditional expectations, the second term on the right-hand side of \eqref{eq:M:n-1} can be computed by 
  \begin{align}
    \label{eq:M:n-1:2}
      &\   \mathbb{E}\bigg[ \hat{M}^n_s\left( (\theta_{(n)},T_{(n+1,m)} ),(l_{(n)},L_{(n+1,m)}) \right) \mathbbm{1}_{ \{ t \geq T_{n+1} \} } \big| \mathcal{F}_t   \bigg] \nonumber \\
      =&\ \mathbb{E}\bigg[  \mathbb{E}\left[\hat{M}^n_s\left( (\theta_{(n)},T_{(n+1,m)} ),(l_{(n)},L_{(n+1,m)}) \right) \mathbbm{1}_{ \{ t \geq T_{n+1} \} } \big| \mathcal{F}_s \right] \big| \mathcal{F}_t \bigg] \nonumber \\
      =&\ \mathbb{E}\bigg[ \int_E\int_{\theta_n}^t  M^{n+1}_s\left((\theta_{(n)},\theta_{n+1}),(l_{(n)},l_{n+1}) \right)\nonumber \\
      &\quad \cdot \hat{\eta}^{n+1}_s\left((\theta_{(n)},\theta_{n+1}),(l_{(n)},l_{n+1}) \right) d\theta_{n+1}\lambda_{n+1}(l_{(n)},dl_{n+1}) \big|\mathcal{F}_t   \bigg] \nonumber\\
      &\ + \mathbb{E}\Bigg[  \bigg(   \sum_{j=n+2}^m \int_{E^{j-n}} \int_{\theta_n}^t \int_{\theta_{n+1}}^s \cdots \int_{\theta_{j-1}}^s M^j_s\left( (\theta_{(n)},\theta_{(n+1,j)}),(l_{(n)},l_{(n+1,j)}) \right) \nonumber \\
      &\quad \cdot  \hat{\eta}^j_s((\theta_{(n)},\theta_{(n+1,j)}),(l_{(n)},l_{(n+1,j)})   \bigg)d\theta_{j}\cdots d\theta_{n+1}
 \prod_{i=n+1}^j \lambda_i(l_{(i-1)},dl_i)     \big| \mathcal{F}_t   \Bigg].
    \end{align}
Likewise, by following the derivation of \eqref{eq:M:n-1:1:term} and using the tower property, the first summand of  \eqref{eq:M:n-1} can be computed by, for $s\geq t\geq \theta_n$
     \begin{align}
    \label{eq:M:n-1:1}
       &\ \mathbb{E}\left[ \hat{M}^{n}_s\left( (\theta_{(n)},T_{(n+1,m)} ),(l_{(n)},L_{(n+1,m)}) \right) \mathbbm{1}_{\{t\geq\theta_n\}}  \big| \mathcal{F}_t  \right] \nonumber\\
       =&\ \mathbb{E}\left[
      \mathbb{E}\left[ \hat{M}^{n}_s\left( (\theta_{(n)},T_{(n+1,m)} ),(l_{(n)},L_{(n+1,m)}) \right) \mathbbm{1}_{ \{  t\geq \theta_n  \} } \big| \mathcal{F}_s \right] \big| \mathcal{F}_t  \right] \nonumber\\
      =&\ \mathbb{E}\bigg[  M^n_s\left(\theta_{(n)},l_{(n)}\right) \hat{\eta}^n_s\left(\theta_{(n)},l_{(n)} \right) + \int_E\int_{\theta_n}^s  M^{n+1}_s\left((\theta_{(n)},\theta_{n+1}),(l_{(n)},l_{n+1}) \right)\nonumber \\
      &\quad \cdot \hat{\eta}^{n+1}_s\left((\theta_{(n)},\theta_{n+1}),(l_{(n)},l_{n+1}) \right) d\theta_{n+1}\lambda_{n+1}(l_{(n)},dl_{n+1}) \big|\mathcal{F}_t   \bigg]  \nonumber\\
      &\ + \mathbb{E}\Bigg[  \bigg(   \sum_{j=n+2}^m \int_{E^{j-n}} \int_{\theta_n}^s \int_{\theta_{n+1}}^s \cdots \int_{\theta_{j-1}}^s M^j_s\left( (\theta_{(n)},\theta_{(n+1,j)}),(l_{(n)},l_{(n+1,j)}) \right) \nonumber \\
      &\quad \cdot  \hat{\eta}^j_s\left((\theta_{(n)},\theta_{(n+1,j)}), (l_{(n)},l_{(n+1,j)})\right) \bigg)d\theta_{j}\cdots d\theta_{n+1}
 \prod_{i=n+1}^j \lambda_i(l_{(i-1)},dl_i)     \big| \mathcal{F}_t   \Bigg]. 
    \end{align}
Combining \eqref{eq:M:n-1}, \eqref{eq:M:n-1:2}, and \eqref{eq:M:n-1:1}, for $s\geq t\geq \theta_n$ with $n=0,,\dots, m-1$, we have
    \begin{align}
    \label{eq:M:n-1:3}
        &\ \mathbb{E}\left[ \hat{M}^{n}_s\left( (\theta_{(n)},T_{(n+1,m)} ),(l_{(n)},L_{(n+1,m)}) \right) \mathbbm{1}_{ \{ T_{n+1}> t\} } \big| \mathcal{F}_t  \right] \nonumber \\
        =&\  \mathbb{E}\bigg[  M^n_s\left(\theta_{(n)},l_{(n)}\right) \hat{\eta}^n_s\left(\theta_{(n)},l_{(n)} \right) + \int_E\int_t^s  M^{n+1}_s\left((\theta_{(n)},\theta_{n+1}),(l_{(n)},l_{n+1}) \right)\nonumber \\
      &\quad \cdot \hat{\eta}^{n+1}_s\left((\theta_{(n)},\theta_{n+1}),(l_{(n)},l_{n+1}) \right) d\theta_{n+1}\lambda_{n+1}(l_{(n)},dl_{n+1}) \big|\mathcal{F}_t   \bigg] \nonumber\\
      &\ + \mathbb{E}\Bigg[  \bigg(   \sum_{j=n+2}^m \int_{E^{j-n}} \int_t^s \int_{\theta_{n+1}}^s \cdots \int_{\theta_{j-1}}^s M^j_s\left( (\theta_{(n)},\theta_{(n+1,j)}),(l_{(n)},l_{(n+1,j)}) \right) \nonumber \\
      &\quad \cdot  \hat{\eta}^j_s\left((\theta_{(n)},\theta_{(n+1,j)}), (l_{(n)},l_{(n+1,j)})\right)   \bigg)d\theta_{j}\cdots d\theta_{n+1}
 \prod_{i=n+1}^j \lambda_j(l_{(i-1)},dl_i)     \big| \mathcal{F}_t   \Bigg].
    \end{align}
 By  \eqref{eq:M:G:F:martingale} and  Lemma \ref{lem:M:inductive}, we have
    \begin{align}
    \label{eq:M:n-1:4}
 &\ \mathbb{E}\Bigg[ \int_E \int_t^s \bigg\{  M^{n+1}_s\left((\theta_{(n)},\theta_{n+1}),(l_{(n)},l_{n+1}) \right) \hat{\eta}^{n+1}_s\left((\theta_{(n)},\theta_{n+1}),(l_{(n)},l_{n+1}) \right) \nonumber \\
 &\quad \cdot d\theta_{n+1}\lambda_{n+1}(l_{(n)},dl_{n+1}) \nonumber \\
 &\quad + \sum_{j=n+2}^m \int_{E^{j-n}} \int_t^s\int_{\theta_{n+1}}^s \cdots \int_{\theta_{j-1}}^s M^j_s\left( (\theta_{(n)},\theta_{(n+1,j)}),(l_{(n)},l_{(n+1,j)}) \right) \nonumber \\
      &\quad \cdot  \hat{\eta}^j_s\left((\theta_{(n)},\theta_{(n+1,j)}), (l_{(n)},l_{(n+1,j)})\right) d\theta_{j}\cdots d\theta_{n+1}
 \prod_{i=n+1}^j \lambda_i(l_{(i-1)},dl_i)   \bigg\}   \big| \mathcal{F}_t   \Bigg] \nonumber \\
 \leq&\ \mathbb{E}\bigg[ \int_E \int_t^s  M^{n+1}_{\theta_{n+1}}\left(\theta_{(n)},\theta_{n+1}),(l_{(n)},l_{n+1}) \right)  \hat{\eta}^{n+1}_{\theta_{n+1}}\left(\theta_{(n)},\theta_{n+1}),(l_{(n)},l_{n+1}) \right)  \nonumber\\
 &\quad d\theta_{n+1}d\lambda_{n+1}(l_{(n)},dl_{n+1}) \big| \mathcal{F}_t \bigg].
    \end{align}
Combining      \eqref{eq:M:n:conditional}, \eqref{eq:M:n-1:3} and \eqref{eq:M:n-1:4}, and using the assumption that \eqref{eq:M:G:F:martingale} is satisfied, we obtain
         \begin{align*}
          &\   \mathbb{E}\left[ M_s \mathbbm{1}_{\{ T_n\leq t < T_{n+1} \} }\big| \mathcal{G}_t\right]\\
          \leq&\ \Bigg[  \frac{\mathbbm{1}_{ \{ \theta_n \leq   t < T_{n+1}  \} }}{  \hat{\eta}^n_t\left(\theta_{(n)},l_{(n)} \right) }  \bigg\{ \mathbb{E}\bigg[  M^n_s\left(\theta_{(n)},l_{(n)}\right) \hat{\eta}^n_s\left(\theta_{(n)},l_{(n)} \right)  \\
          &\ +  \int_E \int_t^s  M^{n+1}_{\theta_{n+1}}\left(\theta_{(n)},\theta_{n+1}),(l_{(n)},l_{n+1}) \right) \hat{\eta}^{n+1}_{\theta_{n+1}}\left(\theta_{(n)},\theta_{n+1}),(l_{(n)},l_{n+1}) \right)  \\
          &\quad d\theta_{n+1}d\lambda_{n+1}(l_{(n)},dl_{n+1}) \big| \mathcal{F}_t \bigg] \bigg\} \Bigg]_{\theta_{(n)}=T_{(n)}, l_{(n)}=L_{(n)} } \\
          \leq&\ M^n_t\left(T_{(n)},L_{(n)} \right)\mathbbm{1}_{ \{ T_n\leq t<T_{n+1}  \} }.
        \end{align*}

Hence, for any $0\leq t\leq s$, we have
    \begin{align*}
         \mathbb{E}[M_s|\mathcal{G}_t] &= \sum_{n=0}^{m-1}\mathbb{E}\left[M_s \mathbbm{1}_{ \{T_n \leq t< T_{n+1} \} } | \mathcal{G}_t \right] + \mathbb{E}[M_s \mathbbm{1}_{\{t\geq T_m \} }|\mathcal{G}_t] \\
         &\leq \sum_{n=0}^{m-1}M^n_t\left(T_{(n)},L_{(n)} \right)\mathbbm{1}_{ \{ T_n\leq t<T_{n+1}  \} } + M^m_t\left(T_{(m)},L_{(m)} \right)\mathbbm{1}_{\{ t \geq T_m \}  } \\
         &= M_t.
    \end{align*}
Therefore,   $(M_t)_{t\geq 0}$ is a $\mathbb{G}$-supermartingale. \hfill $\square$

\section{Proofs of Section \ref{sec:exp}}
\label{sec:finite}
This appendix consists of proofs of statements in Section \ref{sec:exp}.
\subsection{Proof of Theorem \ref{thm:exist:Y}}
\label{sec:pf:thm:exist:Y}

 Theorem~\ref{thm:exist:Y} can be proven by combining Propositions~\ref{pp:Y0:wellposed} and~\ref{pp:Yn:exist} below. We first consider the indexed equation   $Y^{m}(\cdot,\cdot)$, which characterizes $Y$ after the last default time.

\begin{proposition}
\label{pp:Y0:wellposed} Under Assumption \ref{ass:bound}-\ref{ass:density}, the index BSDE \eqref{eq:Ym} admits a unique solution $(Y^{m}(\cdot,\cdot),Z^m(\cdot,\cdot))$, such that $Y^m\in \mathcal{S}(\Delta_n,E^n;\mathbb{R})$ and $Z^m  \in \mathcal{M}^2(\Delta_n,E^n;\mathbb{R}^d)\cap \mathcal{L}^2_{\text{loc}}(\Delta_n,E^n;\mathbb{R}^d)$.
\end{proposition}
\begin{proof}
        We shall verify that the driver of \eqref{eq:Ym} fulfills Assumption A1 in \cite{confortola:briand:QBSDE} for any $\theta_{(m)}\in \Delta_m$ and $l_{(m)} \in E^m$. For notational convenience, we shall omit writing the dependence of $\theta_{(m)}$ and $l_{(m)}$ of the relevant processes and coefficients when no confusion is caused.

For any $z \in \mathbb{R}^d$ and $t \geq \theta_m$, consider the mapping $\pi \mapsto f^m(t,\pi,z)$, which is strictly convex in $\pi$ and coercive as $|\pi| \to \infty$ implies $f^m(t,\pi,z) \to +\infty$. Under these conditions and with $\Pi_m$ being closed and convex, there exists a unique minimizer
 \begin{equation*}
                \pi^{m*}(t,z) := \mathop{\arg\min}_{\pi \in \Pi_m} f^m(t,\pi,z)=\mathop{\arg\min}_{\pi \in \Pi_m} \left| (\sigma_t^m)' \pi - \left(z + \frac{\alpha_t^m}{\gamma}\right) \right|^2,
            \end{equation*}
which corresponds to a Mahalanobis projection onto $\Pi_m$. Then, for any $p\in\Pi_m$, we have
            \begin{equation}
            \label{eq:fm:bd}
            \begin{aligned}
                 \frac{\gamma}{2}\left|(\sigma^m_t)'  \pi^{m*}(t,z) -  \left(z + \frac{\alpha^m_t}{\gamma} \right)  \right|^2   & - (\alpha^m_t)'z -\frac{\left|\alpha^m_t\right|^2}{2\gamma} =    \min_{\pi\in\Pi_m} f^m(t,\pi,z) \\
                 &\leq \frac{\gamma}{2}\left|(\sigma^m_t)' p -  \left(z + \frac{\alpha^m_t}{\gamma} \right)  \right|^2   - (\alpha^m_t)'z -\frac{\left|\alpha^m_t\right|^2}{2\gamma}.
            \end{aligned}
            \end{equation}
      By Assumption \ref{ass:bound}, \eqref{eq:fm:bd} implies the existence of $C>0$ such that
         \begin{align}
        \label{eq:pii*:bound}
            |(\sigma^m_t)' \pi^{m,*}(t,z)| & \leq C(|z|+ 1).
        \end{align}
        for any $z\in \mathbb{R}^d$.  Hence, for any  $z_1,z_2 \in\mathbb{R}^d$ and $t\geq \theta_m$, it holds that
            \begin{align*}
              &\  \left(\rho y - \min_{\pi\in \Pi_m} f^m(t,\pi,z_1)\right) - \left(\rho y - \min_{\pi\in \Pi_m} f^m(t,\pi,z_2) \right) \\
              =&\ \min_{\pi\in \Pi_m} f^m(t,\pi,z_2) -\min_{\pi\in \Pi_m} f^m(t,\pi,z_1) \\
              \leq &\ f^m(t, \pi^{m*}(t,z_1) ,z_2) -  f^m(t, \pi^{m*}(t,z_1) ,z_1) \\
              \leq &\   C|z_1-z_2|\left(1 + |z_1|+|z_2| + |(\sigma^m_t)' \pi^{m*}(t,z_1)| \right)    +  C|z_1-z_2| \\
              \leq &\ C(1 + |z_1|+|z_2|)|z_1-z_2|,
            \end{align*}
    where we have used \eqref{eq:pii*:bound} in the last line, and $C>0$ is a constant independent of $z_1,z_2,t,\theta_{(m)},l_{(m)}$, which changes from line to line. By symmetry, we can also show that the existence of  $C>0$ such that
  \begin{align*}
               \left(\rho y - \min_{\pi\in \Pi_m} f^m(t,\pi,z_2)\right) - \left(\rho y - \min_{\pi\in \Pi_m} f^m(t,\pi,z_1) \right) \leq C(1 + |z_1|+|z_2|)|z_1-z_2|,
            \end{align*}
    for any $z_1,z_2 \in\mathbb{R}^d$ and $t\geq \theta_m$. Therefore, we can deduce that
        \begin{align*}
          \left|\left(\rho y - \min_{\pi\in \Pi_m} f^m(t,\pi,z_1)\right) - \left(\rho y - \min_{\pi\in \Pi_m} f^m(t,\pi,z_2) \right)  \right|  \leq  C(1+|z_1|+|z_2|)|z_1-z_2|,
        \end{align*}
    which verifies Assumption A1 (i) of \cite{confortola:briand:QBSDE}.

   Next, for any     $y_1,y_2,\mathbb{R}$, $z\in \mathbb{R}^d$ and $t\geq \theta_m$,
    \begin{align*}
        (y_1-y_2)\left[-\left(\rho y_1 - \min_{\pi\in \Pi_m} f^m(t,\pi,z)\right) + \left(\rho y_2 - \min_{\pi\in \Pi_m} f^m(t,\pi,z) \right)  \right] = -\rho(y_1-y_2)^2,
    \end{align*}
which verifies  the monotonicity condition (Assumption A1 (ii) of \cite{confortola:briand:QBSDE}).

Finally, we verify that the driver is continuous and satisfies a quadratic growth condition; see Definition 3.1 in \cite{confortola:briand:QBSDE}. It is clear that
    \begin{equation*}
        (y,z) \mapsto \rho y - \min_{\pi\in \Pi_m} f^m(t,\pi,z)
    \end{equation*}
is continuous for any $t\geq \theta_m$. On the other hand, for any $(y,z)\in\mathbb{R}\times \mathbb{R}^d$, by \eqref{eq:pii*:bound}, we have
    \begin{align*}
       &\ \left|  \rho y - \min_{\pi\in \Pi_m} f^m(t,\pi,z)\right|\\
       \leq &\ \rho|y| + \frac{\gamma}{2}|z|^2 + | (\sigma^m_t)' \pi^{m*}(t,z)|\left(\gamma|z| + |\alpha^m_t|  \right) + \frac{\gamma|(\sigma^m_t)' \pi^{m*}(t,z)|^2}{2}   \\
       \leq &\ \rho|y| + C|z|^2 + C|z|(1+|z|) + C(1+|z|)^2 \\
       \leq &\ C(1+|y|+|z|^2).
    \end{align*}
Since the constant $C>0$ appearing in the above estimates are uniform over $\Delta_m\times E^m$, thanks to Assumption \ref{ass:bound}, we conclude by Theorem 3.3 of \cite{confortola:briand:QBSDE} that the indexed infinite horizon BSDE \eqref{eq:Ym} admits a unique solution $(Y^m,Z^m)$, where $Y^m\in \mathcal{S}(\Delta_m,E^m;\mathbb{R})$ and $Z^m  \in \mathcal{M}^2(\Delta_m,E^m;\mathbb{R}^d)\cap \mathcal{L}^2_{\text{loc}}(\Delta_m,E^m;\mathbb{R}^d)$.
\end{proof}

Next, we prove that the indexed BSDE for $Y^{n}(\cdot,\cdot)$ defined in \eqref{eq:Yn}  admits a unique solution for all $n=0,\dots,m-1$ inductively.

\begin{proposition}
\label{pp:Yn:exist}
    Suppose that Assumptions \ref{ass:bound}-\ref{ass:density} hold. For $n=0,\dots,m$, suppose that the indexed BSDE \eqref{eq:Yn} admits a unique  solution $(Y^{n+1},Z^{n+1})$ with $Y^{n+1}\in \mathcal{S}(\Delta_{n+1},E^{n+1};\mathbb{R})$ and  $Z^{n+1}\in \mathcal{M}^2(\Delta_{n+1},E^{n+1};\mathbb{R}^d)\cap \mathcal{L}^2_{\text{loc}}(\Delta_{n+1},E^{n+1};\mathbb{R}^d)$. Then, the equation admits a unique solution  $(Y^n,Z^n)$ such that $Y^{n}\in \mathcal{S}(\Delta_n,E^n;\mathbb{R})$ and  $Z^{n}\in \mathcal{M}^2(\Delta_n,E^n;\mathbb{R}^d)\cap \mathcal{L}^2_{\text{loc}}(\Delta_n,E^n;\mathbb{R}^d)$.
\end{proposition}

\begin{proof}
  We shall construct the solution $(Y^{n},Z^{n})$ by comparison principle and truncation. Again, we shall omit the dependence of $\theta_{(n)}$ and $l_{(n)}$ for notational convenience. We begin by writing the indexed BSDE for $(Y^{n},Z^{n})$ as
        \begin{equation}
        \label{eq:Yn:2}
            \begin{aligned}
                    dY^n_t &=\Bigg( \rho Y^n_t -  \min_{\pi\in\Pi_n} \bigg\{ \frac{\gamma}{2}\left|(\sigma^n_t)'\pi - \left(Z^n_t + \frac{\alpha^n_t}{\gamma} \right) \right|^2 \\
                    &\quad +  \frac{1 }{\gamma}\int_E e^{\gamma\left(Y^{n+1}_t((\theta_{(n)},t),(l_{(n)},l)) - Y^n_t - \pi'\beta^n_t(l) \right) }\lambda_{n+1}(l_{(n)},dl) \bigg\}  \Bigg) dt \\
        &\quad  + \left(  (\alpha^n_t)' Z^{n}_t + \frac{|\alpha^n_t|^2}{2\gamma}  \right)dt + Z^{n}_tdW_t.
            \end{aligned}
        \end{equation}

Due to the presence of the term $e^{-\gamma Y^{n}_t}$, it is not clear whether there is a $C>0$ uniform in $y\in \mathbb{R}$ such that the driver in \eqref{eq:Yn:2} satisfies Assumption A1 (i) in \cite{confortola:briand:QBSDE}.   Instead of handling this term directly,  we apply a truncation argument and replace the exponent by a bounded process. This motivates us to  consider the  following (indexed) infinite horizon BSDE: for $t\geq \theta_n$,
        \begin{equation}
        \label{eq:under:Y}
            d\bunderline{Y}^n_t = \left(\rho \bunderline{Y}^n_t  +(\alpha^n_t)' \bunderline{Z}^n_t + \frac{|\alpha^n_t|^2}{2\gamma}\right)dt + (\bunderline{Z}^n_t)'dW_t.
        \end{equation}
The BSDE \eqref{eq:under:Y} is linear, and thus verifies Assumption A.1 in \cite{confortola:briand:QBSDE}. By Theorem 3.3 therein, and Assumption \ref{ass:bound}, we infer that $\bunderline{Y}^n\in \mathcal{S}(\Delta_n,E^n;\mathbb{R})$, and $\bunderline{Z}^n\in  \mathcal{M}^{2}(\Delta_n,E^n;\mathbb{R}^d)\cap \mathcal{L}^2_{\text{loc}}(\Delta_n,E^n;\mathbb{R}^d)$. Indeed, by a change of measure, one can show that for any $\tau \geq \theta_n$, there exists a measure $\mathbb{Q}^\tau\sim \mathbb{P}$ such that, for any $t \in [\theta_n,\tau]$,
    \begin{equation*}
        \bunderline{Y}^n_t = -\mathbb{E}^{\mathbb{Q}_\tau}\left[\int_t^\tau e^{-\rho(s-t)} \frac{|\alpha^n_s|^2}{2\gamma}  ds \bigg| \mathcal{G}_t \right].
    \end{equation*}
In particular, for any $(\theta_{(n)},l_{(n)})$ and $t\geq \theta_{(n)}$,
    \begin{equation}
    \label{eq:lower:bound:Yn}
        \bunderline{Y}^n_t \geq - \frac{ \|\alpha^n\|^2_{\mathcal{S}(\Delta_n,E^n;\mathbb{R}^m)}}{2\rho\gamma} \ \mathbb{P}\text{-a.s.}.
    \end{equation}

Next, we consider the following indexed infinite horizon BSDE, which is a truncated version of \eqref{eq:Yn:2}:
      \begin{equation}
        \label{eq:Yn:tilde}
            \begin{aligned}
                    d\tilde{Y}^n_t &=\Bigg( \rho \tilde{Y}^n_t -  \min_{\pi\in\Pi_n} \bigg\{ \frac{\gamma}{2}\left[(\sigma^n_t)'\pi -\left(\tilde{Z}^n_t + \frac{\alpha^n_t}{\gamma} \right) \right]^2 \\
                    &\quad +  \frac{1}{\gamma}\int_E e^{\gamma\left(Y^{n+1}_t((\theta_{(n)},t),(l_{(n)},l)) - \tilde{Y}^n_t\vee\bunderline{Y}^n_t - \pi'\beta^n_t(l) \right) }\lambda_{n+1}(l_{(n)},dl) \bigg\}  \Bigg) dt \\
        &\quad  + \left(  (\alpha^n_t)' \tilde{Z}^{n}_t + \frac{|\alpha^n_t|^2}{2\gamma}  \right)dt + \tilde{Z}^{n}_tdW_t \\
        &= \left(\rho \tilde{Y}^n_t - \min_{\pi\in\Pi_n}f^n(t,\pi,\tilde{Y}^n_t\vee \bunderline{Y}^n_t,\tilde{Z}^n) \right)dt + \tilde{Z}^n_tdW_t.
            \end{aligned}
        \end{equation}
If we were able to show that \eqref{eq:Yn:tilde} admits a unique solution such that $\tilde{Y}^{n}\in \mathcal{S}(\Delta_n,E^n;\mathbb{R})$, using the comparison principle of BSDEs\footnote{By the comparison principle for BSDEs, we mean the comparison principle for the truncated equation up to a finite time \( T \) with terminal data $0$, followed by passing to the limit as \( T \to \infty \).}
on \eqref{eq:under:Y} and \eqref{eq:Yn:tilde}, we would be able  to conclude that $\tilde{Y}^{n}_t \geq \bunderline{Y}^n_t$. Then, Equation \eqref{eq:Yn:tilde} is reduced to \eqref{eq:Yn}, and thus $(\tilde{Y}^{n},\tilde{Z}^{n})$ is a solution of the latter.

In virtue of the above discussions, the remaining of the proof is devoted to show the unique existence of solution of \eqref{eq:Yn:tilde}, where we shall verify that Assumption A.1 and Definition 3.1 in \cite{confortola:briand:QBSDE} are fulfilled. Given the processes $Y^{n+1}$ and $\bunderline{Y}^n$, we have $\mathbb{P}$-a.s., for all $(\theta_{(n)},l_{(n)})\in \Delta_n\times E^n$, $(t,y,z_1,z_2)\in [\theta_n,\infty)\times \mathbb{R}\times \mathbb{R}^d\times  \mathbb{R}^d$,
\begin{align}
        \label{eq:quad:growth:1}
 &\   \left(\rho y -\min_{\pi\in\Pi_n} f^n(t,\pi,y\vee \bunderline{Y}^n_t,z_1)   \right)  -  \left(\rho y -\min_{\pi\in\Pi_n}   f^n(t,\pi,y\vee \bunderline{Y}^n_t,z_2) \right) \nonumber \\
 \leq &\ \left(\rho y - f^n\left(t,\tilde{\pi}^{n*}(t,y,z_1),y\vee \bunderline{Y}^n_t,z_1\right)   \right)  -  \left(\rho y -  f^n\left(t,\tilde{\pi}^{n*}(t,y,z_1),y\vee \bunderline{Y}^n_t,z_2\right)   \right) \nonumber
 \\
   \leq &\       \frac{\gamma}{2}\Bigg[\left|(\sigma^n_t)'\tilde{\pi}^{n*}(t,y,z_1) -  \left(z_2+ \frac{\alpha^n_t}{\gamma} \right) \right|^2 \nonumber \\
   &\quad - \left|(\sigma^n_t)'\tilde{\pi}^{n*}(t,y,z_1) -  \left(z_1+ \frac{\alpha^n_t}{\gamma} \right) \right|^2 \Bigg]  +( \alpha^n_t)'(z_2-z_1) \nonumber  \\
   =&\  \frac{\gamma}{2}(z_1-z_2)'\left[2(\sigma^n_t)'\tilde{\pi}^{n*}(t,y,z_1) - \left(z_1+z_2 + \frac{2\alpha^n_t}{\gamma} \right) \right]+( \alpha^n_t)'(z_2-z_1)  \nonumber \\
   \leq&\ C|z_1-z_2|\left(1 + |z_1| +|z_2| +|(\sigma^n_t)'\tilde{\pi}^{n*}(t,y,z_1)|  \right) + C|z_1-z_2|,
\end{align}
where for any $(\theta_{(n)},l_{(n)})\in \Delta_n\times E^n$, $t\geq \theta_n$ and $(y,z)\in \mathbb{R}\times \mathbb{R}^d$,
    \begin{equation*}
        \tilde{\pi}^{n*}(t,y,z) := \arg\min_{\pi\in\Pi_n} f^n\left(t,\pi,y\vee \bunderline{Y}^n_t,z\right).
    \end{equation*}
Again, the minimizer exists uniquely, thanks to the geometric properties of $\Pi_n$, together with the convexity and coerciveness of the mapping $\pi \mapsto f^n(t, \pi, y \vee \bunderline{Y}^n_t, z)$.
    
We claim that there exists $C>0$  such that $\mathbb{P}$-a.s., for all $(\theta_{(n)},l_{(n)})\in \Delta_n\times E^n$, $(t,y,z)\in  [\theta_n,\infty)\times \mathbb{R}\times \mathbb{R}^d$,
   \begin{equation}
   \label{eq:pi:bound:tilde}
       | (\sigma^n_t)'\tilde{\pi}^{n*}(t,y,z)| \leq C(1+|z|).
   \end{equation}
 Indeed, for any $p\in \Pi_n$, we have
    \begin{align*}
         f^n\left(t,\tilde{\pi}^{n*}(t,y,z),y\vee \bunderline{Y}^n_t,z\right) \leq  f^n\left(t,p,y\vee \bunderline{Y}^n_t,z\right) \leq  F^n_1(t,p,z) + F^n_2(t,p,\bunderline{Y}^n_t).
    \end{align*}
Using this, and the uniform boundedness of $\bunderline{Y}^n$ and $Y^{n+1}$, we deduce the existence of a positive constant $C>0$ such that $\mathbb{P}$-a.s., for any $(\theta_{(n)},l_{(n)})\in \Delta_n\times E^n$, $(t,y,z)\in [\theta_n,\infty)\times \mathbb{R}\times \mathbb{R}^d$,
    \begin{align}
    \label{eq:pi:p:bound}
       &\   \frac{\gamma}{2}\left|(\sigma^n_t)'p- \left(z + \frac{\alpha^n_t}{\gamma} \right) \right|^2 + C \nonumber \\
       \geq&\   \frac{\gamma }{2}\left| (\sigma^n_t)'\tilde{\pi}^{n*}(t,y,z) -  \left(z + \frac{\alpha^n_t}{\gamma} \right) \right|^2 \nonumber  \\
       &\ +  \frac{1 }{\gamma}\int_E e^{\gamma\left(Y^{n+1}_t((\theta_{(n)},t),(l_{(n)},l)) - \tilde{Y}^n_t\vee\bunderline{Y}^n_t - \tilde{\pi}^{n*}(t,y,z)'\beta^n_t(l) \right) }\lambda_{n+1}(l_{(n)},dl).
    \end{align}
As the second summand on the right-hand side of \eqref{eq:pi:p:bound} is non-negative, we arrive at \eqref{eq:pi:bound:tilde}.

    Therefore, substituting \eqref{eq:pi:bound:tilde} into     \eqref{eq:quad:growth:1}, we see that the left-hand side of  \eqref{eq:quad:growth:1} is bounded by
        \begin{equation*}
            C(1+|z_1|+|z_2|)|z_1-z_2|.
        \end{equation*}
    By switching the order of subtraction in \eqref{eq:quad:growth:1}, and use $\tilde{\pi}^{n*}(t,y,z_2)$ in place of $\tilde{\pi}^{n*}(t,y,z_1)$, we can show that $\mathbb{P}$-a.s., for all $(\theta_{(n)},l_{(n)})\in \Delta_n\times E^n$, $(t,y,z_1,z_2)\in  [\theta_n,\infty)\times \mathbb{R}\times \mathbb{R}^d\times \mathbb{R}^d$,
    \begin{equation}
         \label{eq:Yn:proof:lipschitz}
\begin{aligned}
   &\  \Bigg|\left(\rho y - \min_{\pi\in\Pi_n} f^n\left(t,\pi,y\vee\bunderline{Y}^n_t,z_1\right)   \right)  - \left(\rho y - \min_{\pi\in\Pi_n}f^n\left(t,\pi,y\vee\bunderline{Y}^n_t,z_2\right)  \right) \Bigg|  \\
  & \leq  C|z_1-z_2|(1+|z_1|+|z_2|),
\end{aligned}
\end{equation}
which thus verifies Assumption A.1 (i) of \cite{confortola:briand:QBSDE}.

Next, we verify the driver satisfies the monotonicity condition. For any $(\theta_{(n)},l_{(n)})\in \Delta_n\times E^n$, $(t,y_1,y_2,z)\in [\theta_n,\infty)\times  \mathbb{R}\times\mathbb{R}\times\mathbb{R}^d$, and $\mathbb{P}$-a.s.,
\begin{align*}
        &\ (y_1-y_2)\Bigg[-\left(\rho y_1  -\min_{\pi\in\Pi_n}f^n(t,\pi,y_1\vee \bunderline{Y}^n_t, z)   \right)  +\left(\rho y_2 -  \min_{\pi\in\Pi_n}f^n(t,\pi,y_2\vee \bunderline{Y}^n_t, z)   \right) \Bigg]   \nonumber \\
   = &\ - \rho(y_1-y_2)^2 + (y_1-y_2) \Bigg[ \min_{\pi\in\Pi_n} \bigg\{ \frac{\gamma}{2}\left|(\sigma^n_t)'\pi -\left(z+ \frac{\alpha^n_t}{\gamma} \right) \right|^2  \\
        &\quad + \frac{1 }{\gamma}\int_E e^{\gamma\left(Y^{n+1}_t((\theta_{(n)},t),(l_{(n)},l)) - y_1\vee\bunderline{Y}^n_t - \pi'\beta^n_t(l) \right) }\lambda_{n+1}(l_{(n)},dl) \bigg\}  \\
   &\quad -\min_{\pi\in\Pi_n} \bigg\{ \frac{\gamma}{2}\left|(\sigma^n_t)'\pi - \left(z+ \frac{\alpha^n_t}{\gamma} \right) \right|^2  \\
        &\quad + \frac{1 }{\gamma}\int_E e^{\gamma\left(Y^{n+1}_t((\theta_{(n)},t),(l_{(n)},l)) - y_2 \vee\bunderline{Y}^n_t - \pi'\beta^n_t(l) \right) }\lambda_{n+1}(l_{(n)},dl) \bigg\} \Bigg] \\
   \leq &\ -\rho(y_1-y_2)^2,
   \end{align*}
since the mapping
    \begin{equation*}
    \begin{aligned}
        &\ y\mapsto \min_{\pi\in\Pi_n} \bigg\{ \frac{\gamma}{2}\left|(\sigma^n_t)'\pi - \left(z+ \frac{\alpha^n_t}{\gamma} \right) \right|^2  \\
        &\quad + \frac{1 }{\gamma}\int_E e^{\gamma\left(Y^{n+1}_t((\theta_{(n)},t),(l_{(n)},l)) - y \vee\bunderline{Y}^n_t - \pi'\beta^n_t(l) \right) }\lambda_{n+1}(l_{(n)},dl) \bigg\}
        \end{aligned}
    \end{equation*}
    is non-increasing. Therefore, the driver satisfies Assumption A.1 (ii) of \cite{confortola:briand:QBSDE}.

 To complete the proof, we verify that the driver satisfies the quadratic growth condition. For any $(\theta_{(n)},l_{(n)})\in \Delta_n\times E^n$, $(t,y,z)\in[\theta_n,\infty)\times\mathbb{R}\times\mathbb{R}^d$, and $\mathbb{P}$-a.s., using \eqref{eq:pi:p:bound}, we have
        \begin{align*}
        &\ \left|\rho y - \min_{\pi\in\Pi_n}f^n(t,\pi,y\vee \bunderline{Y}^n_t, z)    \right|\\
        =&\ \Bigg|\rho y  + ( \alpha^n_t)'z + \frac{|\alpha^n_t|^2}{2\gamma} +   \frac{1}{\gamma}\int_E \lambda_{n+1}(l_{(n)},dl) -   \min_{\pi\in\Pi_n} \bigg\{ \frac{\gamma }{2}\left|(\sigma^n_t)' \pi - \left(z + \frac{\alpha^n_t}{\gamma} \right) \right|^2 \nonumber  \\
                    &\quad + \frac{1 }{\gamma}\int_E e^{\gamma\left(Y^{n+1}_t((\theta_{(n)},t),(l_{(n)},l)) - y\vee\bunderline{Y}^n_t - \pi'\beta^n_t(l) \right) }\lambda_{n+1}(l_{(n)},dl) \bigg\}  \Bigg|\\
        \leq &\ C(1+ |y| + |z|) + \min_{\pi\in\Pi_n}\bigg\{ \frac{\gamma}{2}\left|(\sigma^n_t)'\pi -\left(z + \frac{\alpha^n_t}{\gamma} \right) \right|^2 \nonumber  \\
                    &\quad + \frac{1 }{\gamma}\int_E e^{\gamma\left(Y^{n+1}_t((\theta_{(n)},t),(l_{(n)},l)) - y\vee\bunderline{Y}^n_t - \pi'\beta^n_t(l) \right) }\lambda_{n+1}(l_{(n)},dl) \bigg\}\\
        \leq &\   C(1+ |y| + |z|) +  \frac{\gamma}{2}\left|(\sigma^n_t)'p - \left(z + \frac{\alpha^n_t}{\gamma} \right) \right|^2 +C \\
        \leq &\ C(1+|y|+|z|+|z|^2).
        \end{align*}
    Therefore, the driver satisfies a quadratic growth condition. By Theorem 3.3 of \cite{confortola:briand:QBSDE}, we conclude that \eqref{eq:Yn:tilde} admits a unique solution $(\tilde{Y^{n}},\tilde{Z}^{n})$ such that $\tilde{Y}^{n}\in \mathcal{S}(\Delta_n,E^n;\mathbb{R})$  and $\tilde{Z}^{n}\in  \mathcal{M}^{2}(\Delta_n,E^n;\mathbb{R}^d)\cap \mathcal{L}^2_{\text{loc}}(\Delta_n,E^n;\mathbb{R}^d)$.

     To show the uniqueness, let $(Y^n,Z^n)$ and $(\hat{Y}^n,\hat{Z}^n)$ be two solutions of the  indexed BSDE \eqref{eq:Yn}    such that $Y^n,\hat{Y}^n\in \mathcal{S}(\Delta_n,E^n;\mathbb{R})$ and $\tilde{Z}^n, \hat{Z}^n\in  \mathcal{M}^{2}(\Delta_n,E^n;\mathbb{R}^d)\cap \mathcal{L}^2_{\text{loc}}(\Delta_n,E^n;\mathbb{R}^d)$.  Let also $\delta Y^n:= Y^n - \hat{Y}^n$ and $\delta Z^n := Z^n - \hat{Z}^n$. Following the derivation of \eqref{eq:Yn:proof:lipschitz}, and using the fact that $Y^n,\hat{Y}^n\in\mathcal{S}(\Delta_n,E^n;\mathbb{R})$, there exists $C>0$ such that $\mathbb{P}$-a.s., for any $(\theta_{(n)},l_{(n)})\in \Delta_n\times E^n$, $t\geq \theta_n$,
        \begin{equation}
         \label{eq:Yn:proof:lipschitz:unique}
    \left|  \min_{\pi\in\Pi_n} f^n\left(t,\pi,Y^n_t,Z^n_t\right)    - \min_{\pi\in\Pi_n}f^n\left(t,\pi,Y^n_t,\hat{Z}^n_t\right)    \right|   \leq  C|\delta Z^n_t|(1+|Z^n_t|+|\hat{Z}^n_t|).
 \end{equation}
    On the other hand, using the fact that
    \begin{equation*}
    \begin{aligned}
        &\ y\mapsto \min_{\pi\in\Pi_n} \bigg\{ \frac{\gamma}{2}\left|(\sigma^n_t)'\pi - \left(z+ \frac{\alpha^n_t}{\gamma} \right) \right|^2  \\
        &\quad + \frac{1 }{\gamma}\int_E e^{\gamma\left(Y^{n+1}_t((\theta_{(n)},t),(l_{(n)},l)) - y  - \pi'\beta^n_t(l) \right) }\lambda_{n+1}(l_{(n)},dl) \bigg\}
        \end{aligned}
    \end{equation*}
    is non-increasing, we have $\mathbb{P}$-a.s., for any $(\theta_{(n)},l_{(n)})\in \Delta_n\times E^n$, $t\geq \theta_n$,
    \begin{equation}
             \label{eq:Yn:proof:monoton:unique}
        \begin{aligned}
           &\  \delta Y^n_t\Bigg[-\left(\rho Y^n_t  -\min_{\pi\in\Pi_n}f^n(t,\pi, \bunderline{Y}^n_t, \hat{Z}^n_t)   \right)  +\left(\rho \hat{Y}^n_t -  \min_{\pi\in\Pi_n}f^n(t,\pi,\hat{Y}^n_t, \hat{Z}^n_t)   \right) \Bigg] \\
           \leq&\ -\rho |\delta Y^n_t|^2.
        \end{aligned}
    \end{equation}
Using \eqref{eq:Yn:proof:lipschitz:unique} and \eqref{eq:Yn:proof:monoton:unique}, together with an argument analogous to that in Lemma 3.4 of \cite{confortola:briand:QBSDE}, we conclude that $\mathbb{P}$-a.s., for all $(\theta_{(n)},l_{(n)})\in \Delta_n\times E^n$, $t\geq \theta_n$, $Y^n_t=\hat{Y}^n_t$. 
\end{proof}

 \subsection{Proof of Proposition \ref{pp:Yn:bdd:factor}}
 \label{sec:pf:pp:Yn:bdd:factor}
    Notice that from \eqref{eq:lower:bound:Yn}, we have proven that $Y^n_t \geq -K_Y/\rho$. Henceforth, it suffices to derive an upper bound.
    
    To this end,  we follow a similar truncation argument as in the proof of Theorem \ref{thm:exist:Y}. Define the function $h_Y : \mathbb{R}\to\mathbb{R}$ by 
\begin{equation*}
    h_Y(y) = \max\left\{ - \frac{K_Y}{\rho} , \min\left\{y,\frac{K_Y}{\rho} \right\} \right\}.
    \end{equation*}
    Consider the following truncated version of Equations \eqref{eq:Ym}-\eqref{eq:Yn}: we let $(\hat{Y}^m,\hat{Z}^m) = (Y^m,Z^m)$,
and for $n=m-1,\dots,0$, 
\begin{equation}
\label{eq:Yn:tilde:truncated:factor}
    d\hat{Y}^n_t = \left(\rho \hat{Y}^n_t - \min_{\pi\in\Pi_n}f^n\left(t,\pi, h_Y(\hat{Y}^n_t),\hat{Z}^n_t\right) \right)dt + (\hat{Z}^n_t)'dW_t, 
\end{equation}
where we have omitted writing the dependence of the indexes for notational convenience. Notice that $f^n$ depends on the solution $Y^{n+1}$ of the non-truncated equation \eqref{eq:Yn}. By following the proof of Propositions \ref{pp:Y0:wellposed}-\ref{pp:Yn:exist}, it is clear that the system \eqref{eq:Yn:tilde:truncated:factor} is uniquely solvable. 

We shall show that $|\hat{Y}^n_t|\leq K_Y/\rho$ for all $n=0,\dots,m$, and thus the solution of the truncated system  \eqref{eq:Yn:tilde:truncated:factor}
 solves \eqref{eq:Yn}.  To this end, we introduce the following recursively defined ODEs:
    \begin{equation*}
        \bar{Y}^m_t = \int_t^\infty \left( -\rho \bar{Y}^m_s + \|F^m_1(\cdot,0,0)\|_{\mathcal{S}(\Delta_m,E^m;\mathbb{R})} \right)ds  =   \int_t^\infty  -\rho \bar{Y}^m_s ds, 
    \end{equation*}
and for $n=0,\dots,m-1$,
    \begin{equation*}
        \begin{aligned}
            \bar{Y}^n_t &= \int_t^\infty \left(-\rho \bar{Y}^n_s + \frac{e^{\gamma\left( \bar{Y}^{n+1}_s - h_Y(\bar{Y}^n_s) \right)}}{\gamma}  + \max_{0\leq n\leq m}  \|F^n_1\left(\cdot,0,0\right)\|_{\mathcal{S}(\Delta_m,E^m;\mathbb{R})}  \right)ds \\
            &= \int_t^\infty \left(-\rho \bar{Y}^n_s + \frac{e^{\gamma\left( \bar{Y}^{n+1}_s - h_Y(\bar{Y}^n_s) \right)}}{\gamma}   \right)ds. 
        \end{aligned}
    \end{equation*}
By applying the comparison principle of BSDEs, it is clear that $Y^m_t = \hat{Y}^m_t\leq  \bar{Y}^m_t\leq \bar{Y}^{m-1}_t$ for all $t\geq 0$. Using the fact that $Y^m_t\leq \bar{Y}^m_t$, we also have $ \hat{Y}^{m-1}_t\leq \bar{Y}^{m-1}_t$, again by using the comparison principle. In addition, $|\bar{Y}^m_t|\leq K_Y/\rho$ for $t\geq 0$. Therefore,  $h_Y(\bar{Y}^m_t) = \bar{Y}^m_t$ and
   \begin{align*}
        \bar{Y}^{m-1}_t &= \frac{1}{\gamma} \int_t^\infty e^{-\rho(s-t)+ \gamma(\bar{Y}^m_t - h_Y(\bar{Y}^{m-1}_t)) }  ds \\
        &\leq \frac{1}{\gamma}\int_t^\infty e^{-\rho(s-t)+\gamma(\bar{Y}^m_t - h_Y(\bar{Y}^{m}_t))}  ds \\
        &= \frac{1}{\gamma}\int_t^\infty e^{-\rho(s-t)}ds \leq  \frac{K_Y}{\rho}. 
    \end{align*}
Hence, $\hat{Y}^{m-1}_t \leq \bar{Y}^{m-1}_t \leq K_Y/\rho$ and thus $|\hat{Y}^{m-1}_t|\leq K_Y/\rho$. This implies $h_Y(\hat{Y}^{m-1}_t) = \hat{Y}^{m-1}_t$, and so $\hat{Y}^{m-1}_t=Y^{m-1}_t$. \hfill $\square$


Likewise, for $n< m-1$, we can show inductively that $|\hat{Y}^{n}_t|\leq K_Y/\rho$ and $\hat{Y}^n_t = Y^n_t$. The proof is thus complete. 

\section{Proofs of Section \ref{sec:stochastic:factor}}
\label{sec:pf:factor}
This appendix consists of proofs of statements in Section \ref{sec:stochastic:factor}. 

\subsection{Proof of Theorem \ref{thm:Z:markovian}}
\label{sec:pf:thm:Z:markovian}
In the sequel, we will fix $n=0,\dots,m$ and $(\theta_{(n)},l_{(n)})\in \Delta_n\times E^n$. For notational convenience, we shall omit writing the dependence of the processes on $(\theta_{(n)},l_{(n)})$ when no confusion is caused. For instance, we shall write $Y^n_t$ to mean $Y^n_t(\theta_{(n)},l_{(n)})$, and $Y^{n+1}_t(\theta,l)$ to mean $Y^{n+1}_t((\theta_{(n)},\theta),(l_{(n)},l))$ for any $\theta_n\leq \theta\leq t$. 

Our proof grounds on a truncation argument with a modified comparison principle of BSDEs. The major difference from the existing results lies in the fact that the $m+1$ BSDEs are defined on different parameter indexes. 
 
\noindent\underline{Step 1: Truncating the equations}\\
\indent We first introduce the following truncated equations:
\begin{equation}
\label{eq:Yn:tilde:truncated:factor:Z:m}
    dY^m_t = \left(\rho Y^m_t - \min_{\pi\in\Pi_m}\hat{f}^m\left(\pi,h^m_Z(Z^m_t),\Phi^m_t\right) \right)dt + (Z^m_t)'dW_t,
\end{equation}
and for $n=0,\dots,m-1$,
\begin{equation}
\label{eq:Yn:tilde:truncated:factor:Z}
    dY^n_t = \left(\rho Y^n_t - \min_{\pi\in\Pi_n}\hat{f}^n\left(t,\pi, Y^n_t,h^n_Z(Z^n_t),\Phi^n_t; Y^{n+1} \right) \right)dt + (Z^n_t)'dW_t,
\end{equation}
{Here, we will still write the solution of the truncated BSDEs as $Y^n$. To emphasize that the driver $\hat{f}^n$ in fact depends on the truncated version of $Y^{n+1}$, for $n=0,\dots,m-1$, we adopt the notation that
    \begin{equation}
    \label{eq:hat:f:augmented}
    \begin{aligned}
         \hat{f}^n(t,\pi,y,z,\phi;Y^{n+1})&:=  \hat{F}^n_1(\pi,z,\phi) + \hat{F}^n_2(t,\pi,y,\phi;Y^{n+1}),\\
        \hat{F}^n_2(t,\pi,y,\phi;Y^{n+1})&:= \frac{1}{\gamma}\int_Ee^{\gamma\left(Y^{n+1}_t((\theta_{(n)},t),(l_{(n)},l))-y-\pi'\hat{\beta}^n(\phi,l) \right)}\lambda_{n+1}(l_{(n)},dl).
    \end{aligned} 
    \end{equation}
} Besides,  $h^n_Z:\mathbb{R}^d\to \mathbb{R}^d$ is given by 
    \begin{equation*}
        h^n_Z(z) = \frac{z}{|z|} \min\left\{ K_{Z^n} ,|z| \right\} \mathbbm{1}_{\{z\neq 0 \}}. 
    \end{equation*}

For $n<m$, let $s\in[\theta_{(n)},\theta_{(n+1)})$. For any $\phi\in\mathbb{R}^d$, $p\geq n$ and $t\geq s$, we denote by $\Phi^{p,(n,s,\phi)}$ the solution of the SDE $\Phi^p_\cdot(\theta_{(p)},l_{(p)})$ with the condition $\Phi^{n,(n,s,\phi)}_s(\theta_{(n)},l_{(n)})=\phi$. Clearly, $\Phi_t^{p,(n,s,\phi)} = \Phi_t^{p,(p,\theta_p, \Phi^{p,(n,s,\phi)}_{\theta_p})}$ for $t\geq \theta_p$. We also write $\Phi^{n,s,\phi}_t = \Phi^{n,(n,s,\phi)}_t$. For any $0\leq n \leq p$, we  denote by $(Y^{p,(n,s,\phi)},Z^{p,(n,s,\phi)})$ the solution of the truncated equation \eqref{eq:Yn:tilde:truncated:factor:Z} with $\Phi^p$ replaced by $\Phi^{p,(n,s,\phi)}$. Likewise, we write $(Y^{n,s,\phi},Z^{n,s,\phi}) =(Y^{n,(n,s,\phi)},Z^{n,(n,s,\phi)})$.


Fix $\phi_1,\phi_2\in\mathbb{R}^d$, for any $n\leq p$, $\theta_n\leq s\leq \theta_p\leq t$,   we define
    \begin{equation*}
        \delta Y^{p,(n,s,\phi_1,\phi_2)}_t := Y^{p,(n,s,\phi_1)}_t - Y^{p,(n,s,\phi_2)}_t \text{ and }   \delta Z^{p,(n,s,\phi_1,\phi_2)}_t := Z^{p,(n,s,\phi_1)}_t - Z^{p,(n,s,\phi_2)}_t. 
    \end{equation*}
We also write $(\delta Y^{n,s,\phi_1,\phi_2},\delta Z^{n,s,\phi_1,\phi_2})$ to mean $(\delta Y^{n,(n,s,\phi_1,\phi_2)},\delta Z^{n,(n,s,\phi_1,\phi_2)})$. For any $\theta_n\leq s \leq t$, we have  
    \begin{equation}
    \label{eq:delta:Y:n}
    \begin{aligned}
                d\delta Y^{n,s,\phi_1,\phi_2}_t &= \bigg[ \rho \delta Y^{n,s,\phi_1,\phi_2}_t -  \min_{\pi\in\Pi_n} \hat{f}^{n}\left(t,\pi, Y^{n,s,\phi_1}_t, h^n_Z(Z^{n,s,\phi_1}_t), \Phi^{n,s,\phi_1}_t;Y^{n+1,(n,s,\phi_1)}\right) \\
                &\quad +   \min_{\pi\in\Pi_n} \hat{f}^{n}\left(t,\pi, Y^{n,s,\phi_2}_t, h^n_Z(Z^{n,s,\phi_2}_t), \Phi^{n,s,\phi_2}_t;Y^{n+1,(n,s,\phi_2)}\right) \bigg]dt \\
                &\quad + (\delta Z^{n,s,\phi_1,\phi_2}_t)' dW_t. 
    \end{aligned}
    \end{equation}
Also consider the following ODE: for $0\leq s\leq t$, 
    \begin{equation*}
    \begin{aligned}
                \delta \bar{Y}^{n,s}_t &= \int_t^\infty \bigg[ -\rho \delta\bar{Y}^{n,s}_u + \bigg( \frac{C_\phi C_g(1+C_\varphi)^{m-n}}{C_g-C_\phi} \\
                &\quad +C_\Pi C_\beta (C_g+\rho) \sum_{j=0}^{m-n-1} (1+C_\varphi)^j \bigg)e^{-C_g(u-s)} |\phi_1-\phi_2| \bigg] dt. 
    \end{aligned}
    \end{equation*}
We shall prove that, for any $n=0,\dots,m$, $\theta_n\leq s\leq t $, and $\phi_1,\phi_2\in \mathbb{R}^d$, 
    \begin{equation}
    \label{eq:delta:Y:bound}
        \left|\delta Y^{n,s,\phi_1,\phi_2}_t\right| \leq \delta \bar{Y}^{n,s}_t = \left[\frac{C_\phi C_g(1+C_\varphi)^{m-n}}{(C_g-C_\phi)(C_g+\rho)} + C_\Pi C_\beta \sum_{j=0}^{m-n-1}(1+C_\varphi)^j \right]e^{-C_g(t-s)}|\phi_1-\phi_2|. 
    \end{equation}

\noindent\underline{Step 2: Analysis of drivers}\\
\indent To proceed, we analyze the driver of the equation \eqref{eq:delta:Y:n}. For any $n\leq m$, $\theta_n\leq s\leq t$, $i=1,2$, we let
    \begin{equation*}
        \hat{\pi}^{n,i}_t = \mathop{\arg\min}_{\pi\in\Pi_n}   \hat{f}^{n}\left(t,\pi, Y^{n,s,\phi_i}_t, h^n_Z(Z^{n,s,\phi_i}_t), \Phi^{n,s,\phi_i}_t;Y^{n+1,(n,s,\phi_i)}\right).
    \end{equation*}
Consider 
    \begin{align}
    \label{eq:drift:bound:Z:factor}
      &\  \min_{\pi\in\Pi_n} \hat{f}^{n}\left(t,\pi, Y^{n,s,\phi_1}_t, h^n_Z(Z^{n,s,\phi_1}_t), \Phi^{n,s,\phi_1}_t;Y^{n+1,(n,s,\phi_1)}\right)\nonumber \\
      &\ -    \min_{\pi\in\Pi_n} \hat{f}^{n}\left(t,\pi, Y^{n,s,\phi_2}_t, h^n_Z(Z^{n,s,\phi_2}_t), \Phi^{n,s,\phi_2}_t;Y^{n+1,(n,s,\phi_2)}\right) \nonumber \\
      \leq &\ \hat{F}^{n}_1(\hat{\pi}^{n,2}_t,h^n_Z(Z^{n,s,\phi_1}_t),\Phi^{n,s,\phi_1}_t) - \hat{F}^{n}_1(\hat{\pi}^{n,2}_t,h^n_Z(Z^{n,s,\phi_2}_t),\Phi^{n,s,\phi_2}_t) \nonumber\\
      &\ + \hat{F}^{n}_2(t,\hat{\pi}^{n,2}_t,Y^{n,s,\phi_1}_t,\Phi^{n,s,\phi_1}_t;Y^{n+1,(n,s,\phi_1)})-\hat{F}^{n}_2(t,\hat{\pi}^{n,2}_t,Y^{n,s,\phi_2}_t,\Phi^{n,s,\phi_2}_t;Y^{n+1,(n,s,\phi_2)}) \nonumber \\
      =&\  \tilde{F}^{n}_1(t,\delta Z^{n,s,\phi_1,\phi_2}_t) +\hat{F}^{n}_2(t,\hat{\pi}^{n,2}_t,Y^{n,s,\phi_1}_t,\Phi^{n,s,\phi_1}_t;Y^{n+1,(n,s,\phi_1)})\nonumber  \\
      &\ -\hat{F}^{n}_2(t,\hat{\pi}^{n,2}_t,Y^{n,s,\phi_2}_t,\Phi^{n,s,\phi_2}_t;Y^{n+1,(n,s,\phi_2)}) ,
    \end{align}
where 
    \begin{equation*}
        \begin{aligned}
            \tilde{F}^{n}_1(t,z) &:=  \hat{F}^{n}_1(\hat{\pi}^{n,2}_t,h^n_Z(Z^{n,s,\phi_1}_t),\Phi^{n,s,\phi_1}_t) - \hat{F}^{n}_1(\hat{\pi}^{n,2}_t,h^n_Z(Z^{n,s,\phi_1}_t),\Phi^{n,s,\phi_2}_t) \\
            &\ + \hat{F}^{n}_1(\hat{\pi}^{n,2}_t,h^n_Z(z+Z^{n,s,\phi_2}_t),\Phi^{n,s,\phi_2}_t) -\hat{F}^{n}_1(\hat{\pi}^{n,2}_t,h^n_Z(Z^{n,s,\phi_2}_t),\Phi^{n,s,\phi_2}_t).
        \end{aligned}
    \end{equation*}

Consider the first term of the right-hand side of     \eqref{eq:drift:bound:Z:factor}. By  \eqref{eq:F:lip:factor}, \eqref{eq:ergodic:phi}, and  Assumption \ref{ass:Cphi:Cg},  for $t\geq \theta_n$, we have 
    \begin{align}
    \label{eq:tilde:F:n:ergodic}
        |\tilde{F}^{n}_1(t,0)| &\leq C_\phi(1+|h^n_Z(Z^{n,s,\phi_1}_t)|) \left|\Phi^{n,s,\phi_1}_t-\Phi^{n,s,\phi_2}_t \right| \nonumber\\
        &\leq C_\phi e^{-C_g(t-s)}(1+|h_Z^n(Z^{n,s,\phi_1}_t)|)|\phi_1-\phi_2| \nonumber \\
        &\leq  C_g K_{Z^n} e^{-C_g(t-s)}|\phi_1-\phi_2|. 
    \end{align}
Indeed, it is easy to verify that $(C_g-C_\phi)K_{Z^n} > C_\phi$ when $C_g>C_\phi$, and so $C_\phi(1+K_{Z^n}) \leq C_g K_{Z^n}$.  On other hand, using the boundedness of $\hat{\pi}^{n,2}_t$, $h^n_Z(Z^{n,s,\phi_i}_t)$, $i=1,2$, it is easy to check that there exists $C>0$ with 
    \begin{equation}
    \label{eq:drift:bound:Z:factor:1}
    \begin{aligned}
         &\       \left| \hat{F}^{n}_1(\hat{\pi}^{n,2}_t,h^n_Z(Z^{n,s,\phi_1}_t),\Phi^{n,s,\phi_2}_t) -\hat{F}^{n}_1(\hat{\pi}^{n,2}_t,h^n_Z(Z^{n,s,\phi_2}_t),\Phi^{n,s,\phi_2}_t)\right|\leq   C\left|\delta Z^{n,s,\phi_1,\phi_2}_t \right|.
    \end{aligned}
    \end{equation}
Hence, 
    \begin{equation*}
        \begin{aligned}
              &\       \tilde{F}^{n}_1(t,\delta Z^{n,s,\phi_1,\phi_2}_t) \\
              \leq &\  |\tilde{F}^{n}_1(t,0)| +  \left| \hat{F}^{n}_1(\hat{\pi}^{n,2}_t,h^n_Z(Z^{n,s,\phi_1}_t),\Phi^{n,s,\phi_2}_t) -\hat{F}^{n}_1(\hat{\pi}^{n,2}_t,h^n_Z(Z^{n,s,\phi_2}_t),\Phi^{n,s,\phi_2}_t)\right| \\
              \leq &\ C_g K_{Z^n} e^{-C_g(t-s)} |\phi_1-\phi_2| + C\left|\delta Z^{n,s,\phi_1,\phi_2}_t \right|. 
        \end{aligned}
    \end{equation*}
    
Next, we consider the remaining summands on the right-hand side of \eqref{eq:drift:bound:Z:factor}. By the mean value theorem and the  boundedness of $Y^{n},Y^{n+1},\hat{\pi}^{n,2}$, there exists $C>0$ such that 
\begin{align}
\label{eq:Fn:2:bound:factor}
   &\ \hat{F}^{n}_2\!\left(
        t,\hat{\pi}^{n,2}_t,
        Y^{n,s,\phi_1}_t,\Phi^{n,s,\phi_1}_t;
        Y^{n+1,(n,s,\phi_1)}\right) \nonumber \\
   &\quad - \hat{F}^{n}_2\!\left(
        t,\hat{\pi}^{n,2}_t,
        Y^{n,s,\phi_2}_t,\Phi^{n,s,\phi_2}_t;
        Y^{n+1,(n,s,\phi_2)}\right) \nonumber \\[0.3em]
   &= \frac{1}{\gamma}\int_E 
      \Bigg(
        e^{\,\gamma\Big( 
            Y^{n+1,(n,s,\phi_1)}_t(t,l) 
            - Y^{n,s,\phi_1}_t 
            - (\hat{\pi}^{n,2}_t)'\hat{\beta}^n(\Phi^{n,s,\phi_1}_{t^-},l)
        \Big)} \nonumber \\
   &\qquad\qquad -\,
        e^{\,\gamma\Big( 
            Y^{n+1,(n,s,\phi_2)}_t(t,l) 
            - Y^{n,s,\phi_2}_t 
            - (\hat{\pi}^{n,2}_t)'\hat{\beta}^n(\Phi^{n,s,\phi_2}_{t^-},l)
        \Big)}
      \Bigg)\,\lambda_{n+1}(l_{(n)},dl) \nonumber \\[0.3em]
   &\leq C \int_E \Big(
         \delta Y^{n+1,(n,s,\phi_1,\phi_2)}_t(t,l) 
         - \delta Y^{n,s,\phi_1,\phi_2}_t \nonumber \\
   &\qquad\qquad\qquad
         - (\hat{\pi}^{n,2}_t)'\Big(
             \hat{\beta}^n(\Phi^{n,s,\phi_1}_{t^-},l)
             - \hat{\beta}^n(\Phi^{n,s,\phi_2}_{t^-},l)
           \Big)
      \Big)\,\lambda_{n+1}(l_{(n)},dl) \nonumber \\[0.3em]
   &\leq C \Bigg\{
         \big|\delta Y^{n,s,\phi_1,\phi_2}_t
             - \delta \bar{Y}^{n,s}_t\big| \nonumber \\
   &\qquad + \int_E 
           \Big(
             \delta Y^{n+1,(n,s,\phi_1,\phi_2)}_t(t,l) 
             - \delta \bar{Y}^{n,s}_t \nonumber \\
   &\qquad\qquad
             - (\hat{\pi}^{n,2}_t)'\Big(
                 \hat{\beta}^n(\Phi^{n,s,\phi_1}_{t^-},l)
                 - \hat{\beta}^n(\Phi^{n,s,\phi_2}_{t^-},l)
               \Big)
           \Big)_+\,\lambda_{n+1}(l_{(n)},dl)
      \Bigg\}.
\end{align}

\noindent\underline{Step 3: Comparison principle}\\
\indent In this step, we adapt the proof of Lemma 2.2 in \cite{hu2020systems} to show     \eqref{eq:delta:Y:bound}. Let $G^{n,s,\phi_1,\phi_2}_t := \delta Y^{n,s,\phi_1,\phi_2}_t-\delta\bar{Y}_t^{n,s}$, and for any $t\geq \theta$ and $l\in E$, we let  
\begin{equation*}
\begin{aligned}
        \tilde{G}^{n+1,(n,s,\phi_1,\phi_2)}_t(\theta,l) &:= \delta Y^{n+1,(n,s,\phi_1,\phi_2)}_t(\theta,l) \\
        &\ -(\hat{\pi}^{n,2}_t)'\left(\hat{\beta}^n(\Phi^{n,s,\phi_1}_{t^-},l) - \hat{\beta}^n(\Phi^{n,s,\phi_2}_{t^-},l)\right) -\delta\bar{Y}_t^{n,s}. 
\end{aligned}
\end{equation*}
 By applying It\^o's lemma to the process $(G^{n,s,\phi_1,\phi_2}_t)^2_+ = (G^{n,s,\phi_1,\phi_2}_t)^2\mathbbm{1}_{\{ G^{n,s,\phi_1,\phi_2}_t\geq 0  \}}$, using   \eqref{eq:drift:bound:Z:factor}-\eqref{eq:Fn:2:bound:factor} and H\"older's inequality, for any  $\theta_n\leq s \leq t$,
 \begin{align*}
   &\ (G^{n,s,\phi_1,\phi_2}_t)_+^2 \\[0.3em]
   =&\ \int_t^\infty 2(G^{n,s,\phi_1,\phi_2}_u)_+ \Bigg[
         \min_{\pi\in\Pi_n} 
         \hat{f}^{n}\!\Big(
            t,\pi, Y^{n,s,\phi_1}_u, h^n_Z(Z^{n,s,\phi_1}_u), 
            \Phi^{n,s,\phi_1}_u;\, Y^{n+1,(n,s,\phi_1)}
         \Big) \\
   &\quad - \min_{\pi\in\Pi_n} 
         \hat{f}^{n}\!\Big(
            t,\pi, Y^{n,s,\phi_2}_u, h^n_Z(Z^{n,s,\phi_2}_u), 
            \Phi^{n,s,\phi_2}_u;\, Y^{n+1,(n,s,\phi_2)}
         \Big) - \rho G^{n,s,\phi_1,\phi_2}_u \\
   &\quad 
         - \Bigg(
             \frac{C_\phi C_g (1+C_\varphi)^{m-n}}{C_g-C_\phi} 
             + C_\Pi C_\beta (C_g+\rho) 
               \sum_{j=0}^{m-n-1} (1+C_\varphi)^j
           \Bigg) e^{-C_g(u-s)}|\phi_1-\phi_2|
     \Bigg] du \\
   &\ - \int_t^\infty 
         |\delta Z^{n,s,\phi_1,\phi_2}_u|^2\,
         \mathbbm{1}_{\{G^{n,s,\phi_1,\phi_2}_u \geq 0\}}\,du - \int_t^\infty 
         2(G^{n,s,\phi_1,\phi_2}_u)_+
         (\delta Z^{n,s,\phi_1,\phi_2}_u)'\,dW_u \\[0.7em]
   \leq &\ \int_t^\infty 2(G^{n,s,\phi_1,\phi_2}_u)_+ \Big[
          \tilde{F}^n_1(u,\delta Z^{n,s,\phi_1,\phi_2}_u) \\[-0.3em]
   &\quad + \hat{F}^{n}_2\!\Big(
              u,\hat{\pi}^{n,2}_u,Y^{n,s,\phi_1}_u,\Phi^{n,s,\phi_1}_u;\,
              Y^{n+1,(n,s,\phi_1)}
          \Big) \\
   &\quad - \hat{F}^{n}_2\!\Big(
              u,\hat{\pi}^{n,2}_u,Y^{n,s,\phi_2}_u,\Phi^{n,s,\phi_2}_u;\,
              Y^{n+1,(n,s,\phi_2)}
          \Big) \\
   &\quad - \rho G^{n,s,\phi_1,\phi_2}_u 
          - C_g K_{Z^n} e^{-C_g(u-s)}|\phi_1-\phi_2|
        \Big] du \\
   &\ - \int_t^\infty 
         |\delta Z^{n,s,\phi_1,\phi_2}_u|^2\,
         \mathbbm{1}_{\{G^{n,s,\phi_1,\phi_2}_u \geq 0\}}\,du  - \int_t^\infty 
         2(G^{n,s,\phi_1,\phi_2}_u)_+
         (\delta Z^{n,s,\phi_1,\phi_2}_u)'\,dW_u \\[0.7em]
   \leq &\ \int_t^\infty 2(G^{n,s,\phi_1,\phi_2}_u)_+ \Bigg[
          C\,|\delta Z^{n,s,\phi_1,\phi_2}_u| 
          + C \Bigg(
              \int_E 
                 \big(\tilde{G}^{n+1,(n,s,\phi_1,\phi_2)}_u(u,l)\big)_+ 
                 \lambda_{n+1}(l_{(n)},dl) \\
   &\qquad\qquad + G^{n,s,\phi_1,\phi_2}_u
            \Bigg) 
          - \rho G^{n,s,\phi_1,\phi_2}_u
        \Bigg] du \\
   &\ - \int_t^\infty 
         |\delta Z^{n,s,\phi_1,\phi_2}_u|^2\,
         \mathbbm{1}_{\{G^{n,s,\phi_1,\phi_2}_u \geq 0\}}\,du - \int_t^\infty 
         2(G^{n,s,\phi_1,\phi_2}_u)_+
         (\delta Z^{n,s,\phi_1,\phi_2}_u)'\,dW_u \\[0.7em]
   \leq &\ - \int_t^\infty 
          \mathbbm{1}_{\{G^{n,s,\phi_1,\phi_2}_u\geq 0\}}
          \Big(|\delta Z^{n,s,\phi_1,\phi_2}_u| 
              - C G^{n,s,\phi_1,\phi_2}_u\Big)^2 du \\
   &\ + \int_t^\infty \Bigg[
          C(G^{n,s,\phi_1,\phi_2}_u)^2 
          + \Bigg(
              \int_E 
                \big(\tilde{G}^{n+1,(n,s,\phi_1,\phi_2)}_u(u,l)\big)_+
                \lambda_{n+1}(l_{(n)},dl)
            \Bigg)^2
        \Bigg] du \\
   &\ - \int_t^\infty 
         2(G^{n,s,\phi_1,\phi_2}_u)_+
         (\delta Z^{n,s,\phi_1,\phi_2}_u)'\,dW_u.
\end{align*}
where the constant $C>0$ changes from line to line. Notice that we have used the fact that $(x)_+|x| = (x)_+x$ for any $x\in \mathbb{R}$ in the second-to-last inequality. Therefore, there exists $C>0$ such that for any $t\geq s$,   
    \begin{align}
    \label{eq:G:n:factor}
       &\  \mathbb{E}\left[  |(G^{n,s,\phi_1,\phi_2}_t(\theta_{(n)},l_{(n)}))_+|^2 \right] \nonumber\\
       \leq&\  C\int_t^\infty \Bigg( \mathbb{E}\left[ |(G^{n,s,\phi_1,\phi_2}_u(\theta_{(n)},l_{(n)}))_+|^2\right]\nonumber \\
       &\quad + \mathbb{E}\left[  \sup_{ l\in  E}|(\tilde{G}^{n+1,(n,s,\phi_1,\phi_2)}_u((\theta_{(n)},u), (l_{(n)},l) ))_+|^2 \right] \Bigg) du.
            \end{align}
Likewise, for $n=m$, there exists $C>0$ such that, for any $t\geq s$, 
    \begin{equation}
        \label{eq:G:m:factor}
    \begin{aligned}
    &\ \mathbb{E}\left[   |(G^{m,s,\phi_1,\phi_2}_t(\theta_{(m)},l_{(m)}))_+|^2 \right]  
    \leq  C\int_t^\infty  \mathbb{E}\left[   |(G^{m,s,\phi_1,\phi_2}_u(\theta_{(m)},l_{(m)}))_+|^2 \right] du. 
    \end{aligned}
    \end{equation}
 
Using the estimates \eqref{eq:G:n:factor} and \eqref{eq:G:m:factor}, we can upper bound $\delta Y^{n,s,\phi_1,\phi_2}$ recursively as follows. For $n=m$, by applying Gr\"onwall's inequality to \eqref{eq:G:m:factor}, we deduce that, for any $(\theta_{(m)},l_{(m)})\in\Delta_m\times E^m$ and $t\geq s\geq \theta_m$, $G^{m,s,\phi_1,\phi_2}_t(\theta_{(m)},l_{(m)})=0$, and thus 
    \begin{equation}
    \label{eq:delta:Y:m:factor}
        \delta Y^{m,s}_t(\theta_{(m)},l_{(m)}) \leq \delta\bar{Y}^{m,s}_t = \frac{C_\phi C_g e^{-C_g(t-s)}}{(C_g-C_\phi)(C_g+\rho)}|\phi_1-\phi_2|. 
    \end{equation}

For $n=m-1$, using the established bound \eqref{eq:delta:Y:m:factor}, we have, for any $(\theta_{(m-1)},l_{(m-1)},l)\in \Delta_{m-1}\times E^{m-1}\times E$ and $\theta_{m-1}\leq s\leq t$,
    \begin{align*}
       &\  \delta Y^{m,(m-1,s,\phi_1,\phi_2)}_t((\theta_{(m-1)},t),(l_{(m-1)},l)) \\
       &\   -(\hat{\pi}^{m-1,2}_t)'\bigg(\hat{\beta}^{m-1}(\Phi^{m-1,s,\phi_1}_{t^-}(\theta_{(m-1)},l_{(m-1)},l)   - \hat{\beta}^{m-1}(\Phi^{m-1,s,\phi_2}_{t^-}(\theta_{(m-1)},l_{(m-1)},l)\bigg) \\
       \leq &\  \delta Y^{m,t,\Phi^{m,(m-1,s,\phi_1)}_t,\Phi^{m,(m-1,s,\phi_2)}_t}_t((\theta_{(m-1)},t),(l_{(m-1)},l))  \\
       &\ + C_\Pi C_\beta\left| \Phi^{m-1,s,\phi_1}_{t^-}(\theta_{(m-1)},l_{(m-1)})-\Phi^{m-1,s,\phi_2}_{t^-}(\theta_{(m-1)},l_{(m-1)}) \right| \\
       \leq &\ \frac{C_\phi C_g}{(C_g-C_\phi)(C_g+\rho)}\bigg|\Phi^{m,(m-1,s,\phi_1)}_t((\theta_{(m-1)},t),(l_{(m-1)},l))\\
       &\quad -\Phi^{m,(m-1,s,\phi_2)}_t((\theta_{(m-1)},t),(l_{(m-1)},l)) \bigg| +C_{\Pi}C_\beta e^{-C_g(t-s)}|\phi_1-\phi_2|  \\
       \leq &\  \frac{C_\phi C_g}{(C_g-C_\phi)(C_g+\rho)}\Bigg(  \bigg|\Phi^{m-1,s,\phi_1}_{t^-}(\theta_{(m-1)},l_{(m-1)})  -\Phi^{m-1,s,\phi_2}_{t^-}(\theta_{(m-1)},l_{(m-1)})) \bigg| \\
       &\quad + \left|\varphi^m\left(\Phi^{m-1,s,\phi_1}_{t^-}(\theta_{(m-1)},l_{(m-1)}) \right) - \varphi^m\left(\Phi^{m-1,s,\phi_2}_{t^-}(\theta_{(m-1)},l_{(m-1)})) \right) \right| \Bigg) \\
       &\ + C_\Pi C_\beta e^{-C_g(t-s)}|\phi_1-\phi_2| \\
       \leq&\ \left[\frac{C_\phi C_g(1+C_\varphi)}{(C_g-C_\phi)(C_g+\rho)}+ C_\Pi C_\beta \right]|\phi_1-\phi_2|e^{-C_g(t-s)}  = \delta \bar{Y}^{m-1,s}_t. 
    \end{align*}
This implies that 
    \begin{equation*}
        (\tilde{G}^{m,(m-1,s,\phi_1,\phi_2)}_t((\theta_{(m-1)},t), (l_{(m-1)},l) ))_+ = 0
    \end{equation*}
for any $(\theta_{(m-1)},l_{(m-1)},l)\in \Delta_{m-1}\times E^{m-1}\times E$, $t\geq s \geq \theta_{m-1}$. Hence, the inequality \eqref{eq:G:n:factor} for $n=m-1$ is reduced to 
    \begin{equation*}
    \begin{aligned}
       &\  \mathbb{E}\left[   |(G^{m-1,s,\phi_1,\phi_2}_t(\theta_{(m-1)},l_{(m-1)}))_+|^2 \right] 
       \leq  C\int_t^\infty  \mathbb{E}\left[  |(G^{m-1,s,\phi_1,\phi_2}_u(\theta_{(m-1)},l_{(m-1)}))_+|^2\right]  du. 
            \end{aligned}
    \end{equation*}
By Gr\"onwall's inequality, we obtain the bound 
     \begin{equation}
    \label{eq:delta:Y:m-1:factor}
        \delta Y^{m-1,s}_t(\theta_{(m-1)},l_{(m-1)}) \leq \delta\bar{Y}^{m-1,s}_t = \left[   \frac{C_\phi C_g(1+C_\varphi)}{(C_g-C_\phi)(C_g+\rho)} + C_\Pi C_\beta  \right]e^{-C_g(t-s)}|\phi_1-\phi_2|, 
    \end{equation}
for any $(\theta_{(m-1)},l_{(m-1)})\in \Delta_{m-1}\times E^{m-1}$ and $t\geq s \geq \theta_{m-1}$. Applying the above argument inductively, we have $\delta Y^{n,s}_t(\theta_{(n)},l_{(n)}) \leq \delta\bar{Y}^{n,s}_t$ for any $n=0,\dots,m$, $(\theta_{(n)},l_{(n)})\in \Delta_{n}\times E^{n}$ and $t\geq s \geq \theta_{n}$.  By symmetry, we arrive at \eqref{eq:delta:Y:bound}. \\

\noindent\underline{Step 4: Markovian solution and bounds for $Z^n$:}\\
\indent The fact that the solution component $Y^{n,s,\phi}$ admits a Markovian representation is a consequence of Theorem 4.1 in \cite{el1997backward}; see also \cite{liang:2017:bsde,hu2020systems}. Thus, for any $n=0,\dots,m$, $(\theta_{(n)},l_{(n)})\in \Delta_n\times E^n$, and $t\geq \theta_n$,  we can write $Y^{n,\theta_n,\phi}_t(\theta_{(n)},l_{(n)}) = y^n( \Phi^{n,\theta_n,\phi}_t(\theta_{(n)},l_{(n)}))$. The Lipschitz property \eqref{eq:y:z:markov:bound:lip} follows from  \eqref{eq:delta:Y:bound} and the observation that 
    \begin{equation*}
        \left| y^n(\phi_1) - y^n(\phi_2)  \right| = \left|\delta Y^{n,\theta_n,\phi_1,\phi_2}_{\theta_n}(\theta_{(n)},l_{(n)})\right|. 
    \end{equation*}
In addition, there exists a measurable function $z^n:\mathbb{R}^d\to\mathbb{R}^d$ such that $Z^{n,\theta_n,\phi}_t(\theta_{(n)},l_{(n)}) = z^n( \Phi^{n,\theta_n,\phi}_t(\theta_{(n)},l_{(n)}))$. By Corollary 4.1 of \cite{el1997backward}, we have
    \begin{equation*}
       ( \kappa^n(\theta_{(n)},l_{(n)}))' \nabla_\phi y^n(\Phi^{n,\theta_n,\phi}_t(\theta_{(n)}),l_{(n)}) = z^n( \Phi^{n,\theta_n,\phi}_t(\theta_{(n)},l_{(n)})). 
    \end{equation*}
Since $|\kappa^n(\theta_{(n)},l_{(n)})|=1$, we arrive at the bounds \eqref{eq:y:z:markov:bound:lip}. Using this, we see that the truncated equations \eqref{eq:Yn:tilde:truncated:factor:Z} and \eqref{eq:Yn:tilde:truncated:factor:Z:m} are reduced to \eqref{eq:Yn} and \eqref{eq:Ym} with drivers  \eqref{eq:factor:drivers_2} and \eqref{eq:factor:drivers_1}, respectively. The proof is thus complete. \hfill $\square$

\subsection{Proof of Proposition \ref{pp:Delta:Y}}
\label{sec:pf:pp:Delta:Y}
    We shall prove the estimates recursively.  We define 
    \begin{equation*}
    \begin{aligned}
        \Delta Y^{m}_t((\theta_{(m-1)},\theta),(l_{(m-1)},l)) &:= Y^m_t((\theta_{(m-1)},\theta),(l_{(m-1)},l)) - Y^{m-1}_t\left(\theta_{(m-1)},l_{(m-1)} \right), \\
              \Delta Z^{m}_t((\theta_{(m-1)},\theta),(l_{(m-1)},l)) &:= Z^m_t((\theta_{(m-1)},\theta),(l_{(m-1)},l)) - Z^{m-1}_t\left(\theta_{(m-1)},l_{(m-1)} \right),
    \end{aligned}
    \end{equation*}
for any $((\theta_{(m-1)},\theta),(l_{(m-1)},l))\in \Delta_{m}\times E^{m}$ and  $t\geq \theta$.
For notational convenience, we will omit writing the dependence of $(\theta_{(m-1)},l_{(m-1)})$ in the remaining proof. 

By It\^o's lemma, for any $t\leq \tau$,
    \begin{align}
    \label{eq:DYm:1}
        \Delta Y^m_t(\theta,l) &= \Delta Y^m_\tau(\theta,l) +  \int_t^\tau \bigg( -\rho \Delta Y^m_u(\theta,l) +\min_{\pi\in\Pi_m} \hat{F}^m_1\left(\pi, Z^m_u(\theta,l), \Phi^m_u(\theta,l)\right) \nonumber  \\
        &\quad  -\min_{\pi\in\Pi_{m-1}}\left\{ \hat{F}^{m-1}_1\left(\pi, Z^{m-1}_u, \Phi^{m-1}_u\right) + \hat{F}^{m-1}_2\left(u,\pi, Y^{m-1}_u, \Phi^{m-1}_u\right) \right\}  \bigg)du \nonumber \\
        &\ -\int_t^\infty \Delta Z^m_u(\theta,l) dW_u.
    \end{align}
For any $u\geq t\geq \theta$, let
    \begin{equation*}
        \pi^{m-1}_u := \mathop{\arg\min}_{\pi \in \Pi_{m-1}} \left\{ \hat{F}^{m-1}_1\left(\pi, Z^{m-1}_u, \Phi^{m-1}_u\right) + \hat{F}^{m-1}_2\left(u,\pi, Y^{m-1}_u, \Phi^{m-1}_u\right) \right\}. 
    \end{equation*}
Using  Assumptions \ref{ass:factor}-\ref{ass:pi:compact} and Theorem \ref{thm:Z:markovian}, there exist $C_{m,1},C_{m,2},C_{m,3} >0$  independent of $\rho$ such that
    \begin{align}
    \label{eq:DYm:2}
      &\   \min_{\pi\in\Pi_m} \hat{F}^m_1\left(\pi, Z^m_u(\theta,l), \Phi^m_u(\theta,l)\right) \nonumber \\
        & -   \min_{\pi\in\Pi_{m-1}}\left\{ \hat{F}^{m-1}_1\left(\pi, Z^{m-1}_u, \Phi^{m-1}_u\right) + \hat{F}^{m-1}_2\left(u,\pi, Y^{m-1}_u, \Phi^{m-1}_u\right) \right\} \nonumber \\ 
        \leq &\   \hat{F}^m_1\left(0, Z^m_u(\theta,l), \Phi^m_u(\theta,l)\right) \nonumber \\
        & -   \hat{F}^{m-1}_1\left(\pi^{m-1}_u, Z^{m-1}_u, \Phi^{m-1}_u\right) - \hat{F}^{m-1}_2\left(u,\pi_u^{m-1}, Y^{m-1}_u, \Phi^{m-1}_u\right) \nonumber \\
        \leq &\ C_{m,1} - C_{m,2}\int_E e^{\gamma\left(Y^{m}_u(u,l') - Y^{m-1}_u \right)} \lambda_m(dl') \nonumber \\
        \leq &\ C_{m,1} - C_{m,2}\int_E \left[1+\gamma\left( Y^m_u(u,l') - Y^{m-1}_u  \right)\right]\lambda_m(dl') \nonumber \\
        =&\ C_{m,1}-C_{m,2} - \gamma C_{m,2}\left( \Delta Y^m_u(\theta,l) + \underline{\int_E \left( Y^m_u(u,l') - Y^m_u(\theta,l)  \right)\lambda_m(dl') } \right) \nonumber\\
        \leq&\ C_{m,1}-C_{m,2}  +\gamma C_{m,2}\left[ - \Delta Y^m_u(\theta,l) + C_{m,3} \int_E \left|\Phi^m_u(u,l') - \Phi^m_u(\theta,l)  \right| \lambda_m(dl')\right]. 
    \end{align}
Specifically, we have 
    \begin{equation*}
        C_{m,1} = \frac{\gamma}{2}(K^m_Z)^2 , \ C_{m,2} =\frac{1}{\gamma} e^{-C_\Pi K_\beta}, \ C_{m,3} = K^m_Z. 
    \end{equation*}
Note that the underlying term is the extra term we need to bound due to index mismatch. Combining \eqref{eq:DYm:1}-\eqref{eq:DYm:2}, we obtain 
    \begin{align*}
        \Delta Y^m_t(\theta,l) &\leq \Delta Y^m_\tau(\theta,l) + \int_t^\tau\bigg( C_{m,1}-C_{m,2} + \gamma C_{m,2}C_{m,3} \int_E \left|\Phi^m_u(u,l') - \Phi^m_u(\theta,l)  \right| \lambda_m(dl') \\
        &\qquad-(\rho+\gamma C_{m,2}) \Delta Y^m_u(\theta,l)   \bigg)du  -\int_t^\tau \Delta Z^m_u(\theta,l) dW_u.
    \end{align*}
By applying the comparison principle of BSDEs,  for any $((\theta_{(m-1)},\theta),(l_{(m-1)},l))\in \Delta_{m}\times E^{m}$ and  $\theta \leq t\leq \tau$, we have 
    \begin{equation*}
        \Delta Y^m_t(\theta,l) \leq \Delta \bar{Y}^m_t(\theta,l),
    \end{equation*}
where the latter is the solution of the following BSDE:
     \begin{align}
     \label{eq:Delta:Y:m:BSDE}
        \Delta \bar{Y}^m_t(\theta,l) &= \Delta \bar{Y}^m_\tau(\theta,l) + \int_t^\tau\bigg( C_{m,1}-C_{m,2} + \gamma C_{m,2}C_{m,3} \int_E \left|\Phi^m_u(u,l') - \Phi^m_u(\theta,l)  \right| \lambda_m(dl') \nonumber \\
        &\qquad-(\rho+\gamma C_{m,2}) \Delta \bar{Y}^m_u(\theta,l)   \bigg)du  -\int_t^\tau \Delta \bar{Z}^m_u(\theta,l) dW_u.
    \end{align}
It is clear that the BSDE \eqref{eq:Delta:Y:m:BSDE} has a unique solution.
   Solving the equation yields 
    \begin{align*}
        \Delta \bar{Y}^m_t(\theta,l) &= e^{-(\rho + \gamma C_{m,2})(\tau-t)} \mathbb{E}\left[\Delta Y^m_\tau(\theta,l) | \mathcal{F}_t \right] \nonumber \\
        &\ + \int_t^\tau e^{-(\rho+\gamma C_{m,2})(u-t)}\bigg(C_{m,1}-C_{m,2} \nonumber \\
        &\quad + \gamma C_{m,2}C_{m,3} \int_E \mathbb{E}\left[ \left|\Phi^m_u(u,l') - \Phi^m_u(\theta,l)  \right| \mid \mathcal{F}_t \right] \bigg)\lambda_m(dl') du. 
    \end{align*}
By passing to the limit $\tau\to\infty$, we obtain
\begin{equation}
\begin{aligned}
    \label{eq:DYm:bound:upper:general}
    & \ \ \ \ Y^{m}_t(\theta,l) - Y^{m-1}_t \\
    &\leq \frac{C_{m,1}-C_{m,2}}{C_{m,2}} + \gamma C_{m,2}C_{m,3} \int_t^{\infty} e^{-\gamma C_{m,2}(u-t)} \int_E \mathbb{E}\left[\left|\Phi^m_u(u,l') - \Phi^m_u(\theta,l) \right|  \big|\mathcal{F}_t \right] \lambda_m(dl')du. 
\end{aligned}
\end{equation}

To proceed, we take $\theta=t$ in  \eqref{eq:DYm:bound:upper:general}, such that 
    \begin{align}
        \label{eq:DYm:bound:upper}
           &\ \ \ \  Y^m_t\left(t,l\right) - Y^{m-1}_t \nonumber  \\
           &\leq  \frac{C_{m,1}-C_{m,2}}{C_{m,2}} + \gamma C_{m,2}C_{m,3} \int_t^\infty e^{-\gamma C_{m,2}(u-t)}  \int_E\mathbb{E}\left[\left|\Phi^m_u(u,l') - \Phi^m_u(t,l) \right| \mid\mathcal{F}_t\right] \lambda_m(dl') du    \nonumber\\
            &=   \frac{C_{m,1}-C_{m,2}}{C_{m,2}} + \gamma C_{m,2}C_{m,3}\int_t^\infty e^{-\gamma C_{m,2}(u-t)} \nonumber\\
            &\quad\cdot \int_E\mathbb{E}\left[\left| \Phi^{m-1}_u +\varphi^{m-1}(\Phi^{m-1}_u)  - \Phi^m_u(t,l) \right| \mid\mathcal{F}_t\right] \lambda_m(dl') du  .
    \end{align}
Applying It\^o's lemma to $|\Phi^{m-1}_s-\Phi^m_s(t,l)|^2$, $t\leq s\leq u$, and noticing that $\Phi^{m-1}_t(t,l)-\Phi^{m-1}_t=\varphi^{m-1}(\Phi^{m-1}_t)$, we have             
    \begin{align*}
       &\  \mathbb{E}[|\Phi^m_u(t,l) - \Phi^{m-1}_u |^2 |\mathcal{F}_t ] \\
       =&\  \mathbb{E}\bigg[ \left|  \Phi^m_t(t,l) - \Phi^{m-1}_t \right|^2 + \int_t^u \bigg( 2\left(\Phi^m_s(t,l) - \Phi^{m-1}_s \right)'\left(g^{m-1}(\Phi^m_s(t,l)) - g^{m-1}(\Phi^{m-1}_s) \right) \\
       &\quad + 2\left(\Phi^m_s(t,l) - \Phi^{m-1}_s \right)'\left( g^m(\Phi^m_s(t,l),t,l) - g^{m-1}(\Phi^m_s(t,l)) \right) \\
       &\quad + |\kappa^{m-1}-\kappa^m(t,l)|^2 \bigg)ds \mid \mathcal{F}_t    \bigg] \\
       \leq&\ \mathbb{E}\bigg[\left|  \varphi^{m-1}(\Phi^{m-1}_t) \right|^2 + \int_t^u \bigg( C_g^{-1}\left| g^m(\Phi^m_s(t,l),t,l) - g^{m-1}(\Phi^{m-1}_s(t,l)) \right|^2  \\
       &\quad + |\kappa^{m-1}-\kappa^m(t,l)|^2 \bigg)ds \mid \mathcal{F}_t    \bigg],
    \end{align*}
where the last line follows from \eqref{eq:dissipative:gn} and Young's inequality. By Gr\"onwall's inequality, and the boundedness of $\varphi^{m-1}$ and $g^m(\cdot,t,l)-g^{m-1}(\cdot)$, for any $u\geq t\geq 0$, 
    \begin{equation*}
        \mathbb{E}[|\Phi^m_u(t,l) - \Phi^{m-1}_u |^2 |\mathcal{F}_t ] \leq K_\varphi^2 + (C_g^{-1}D_g^2+4) (u-t). 
    \end{equation*}
Hence, 
    \begin{align}
        \label{eq:difference:Phi:2}
        \mathbb{E}\left[\left|\Phi^m_u(u,l') - \Phi^m_u(t,l) \right| \mid\mathcal{F}_t\right] &\leq \mathbb{E}[|\varphi^{m-1}(\Phi^{m-1}_u)| \mid \mathcal{F}_t ] +  \mathbb{E}[|\Phi^m_u(t,l) - \Phi^{m-1}_u | |\mathcal{F}_t ] \nonumber \\
        &\leq K_\varphi +  \sqrt{K_\varphi^2 + (C_g^{-1}D_g^2 + 4)(u-t) } \nonumber \\
        &\leq 2K_\varphi + \sqrt{ (C_g^{-1}D_g^2 + 4)(u-t) }.
    \end{align}
Therefore, substituting  \eqref{eq:difference:Phi:2} into \eqref{eq:DYm:bound:upper},  we obtain $Y^m_t(t,l) - Y^{m-1}_t \leq K_m$ for any $\rho>0$, and all $((\theta_{(m-1)},\theta),(l_{(m-1)},l))\in \Delta_m\times E^m$. Collecting the above constants, we deduce \eqref{eq:Delta:Y:n:bound:stronger} for $n=m$.

Using the uniform bound \eqref{eq:Delta:Y:n:bound:stronger} for $n=m$ we can repeat the above argument to establish the bound for $n=m-1$.  The remaining cases then follow by induction, and we omit the details. \hfill $\square$

\subsection{Proof of Theorem \ref{thm:bound:pi}}
\label{sec:pf:thm:bound:pi}
       We prove the statement by induction. Let $C_\Pi$ be the solution of \eqref{eq:Cpi:solution}, and for $n=0,\dots,m$, let $ (Y^{n,C_{\Pi}},Z^{n,C_\Pi})$ be the solution of the following infinite-horizon BSDE:
        \begin{equation*}
            dY^{n,C_\Pi}_t =\left( \rho Y^{n,C_\Pi}_t - \min_{\pi \in \Pi_n, |\pi|\leq C_\Pi } \hat{f}^n(t,\pi, Y^{n,C_\Pi}_t ,Z^{n,C_\Pi}_t,\Phi^n_t;Y^{n+1}) \right)dt + (Z^{n,C_\Pi}_t)'dW_t,
        \end{equation*}
    where we have adopted the convention for the drivers $\hat{f}^n$ as in the proof of Theorem \ref{thm:Z:markovian}, and that $\hat{f}^m(t,\pi,y_1,z,\phi;y_2) = \hat{f}^m(t,\pi,z,\phi)$, where $Y^{n+1}$ is the solution of the infinite-horizon BSDE \eqref{eq:Yn} corresponding to the index $n+1$ with the driver \eqref{eq:factor:drivers_2} in the absence of the constraint $|\pi|\leq C_\Pi$. Using the induction assumption $(Y^{n+1,C_\Pi},Z^{n+1,C_\Pi}) = (Y^{n+1},Z^{n+1})$, $n=0,\dots,m-1$,  it is clear that the equation above admits a unique solution $(Y^{n,C_\Pi},Z^{n,C_\Pi})$. We claim that $(Y^{n,C_\Pi},Z^{n,C_\Pi})=(Y^n,Z^n)$. 
    
   To this end, consider another BSDE:
           \begin{equation*}
            d\tilde{Y}^{n}_t =\left( \rho \tilde{Y}^{n}_t - \min_{\pi \in \Pi_n} \hat{f}^n(t,\pi, Y^{n,C_\Pi}_t , Z^{n,C_\Pi}_t,\Phi^n_t;Y^{n+1}) \right)dt + (\tilde{Z}^{n}_t)'dW_t,
        \end{equation*}  
    which admits a unique solution since $Y^{n,C_\Pi},Y^{n+1}$ are bounded, and the driver clearly verifies Assumption A1 (i)-(ii) of \cite{confortola:briand:QBSDE}. 

    By the estimates \eqref{eq:y:z:markov:bound:lip}, \eqref{eq:Delta:Y:n:bound:stronger}, the induction assumption, and following the derivation of \eqref{eq:pi:bound}, the minimizer
        \begin{equation*} 
         \pi^{n,C_\Pi}_t :=   \min_{\pi \in \Pi_n} \hat{f}^n(t,\pi, Y^{n,C_\Pi}_t ,Z^{n,C_\Pi}_t,\Phi^n_t;Y^{n+1}) =  \min_{\pi \in \Pi_n} \hat{f}^n(t,\pi, Y^{n,C_\Pi}_t , Z^{n,C_\Pi}_t,\Phi^n_t;Y^{n+1,C_\Pi}) 
        \end{equation*}
    satisfies
        \begin{align*}
            |\pi^{n,C_\Pi}_t| &\leq \frac{1}{\sigma_{\min}}\left( \frac{\sqrt{2}}{\gamma}e^{\frac{\gamma}{2} K_{\Delta Y^{n+1}}} \mathbbm{1}_{\{n\neq m\}} + 2\left|Z^{n,C_\Pi}_t - \frac{\hat{\alpha}^n(\Phi^n_t)}{\gamma} \right| \right) \\
            &\leq \frac{1}{\sigma_{\min}}\left( \frac{\sqrt{2}}{\gamma}e^{\frac{\gamma}{2} K_{\Delta Y^{n+1}}} \mathbbm{1}_{\{n\neq m\}}+ \frac{2\|\hat{\alpha}^n\|}{\gamma} + 2K_{Z^n}\right) \leq C_\Pi,
        \end{align*}
    since $Y^{n+1,C_\Pi}_t(t,l) - Y^{n,C_\Pi}_t \leq K_{\Delta Y^{n+1}}$, thanks to Proposition \ref{pp:Delta:Y}. Therefore, 
           \begin{align*}
            d\tilde{Y}^{n}_t &=\left( \rho \tilde{Y}^{n}_t - \min_{\pi \in \Pi_n} \hat{f}^n(t,\pi, Y^{n,C_\Pi}_t , Z^{n,C_\Pi}_t,\Phi^n_t;Y^{n+1}) \right)dt + (\tilde{Z}^{n}_t)'dW_t \\
            &= \left( \rho \tilde{Y}^{n}_t - \min_{\pi \in \Pi_n,  |\pi|\leq C_\Pi } \hat{f}^n(t,\pi, Y^{n,C_\Pi}_t , Z^{n,C_\Pi}_t,\Phi^n_t;Y^{n+1}) \right)dt + (\tilde{Z}^{n}_t)'dW_t
        \end{align*}
    By the uniqueness of the solution of the equation for $(Y^{n,C_\Pi},Z^{n,C_\Pi})$, we have $(Y^{n,C_\Pi},Z^{n,C_\Pi}) = (\tilde{Y}^{n},\tilde{Z}^{n})$. Hence, $(Y^{n,C_\Pi},Z^{n,C_\Pi})$ satisfies 
     \begin{equation*}
            dY^{n,C_\Pi}_t =\left( \rho Y^{n,C_\Pi}_t - \min_{\pi \in \Pi_n} \hat{f}^n(t,\pi, Y^{n,C_\Pi}_t ,Z^{n,C_\Pi}_t,\Phi^n_t;Y^{n+1}) \right)dt + (Z^{n,C_\Pi}_t)'dW_t.
        \end{equation*}
    Finally, by the uniqueness of solutions of the non-truncated BSDE $(Y^n,Z^n)$, we have $\pi^{n,C_\Pi}_t=\pi^{*n}_t$, $(Y^n,Z^n) = (Y^{n,C_\Pi},Z^{n,C_\Pi})$, and $|\pi^{*n}_t|\leq C_\Pi$. Note that the above argument holds true for $n=m$, the proof is thus complete. \hfill $\square$

\subsection{Proof of Lemma \ref{lem:exp:estimate:factor}}
\label{sec:pf:lem:exp:estimate:factor}
    We first show that there exist $C_1>0,C_2\geq 0$ such that, for any $n=0,\dots,m-1$, $(\theta_{(n)},l_{(n)})\in \Delta_n\times E^n$ and $\phi\in \mathbb{R}^d$, 
    \begin{equation}
    \label{eq:g:coercive}
              \left( g^n(\phi,\theta_{(n)},l_{(n)}) \right)'\phi \leq -C_1|\phi|^2 + C_2. 
            \end{equation}
    Using Assumption \ref{ass:ergodic:kappa:g}, we have, for any $n=0,\dots,m$ and $(\theta_{(n)},l_{(n)})\in \Delta_n\times E^n$,
        \begin{align*}
            \left|g^n(0,\theta_{(n)},l_{(n)})\right| 
            &\leq \sum_{k=1}^m \left|g^k(0,\theta_{(k)},l_{(k)}) - g^{k-1}(0,\theta_{(k-1)},l_{(k-1)}) \right| + |g^0(0,0)| =: C_0.
        \end{align*}
    Using this, Assumption \ref{ass:lip:diss:factor}, and Young's inequality, for any $\phi\in \mathbb{R}^d$, $n=0,\dots,m$ and $(\theta_{(n)},l_{(n)})\in \Delta_n\times E^n$,
        \begin{align*}
            g^n(\phi,\theta_{(n)},l_{(n)})'\phi &\leq -C_g|\phi|^2 +   g^n(0,\theta_{(n)},l_{(n)})'\phi \\
            &\leq -C_g|\phi|^2  + \frac{C_g}{2}|\phi|^2 + \frac{1}{2C_g} \left|g^n(0,\theta_{(n)},l_{(n)})\right|^2 \\
            &\leq  -\frac{C_g}{2}|\phi|^2 + \frac{C_0}{2C_g},
        \end{align*}
  and thus \eqref{eq:g:coercive} follows with $C_1 := \frac{C_g}{2}$ and $C_2 := \frac{C_0}{2C_g}$. 
  

    For $\varepsilon>0$, we define the process $P^\varepsilon_t :=e^{\varepsilon |\Phi^{m}_t|^2}$, $t\geq \theta_{m}$. We claim that  there exists $C_\varepsilon>0$ such that, for any $(\theta_{(m)},l_{(m)})\in \Delta_m\times E^m$ and $t\geq \theta_m$, 
    \begin{equation}
    \label{eq:exp:sq:bound}
\mathbb{E}\left[e^{\varepsilon \left|\Phi^m_t\right|^2}\right] \le C_\varepsilon < \infty
\end{equation}
whenever $\varepsilon>0$ is sufficiently small. Then, the upper bound  follows from the elementary inequality that $c|\Phi^{m}_{t}| \leq \varepsilon|\Phi^{m}_t|^2 + c^2/(4\varepsilon)$.
    
We proceed to establish \eqref{eq:exp:sq:bound}. Applying It\^o's lemma to $P^\varepsilon_t$, we have 
\begin{align}
\label{eq:ito-explicit-kappa}
dP_t^\varepsilon
&= P_t^\varepsilon\Big(2\varepsilon( \Phi^{m}_t)'g^{m}(\Phi^{m}_t)
    + \varepsilon      + 2\varepsilon^2|\Phi^{m}_t|^2\Big)\,dt    +  2\varepsilon\,P_t^\varepsilon  (\Phi^{m}_t)'\kappa^{m} dW_t.
\end{align}
By taking conditional expectation on \eqref{eq:ito-explicit-kappa} and using \eqref{eq:g:coercive}, we have, for any $t\geq\theta_m$,\footnote{The fact that $\mathbb{E}[\int_{\theta_m}^t P^\varepsilon_s (\Phi^m_s)'\kappa^m\,dW_s \mid \mathcal{F}_{\theta_m}] = 0$ follows from the square-integrability of the integrand, which holds because $\mathbb{E}[\int_{\theta_m}^t (P_s^{2\varepsilon}+|\Phi^m_s|^2)\,ds] < \infty$. The latter can be shown using the same dissipativity and Gr\"onwall-type argument together with a standard localization.}  
\begin{align*}
\label{eq:stopped-cond}
& \ \ \ \ \mathbb{E}\big[P_t^\varepsilon \big|\mathcal{F}_{\theta_{m}}\big] \nonumber \\
&= P_{\theta_{m}}^\varepsilon  + \mathbb{E}\!\left[\left.\int_{\theta_{m}}^{t}
  P_s^\varepsilon\Big(2\varepsilon( \Phi^{m}_s)' g^{m}(\Phi^{m}_s)
  + \varepsilon + 2\varepsilon^2|\Phi^{m}_s|^2\Big)\,ds
  \right|\mathcal{F}_{\theta_{m}}\right] \nonumber \\
  &\leq P_{\theta_{m}}^\varepsilon  + \mathbb{E}\!\left[\left.\int_{\theta_{m}}^{t}
  P_s^\varepsilon\Big(-2\varepsilon(C_1 - \varepsilon)|\Phi^{m}_s|^2 + \varepsilon(2 C_2
  + 1)\Big)\,ds
  \right|\mathcal{F}_{\theta_{m}}\right] \nonumber\\
  &\le P_{\theta_{m}}^\varepsilon
+ \mathbb{E}\!\Bigg[\Big.\int_{\theta_{m}}^{t} 
\Big(-2(C_1-\varepsilon)(P_s^\varepsilon-1) + \varepsilon(2C_2+1) P_s^\varepsilon \Big)\,ds
\Big| \mathcal{F}_{\theta_{m}}\Bigg] \nonumber\\
&= P_{\theta_{m}}^\varepsilon
+ \int_{\theta_{m}}^{t} \Big( 
-\left[2(C_1-\varepsilon)-\varepsilon(2C_2+1)\right]  \mathbb{E}\left[P^{\varepsilon}_s| \mathcal{F}_{\theta_m} \right] 
+ 2(C_1-\varepsilon) \Big) ds  
\end{align*}
where we have used the inequality $x e^{\varepsilon x}\ge \frac{1}{\varepsilon}(e^{\varepsilon x}-1)$ for $x\ge0$.

By choosing $\varepsilon>0$ small enough such that $c_\varepsilon:= 2(C_1-\varepsilon) -\varepsilon(2C_2+1)>0$, we have, by Gr\"onwall's inequality, 
\begin{align*}
\mathbb{E}\big[P_{t}^\varepsilon \,\big|\,\mathcal{F}_{\theta_{m}}\big]
&\le e^{-c_\varepsilon(t-\theta_{m})} P_{\theta_{m}}^\varepsilon
+ \frac{2(C_1-\varepsilon)}{c_\varepsilon} \left( 1 - e^{-c_\varepsilon(t-\theta_{m})} \right) \\
&\leq e^{-c_\varepsilon(t-\theta_{m})} e^{\varepsilon|\Phi^m_{\theta_m}|^2} 
+ \frac{2(C_1-\varepsilon)}{c_\varepsilon}.  
\end{align*}
Iterating this argument recursively on $e^{\varepsilon |\Phi^n_{\theta_n}|^2}$ for $n=m-1,\dots,0$  and using the boundedness of $\varphi^{n}$, we obtain \eqref{eq:exp:sq:bound} for $n\leq m$. 

Finally, the lower bound is a direct consequence of Jensen's inequality: for any $n=0,\dots,m$, $(\theta_{(n)},l_{(n)})\in\Delta_n\times E^n$, and  $t\geq \theta_n$, 
    \begin{equation*}
        \mathbb{E}\left[e^{-c|\Phi^{n}_{t}(\theta_{(n)},l_{(n)})|} \right] \geq \left(   \mathbb{E}\left[e^{c|\Phi^{n}_{t}(\theta_{(n)},l_{(n)})|} \right] \right)^{-1} \geq \frac{1}{K_c}. 
    \end{equation*}
\hfill $\square$

 \subsection{Proof of Proposition \ref{pp:risk:sensitive:rho}}
 \label{sec:pf:pp:risk:sensitive:rho}
 We take the sequence $(\rho_i)_{i=1}^\infty$ as specified in the convergence \eqref{eq:ym:converge}. Since 
\begin{align*}
\mathbb{E}\left[\left. e^{-\gamma\left( X^{\pi^*,\rho_i}_T - X^{\pi^*,\rho_i}_{T_m} \right)} \,\right|\,\mathbbm{1}_{ \{T\geq T_m\}} \right]
= \frac{\mathbb{E}\left[ e^{-\gamma\left( X^{\pi^*,\rho_i}_T - X^{\pi^*,\rho_i}_{T_m} \right)}\mathbbm{1}_{ \{T\geq T_m\}} \right]}
       {\mathbb{P}(T\geq T_m)},
\end{align*}
and $T_m<\infty$ a.s., it is therefore sufficient to study the expectation 
\[
\mathbb{E}\left[ e^{-\gamma\left( X^{\pi^*,\rho_i}_T - X^{\pi^*,\rho_i}_{T_m} \right)}\mathbbm{1}_{ \{T\geq T_m\}} \right],
\]
which can be written as 
\begin{align*}
&\ \mathbb{E}\left[ e^{-\gamma\left( X^{\pi^*,\rho_i}_T - X^{\pi^*,\rho_i}_{T_m} \right)} \mathbbm{1}_{ \{T\geq T_m\}} \right] \\
=&\ \mathbb{E}\left[ e^{-\gamma\left( X^{\pi^*,\rho_i,m}_T(T_{(m)},L_{(m)}) - X^{\pi^*,\rho_i,m}_{T_m}(T_{(m)},L_{(m)}) \right)} \mathbbm{1}_{ \{T\geq T_m\}} \right] \\
=&\ \mathbb{E}\left[\int_{\Delta_m\times E^m} 
e^{-\gamma\left( X^{\pi^*,\rho_i,m}_T(\theta_{(m)},l_{(m)})-X^{\pi^*,\rho_i,m}_{\theta_m}(\theta_{(m)},l_{(m)}) \right)} 
\,\eta_T(\theta_{(m)},l_{(m)}) \,
\mathbbm{1}_{\{T \geq \theta_m \} } \, 
d\theta_{(m)}\, \boldsymbol{\lambda}(dl_{(m)}) \right].
\end{align*}

By \eqref{eq:Xn}, we have 
     \begin{align*}
        &\ e^{-\gamma \left(X^{\pi^*,\rho_i,m}_T(\theta_{(m)},l_{(m)})- X^{\pi^*,\rho_i,m}_{\theta_m}(\theta_{(m)},l_{(m)})\right)} \\
       =&\ \exp\bigg(\gamma \int_{\theta_m}^T \bigg[  \frac{\gamma}{2}\left|\pi^{*m}_t(\theta_{(m)},l_{(m)})'\hat{\sigma}^m(\Phi^m_t(\theta_{(m)},l_{(m)}))\right|^2  \\
       &\quad -\pi^{*m}_t(\theta_{(m)},l_{(m)})'(\hat{\sigma}^m\hat{\alpha}^m)(\Phi^m_t(\theta_{(m)},l_{(m)}))  \bigg]dt \bigg)  \\
       &\ \cdot \mathcal{E}_{\theta_m,T}\left( -\int_{\theta_m}^\cdot \gamma\pi^{*m}_t(\theta_{(m)},l_{(m)})'\hat{\sigma}^m(\Phi^m_t(\theta_{(m)},l_{(m)}))dW_t \right) \\
       =&\ e^{\int_{\theta_m}^T \mathcal{L}(\pi^{*m}_t(\theta_{(m)},l_{(m)}),\Phi^m_t(\theta_{(m)},l_{(m)})) dt }\mathcal{E}_{\theta_m,T}\left( -\int_{\theta_m}^\cdot \gamma\pi^{*m}_t(\theta_{(m)},l_{(m)})'\hat{\sigma}^m(\Phi^m_t(\theta_{(m)},l_{(m)}))dW_t \right),
    \end{align*}
where $\mathcal{L}: \mathbb{R}^m\times \mathbb{R}^d \to \mathbb{R}$ is given by
    \begin{equation*}
        \mathcal{L}(\pi, \phi) := \frac{\gamma^2}  {2}|\pi'\hat{\sigma}^m(\phi)|^2 - \gamma \pi'\hat{\sigma}^m(\phi)\hat{\alpha}^m(\phi). 
    \end{equation*}
Hence, 
    \begin{align}
    \label{eq:growth:LT:1}
       &\  \mathbb{E}\left[ e^{-\gamma\left( X^{\pi^*,\rho_i}_T - X^{\pi^*,\rho_i}_{T_m} \right)} \mathbbm{1}_{ \{T\geq T_m\}} \right]  \nonumber\\
       =&\ \mathbb{E}\Bigg[ \int_{\Delta_m\times E^m}   
 \eta_T(\theta_{(m)},l_{(m)}) \mathbbm{1}_{\{T \geq \theta_m \} } e^{\int_{\theta_m}^T \mathcal{L}(\pi^{*m}_t(\theta_{(m)},l_{(m)}),\Phi^m_t(\theta_{(m)},l_{(m)})) dt }\nonumber \\
 &\quad \cdot \mathcal{E}_{\theta_m,T}\left( -\int_{\theta_m}^\cdot \gamma\pi^{*m}_t(\theta_{(m)},l_{(m)})'\hat{\sigma}^m(\Phi^m_t(\theta_{(m)},l_{(m)}))dW_t \right) d\theta_{(m)} \boldsymbol{\lambda}(dl_{(m)}) \Bigg].
    \end{align}

Next, we consider the following expectation:
    \begin{align}
    \label{eq:growth:LT:2}
        &\ \mathbb{E}\left[ e^{-\gamma \left( Y^{\rho_i}_T - Y^{\rho_i}_{T_m} - \int_{T_m}^T \rho_i Y^{\rho_i}_s\,ds \right)} \mathcal{E}_{T_m,T}\left( \gamma\int_{T_m}^\cdot \left( (Z^{\rho_i}_t)' - (\pi^*_t)'\sigma_t \right)dW_t \right) \mathbbm{1}_{\{T\geq T_m\}}  \right]\nonumber \\
        =&\  \mathbb{E}\Bigg[\int_{\Delta_m\times E^m} \eta_T(\theta_{(m)},l_{(m)}) \mathbbm{1}_{\{T \geq \theta_m \} }   e^{-\gamma\left(Y^{m,\rho_i}_T(\theta_{(m)},l_{(m)}) - Y^{m,\rho_i}_{\theta_m}(\theta_{(m)},l_{(m)}) - \rho_i\int_{T_m}^T Y^{m,\rho_i}_s(\theta_{(m)},l_{(m)}) \right)  } \nonumber  \\
        &\quad\cdot \mathcal{E}_{\theta_m,T}\left( \gamma\int_{\theta_m}^\cdot \left( Z^{m,\rho_i}_t(\theta_{(m)},l_{(m)})' - [(\pi^{*m}_t)'\hat{\sigma}^m](\theta_{(m)},l_{(m)}) \right)dW_t\right) d\theta_{(m)}\boldsymbol{\lambda}(dl_{(m)}) \Bigg]. 
    \end{align}
Using \eqref{eq:Ym}, it is straightforward to check that
    \begin{align}
    \label{eq:growth:LT:3}
        &\  e^{-\gamma\left(Y^{m,\rho_i}_T(\theta_{(m)},l_{(m)}) - Y^{m,\rho_i}_{\theta_m}(\theta_{(m)},l_{(m)}) - \rho_i\int_{T_m}^T Y^{m,\rho_i}_s(\theta_{(m)},l_{(m)})ds \right)  } \nonumber \\
        &\quad \cdot \mathcal{E}_{\theta_m,T}\left( \gamma\int_{\theta_m}^\cdot \left( Z^{m,\rho_i}_t(\theta_{(m)},l_{(m)})' - [(\pi^{*m}_t)'\hat{\sigma}^m](\theta_{(m)},l_{(m)}) \right)dW_t\right) \nonumber \\
        =&\ e^{\int_{\theta_m}^T \mathcal{L}(\pi^{*m}_t(\theta_{(m)},l_{(m)}),\Phi^m_t(\theta_{(m)},l_{(m)})) dt } \nonumber \\
 &\quad \cdot \mathcal{E}_{\theta_m,T}\left( -\int_{\theta_m}^\cdot \gamma\pi^{*m}_t(\theta_{(m)},l_{(m)})'\hat{\sigma}^m(\Phi^m_t(\theta_{(m)},l_{(m)}))dW_t \right). 
    \end{align}
Combining \eqref{eq:growth:LT:1}-\eqref{eq:growth:LT:3}, we see that 
    \begin{align*}
      &\   \mathbb{E}\left[ e^{-\gamma\left( X^{\pi^*,\rho_i}_T - X^{\pi^*,\rho_i}_{T_m} \right)} \mathbbm{1}_{ \{T\geq T_m\}} \right] \nonumber\\
      =&\   \mathbb{E}\left[ e^{-\gamma \left( Y^{\rho_i}_T - Y^{\rho_i}_{T_m} - \rho_i\int_{T_m}^T Y^{\rho_i}_sds \right)} \mathcal{E}_{T_m,T}\left( \gamma\int_{T_m}^\cdot \left( (Z^{\rho_i}_t)' - (\pi^*_t)'\sigma_t \right)dW_t \right) \mathbbm{1}_{\{T\geq T_m\}}  \right] \nonumber \\
      =&\  \mathbb{E}\bigg[\int_{\Delta_m\times E^m}  \psi_T^{\rho_i}(\theta_{(m)},l_{(m)}) \eta_T(\theta_{(m)},l_{(m)}) e^{-\gamma\left(\bar{y}^{m,\rho_i}(\Phi^m_T(\theta_{(m)},l_{(m)}))  - \bar{y}^{m,\rho_i}(\Phi^m_{\theta_m}(\theta_{(m)},l_{(m)}) ) \right)  }  \nonumber \\
      &\quad \cdot e^{\gamma\int_{T_m}^T \rho_i y^{m,\rho_i}(\Phi^m_s(\theta_{(m)},l_{(m)}) }  \mathbbm{1}_{\{T\geq \theta_m\}} d\theta_{(m)}\boldsymbol{\lambda}(dl_{(m)})\bigg],
    \end{align*}
where  $\bar{y}^{m,\rho}(\phi):= y^{m,\rho}(\phi)-y^{m,\rho}(\hat{\phi}_m)$, $\phi\in \mathbb{R}^d$, and 
    \begin{equation*}
       \psi_T^{\rho_i}(\theta_{(m)},l_{(m)}) :=  \mathcal{E}_{\theta_m,T}\left( \gamma\int_{\theta_m}^\cdot \left( z^{m,\rho_i}_t(\Phi^m_t(\theta_{(m)},l_{(m)}))' - [(\pi^{*m}_t)'\hat{\sigma}^m](\theta_{(m)},l_{(m)}) \right)dW_t\right)
    \end{equation*}
is a uniformly integrable martingale, thanks to the boundedness and admissibility of $\pi^*$.

{\color{black}Using the uniform-in-$\rho$ linear growth property \eqref{eq:linear:growth:m} of $\bar{y}^{m,\rho}$ , the uniform integrability of $\Phi^m_t(\theta_{(m)},l_{(m)})$ (Lemma \ref{lem:exp:estimate:factor}), and the convergence $\bar{y}^{m,\rho_i}(\cdot)\to \bar{y}^m(\cdot)$, $\rho_iy^{m,\rho_i}(\Phi^m_t) = \rho_i(y^{m,\rho_i}(\Phi^m_t)-\bar{y}^m(\hat{\phi}_m)) + \rho_i\bar{y}^m(\hat{\phi}_m) \to \varrho_m$, $z^{m,\rho_i}(\cdot)\to \bar{z}^m(\cdot)$, and the uniform bound \eqref{eq:y:z:markov:bound:lip}, we have, by Vitali convergence, 
  \begin{align}
    \label{eq:growth:LT:4}
      &\ \lim_{i\to \infty}  \mathbb{E}\left[ e^{-\gamma\left( X^{\pi^*,\rho_i}_T - X^{\pi^*,\rho_i}_{T_m} \right)} \mathbbm{1}_{ \{T\geq T_m\}} \right] \nonumber\\
      =&\  e^{\gamma\varrho_m(T-T_m)}\mathbb{E}\bigg[\int_{\Delta_m\times E^m}  \psi_T(\theta_{(m)},l_{(m)})  e^{-\gamma\left(\bar{y}^{m}(\Phi^m_T(\theta_{(m)},l_{(m)}))  - \bar{y}^{m}(\Phi^m_{\theta_m}(\theta_{(m)},l_{(m)}) ) \right)  }  \nonumber \\
      &\quad \cdot  \eta_T(\theta_{(m)},l_{(m)})  \mathbbm{1}_{\{T\geq \theta_m\}} d\theta_{(m)}\boldsymbol{\lambda}(dl_{(m)})\bigg],
    \end{align}
where 
  \begin{equation*}
    \label{eq:DD:growth}
       \psi_T(\theta_{(m)},l_{(m)}) :=  \mathcal{E}_{\theta_m,T}\left( \gamma\int_{\theta_m}^\cdot \left( \mathcal{Z}^m_t(\theta_{(m)},l_{(m)})' - [(\uppi^{*m}_t)'\hat{\sigma}^m](\theta_{(m)},l_{(m)}) \right)dW_t\right),
    \end{equation*}
and 
    \begin{equation*}
        \uppi^{m*}_t:=\mathop{\arg\min}_{\pi\in\Pi_m}  \hat{f}^m\left(\pi,\mathcal{Z}^{m}_t(\theta_{(m)},l_{(m)}),\Phi^m_t(\theta_{(m)},l_{(m)})\right).
    \end{equation*}
By the boundedness of $\mathcal{Z}^m$ and $ \uppi^{m*}_t$, $\psi_T$ is a uniformly integrable martingale. 
} 


We proceed to show that, there exist $0<c<C<\infty$ such that, for any $T>0$, 
    \begin{equation}
    \label{eq:psi:eta:eY:bounds}
        \begin{aligned}
          c<    & \mathbb{E}\bigg[\int_{\Delta_m\times E^m}  \psi_T(\theta_{(m)},l_{(m)}) \eta_T(\theta_{(m)},l_{(m)}) e^{-\gamma\left(\bar{y}^m(\Phi^m_T(\theta_{(m)},l_{(m)}))  - \bar{y}^m(\Phi^m_{\theta_m}(\theta_{(m)},l_{(m)}) ) \right)  }  \\
      &\quad \cdot  \mathbbm{1}_{\{T\geq \theta_m\}} d\theta_{(m)}\boldsymbol{\lambda}(dl_{(m)})\bigg] < C. 
        \end{aligned}
    \end{equation}
To this end, we define a measure $\mathbb{Q}^{\psi}$ by 
        \begin{equation*}
            \frac{d\mathbb{Q}^\psi}{d\mathbb{P}}\bigg|_{\mathcal{F}_{\theta_m}} = \psi_T(\theta_{(m)},l_{(m)}). 
        \end{equation*}
By the linear growth property of the Markovian representation of $\mathcal{Y}^m_T(\theta_{(m)},l_{(m)})$ and the ergodicity condition of $\Phi^m$, there exists $C>0$ independent of $T$ such that
    \begin{equation}
    \label{eq:eY:Q:1}
    \begin{aligned}
        \frac{1}{C}e^{-C|\Phi^m_{\theta_m}(\theta_{(m)},l_{(m)})|} &\leq \mathbb{E}^{\mathbb{Q}^\psi}\left[   e^{-\gamma \left(\bar{y}^m(\Phi^m_T(\theta_{(m)},l_{(m)}))  -\bar{y}^m(\Phi^m_{\theta_m}(\theta_{(m)},l_{(m)}) ) \right)  } \bigg|\mathcal{F}_{\theta_m} \right] \\
        &\leq C e^{C|\Phi^m_{\theta_m}(\theta_{(m)},l_{(m)})|}  
    \end{aligned}
    \end{equation}
for any $(\theta_{(m)},l_{(m)})\in \Delta_m\times E^m$; see Proposition B.1 in \cite{hu2020systems}. By Lemma \ref{lem:exp:estimate:factor}, there exists $K>0$ independent of $T$ such that  
    \begin{equation}
        \label{eq:EY:growth}
        0 < K^{-1} \leq \mathbb{E}\left[\mathbb{E}^{\mathbb{Q}^\psi}\left[   e^{-\gamma \left(\bar{y}^m(\Phi^m_T(\theta_{(m)},l_{(m)})  -\bar{y}^m(\Phi^m_{\theta_m}(\theta_{(m)},l_{(m)}) ) \right)  } \bigg|\mathcal{F}_{\theta_m} \right] \right] \leq K. 
    \end{equation}
Using this and the upper bound for $\eta$, we have
    \begin{align}
    \label{eq:psi:eta:eY:growth:split}
      &\   \mathbb{E}\left[ \psi_T(\theta_{(m)},l_{(m)}) \eta_T(\theta_{(m)},l_{(m)}) e^{-\gamma\left(\bar{y}^m(\Phi^m_T(\theta_{(m)},l_{(m)})  -\bar{y}^m(\Phi^m_{\theta_m}(\theta_{(m)},l_{(m)}) ) \right)  }    \mathbbm{1}_{\{T\geq \theta_m\}} \big| \mathcal{F}_{\theta_m}  \right] \nonumber \\
       \leq&\ \mathbbm{1}_{ \{T \geq \theta_m \} } H_T(\theta_{(m)},l_{(m)}) \mathbb{E}\left[\mathbb{E}^{\mathbb{Q}^\psi}\left[e^{-\gamma\left(\bar{y}^m(\Phi^m_T(\theta_{(m)},l_{(m)})  -\bar{y}^m(\Phi^m_{\theta_m}(\theta_{(m)},l_{(m)}) ) \right)  }    \mathbbm{1}_{\{T\geq \theta_m\}} \big| \mathcal{F}_{\theta_m}  \right]\right]\nonumber \\
       \leq &\ K\mathbbm{1}_{ \{T \geq \theta_m \} } H_T(\theta_{(m)},l_{(m)}) .
    \end{align}
By \eqref{eq:eta:condition:growth},  we obtain the upper bound in \eqref{eq:psi:eta:eY:bounds}. 

Likewise, by \eqref{eq:EY:growth} and the lower bound for $\eta$, we have 
\begin{align*}
    \label{eq:psi:eta:eY:growth:split}
      &\   \mathbb{E}\left[\mathbb{E}\left[ \psi_T(\theta_{(m)},l_{(m)}) \eta_T(\theta_{(m)},l_{(m)}) e^{-\gamma\left(\bar{y}^m(\Phi^m_T(\theta_{(m)},l_{(m)})  -\bar{y}^m(\Phi^m_{\theta_m}(\theta_{(m)},l_{(m)}) ) \right)  }    \mathbbm{1}_{\{T\geq \theta_m\}} \big| \mathcal{F}_{\theta_m}  \right]\right] \nonumber \\
       \geq&\ \mathbbm{1}_{ \{T \geq \theta_m \} } h_T(\theta_{(m)},l_{(m)}) \mathbb{E}\left[\mathbb{E}^{\mathbb{Q}^\psi}\left[e^{-\gamma\left(\bar{y}^m(\Phi^m_T(\theta_{(m)},l_{(m)})  -\bar{y}^m(\Phi^m_{\theta_m}(\theta_{(m)},l_{(m)}) ) \right)  }    \mathbbm{1}_{\{T\geq \theta_m\}} \big| \mathcal{F}_{\theta_m}  \right]\right]\nonumber \\
       \geq &\ K^{-1}\mathbbm{1}_{ \{T \geq \theta_m \} } h_T(\theta_{(m)},l_{(m)}) .
    \end{align*}
The lower bound in \eqref{eq:psi:eta:eY:bounds} follows from \eqref{eq:eta:condition:growth}. 


Finally, by \eqref{eq:psi:eta:eY:bounds}, we have 
    \begin{equation*}
        \begin{aligned}
             &\ \lim_{T\to\infty}   \frac{1}{T} \log \mathbb{E}\bigg[  \int_{\Delta_m\times E^m} \psi_T(\theta_{(m)},l_{(m)}) \eta_T(\theta_{(m)},l_{(m)}) e^{-\gamma\left(\mathcal{Y}^m_T(\theta_{(m)},l_{(m)}) - \mathcal{Y}^m_{\theta_m}(\theta_{(m)},l_{(m)}) \right)  }   \nonumber \\
      &\quad \cdot  \mathbbm{1}_{\{T\geq \theta_m\}} d\theta_{(m)}\boldsymbol{\lambda}(dl_{(m)}) \bigg]= 0.
        \end{aligned}
    \end{equation*}
Using this and \eqref{eq:growth:LT:4}, we arrive at the result. \hfill $\square$

\subsection{Proof of Proposition \ref{pp:lower:bound}}
\label{sec:pf:pp:lower:bound}
     {\color{black}  Consider the perturbed functions $\bar{y}^{n,\rho}(\phi)= y^{n,\rho}(\phi)-y^{n,\rho}(\hat{\phi}_n)$, $n=0,\dots,m$, $\phi\in\mathbb{R}^d$, and the associated process $\bar{Y}^{n,\rho}_t(\theta_{(n)},l_{(n)}):= \bar{y}^{n,\rho}(\Phi^n_t(\theta_{(n)},l_{(n)}))$ that satisfies \eqref{eq:Yn:peturb}. By Theorem \ref{thm:Z:markovian}, the function $\bar{y}^{n,\rho}$ satisfies the  uniform-in-$\rho$ linear growth and Lipschitz property: for any $\phi,\phi_1,\phi_2\in \mathbb{R}^d$ and $\rho>0$,
   \begin{equation}
    \label{eq:linear:growth:delta}
            \begin{aligned}
                    &   | \bar{y}^{n,\rho}(\phi) | \leq K_{Z^n}|\phi - \hat{\phi}_n| \leq K_{Z^n}(|\phi|+|\hat{\phi}_n|), \ | \bar{y}^{n,\rho}(\phi_1) -  \bar{y}^{n,\rho}(\phi_2)| \leq K_{Z^n}|\phi_1-\phi_2|.
            \end{aligned}
    \end{equation}
    On the other hand, by Theorem \ref{thm:Z:markovian} and Proposition \ref{pp:Delta:Y}, we have, for any $n=0,\dots,m-1$ and $\rho>0$,
        \begin{align*}
             e^{y^{n+1,\rho}(\hat{\phi}_{n+1}) - y^{n,\rho}(\hat{\phi}_n)} & = e^{y^{n+1,\rho}(\hat{\phi}_{n+1}) - y^{n+1,\rho}(\hat{\phi}_n +\varphi^n(\hat{\phi}_n))  +y^{n+1,\rho}(\hat{\phi}_n +\varphi^n(\hat{\phi}_n)) - y^{n,\rho}(\hat{\phi}_n)} \\
             &\leq e^{K_{Z^{n+1}}(1+|\hat{\phi}_{n+1}| + (1+K_\varphi)|\hat{\phi}_n|  ) + K_{\Delta Y^{n+1}}}.
        \end{align*}

    Using \eqref{eq:linear:growth:delta}, $\rho |y^{n,\rho}(\hat{\phi}_n) | \leq K_Y$, and a standard diagonal argument,  there exists a sequence $(\rho_i)_{i=1}^\infty$ with $\rho_i\downarrow 0$ such that, for $\phi$ in a dense subset of $\mathbb{R}^d$, 
        \begin{equation}
        \label{eq:y:converge}
            \lim_{i\to \infty}\rho_i y^{n,\rho_i}(\hat{\phi}_n) = \varrho_n \in \mathbb{R}, \ \lim_{i\to\infty} \bar{y}^{n,\rho_i}(\phi) = \bar{y}^n({\phi}),
        \end{equation}
    for any $n=0,\dots,m$, and  
        \begin{equation}
                    \label{eq:y:converge:2}
                \lim_{i\to \infty} e^{\gamma(y^{n+1,\rho_i}(\hat{\phi}_{n+1}) - y^{n,\rho_i}(\hat{\phi}_n))} = \delta_n \in\left[0,  e^{\gamma K_{Z^{n+1}}(1+|\hat{\phi}_{n+1}| + (1+K_\varphi)|\hat{\phi}_n|  ) + \gamma K_{\Delta Y^{n+1}}}\right] 
        \end{equation}
   for $n=0,\dots,m-1$.  By the uniform Lipschitz property \eqref{eq:linear:growth:delta}, the convergence can be extended to the entire domain $\mathbb{R}^{d}$. In addition, for any $\rho>0$ and $n=0,\dots,m-1$, using  \eqref{eq:linear:growth:delta} and     \eqref{eq:Delta:Y:n:bound:stronger}, 
     \begin{align*}
            \rho \left(y^{n+1,\rho}(\hat{\phi}_{n+1})  - y^{n,\rho}(\hat{\phi}_n) \right) &\leq   \rho \big(y^{n+1,\rho}(\hat{\phi}_{n+1}) -y^{n+1,\rho}(\hat{\phi}_{n}+\varphi^{n}(\hat{\phi}_{n}))  \\
            &\  + y^{n+1,\rho}(\hat{\phi}_{n}+\varphi^{n}(\hat{\phi}_{n})) - y^{n,\rho}(\hat{\phi}_n)\big) \\
            &\leq C\rho( 1+ |\hat{\phi}_n| + |\hat{\phi}_{n+1}|) + \rho K_{\Delta Y^{n+1}} \to 0 
        \end{align*}
    as $\rho\to 0$. Hence, we have $\varrho_{n+1}-\varrho_n\leq 0$. 
 The convergence \eqref{eq:y:converge} then indicates the existence of a sequence $(\rho_i)_{i=1}^\infty$, $\rho_i\downarrow 0$, such that $$\lim_{i\to \infty} \bar{Y}^{n,\rho_i}_t(\theta_{(n)},l_{(n)}) = \mathcal{Y}^n_t(\theta_{(n)},l_{(n)}) = \bar{y}^n\left(\Phi^n_t(\theta_{(n)},l_{(n)}) \right).$$ Likewise, using the uniform boundedness of $Z^{n,\rho}$ in $\rho$ (see Theorem \ref{thm:Z:markovian}), it is standard to show the existence of a sequence of functions $\bar{z}^n :\mathbb{R}^{d}\to \mathbb{R}^{d}$, $n=0,\dots,m$, such that $\lim_{i\to\infty} Z^{n,\rho_i}_t(\theta_{(n)},l_{(n)})  = \mathcal{Z}^n_t(\theta_{(n)},l_{(n)}) = \bar{z}^n(\Phi^n_t(\theta_{(n)},l_{(n)}))$. The tuple $\big( \mathcal{Y}^n_t(\theta_{(n)},l_{(n)}),$ $\mathcal{Z}^n_t(\theta_{(n)},l_{(n)}),\varrho_n,\delta_n)_{n=0}^m$ is then the solution of the following system of ergodic BSDEs: for $n=0,\dots,{m-1}$,  $(\theta_{(n)},l_{(n)})\in \Delta_n\times E^n$, and $t\geq \theta_n$,
    \begin{align*}
              d\mathcal{Y}^{n}_t(\theta_{(n)},l_{(n)}) &= \bigg(\varrho_{n+1} -\min_{\pi\in\Pi_{n}} \bigg\{\hat{F}^{n}_1(\pi,\mathcal{Z}^{n}_t ,\Phi^{n}_t)  \\
              &\quad + \frac{\delta_{n}}{\gamma}\int_E e^{\gamma(\mathcal{Y}^{n+1}_t(t,l) - \mathcal{Y}^{n}_t  - \pi'\hat{\beta}^n(\Phi^{n}_{t-},l)) } \bigg\} \lambda_{n+1}(dl)  \bigg)dt + (\mathcal{Z}^{n}_t)'dW_t,
        \end{align*}
    where we have taken the convention $\delta_{m}\equiv 0$, and omitted writing the dependence of the index $(\theta_{(n)},l_{(n)})$ for notational convenience. 

      Fix $n=0,\dots,m-1$, and assume the contrary that $\delta_n =0$. Then,
    for any $t\leq \tau$, using the fact that $\varrho_{n+1}\leq \varrho_n$, we have 
        \begin{align*}
           & \ \ \ \  \mathcal{Y}^n_t(\theta_{(n)},l_{(n)}) - \mathcal{Y}^{n+1}_t(\theta_{(n+1)},l_{(n+1)})\\
           &=  \mathcal{Y}^n_\tau - \mathcal{Y}^{n+1}_\tau + \int_t^\tau \Bigg( \varrho_{n+1} - \varrho_n + \min_{n\in\Pi_n} \hat{F}^{n}_1(\pi,\mathcal{Z}^n_s,\Phi^n_s)- \min_{\pi\in\Pi_{n+1}} \bigg\{\hat{F}^{n+1}_1(\pi,\mathcal{Z}^{n+1}_s,\Phi^{n+1}_s)  \\
              &\qquad + \frac{\delta_{n+1}}{\gamma}\int_E e^{\gamma(\mathcal{Y}^{n+2}_t(s,l) - \mathcal{Y}^{n+1}_s - \pi'\hat{\beta}^n(\Phi^{n+1}_{s-},l)) } \bigg\} \lambda_{n+2}(dl)  \Bigg)ds +\int_t^\tau \left(\mathcal{Z}^{n+1}_s - \mathcal{Z}^n_s \right)'   dW_s \\
              &\leq  \mathcal{Y}^n_\tau - \mathcal{Y}^{n+1}_\tau + \int_t^\tau \Bigg( \min_{\pi\in\Pi_n} \hat{F}^{n}_1(\pi,\mathcal{Z}^n_s,\Phi^n_s)- \min_{\pi\in\Pi_{n+1}} \hat{F}^{n+1}_1(\pi,\mathcal{Z}^{n+1}_s,\Phi^{n+1}_s)  \Bigg)ds \\
              &\quad +\int_t^\tau \left(\mathcal{Z}^{n+1}_s - \mathcal{Z}^n_s \right)'   dW_s. 
        \end{align*}
Using Assumption \ref{ass:extra:ergodic:alpha},       \begin{align*}
           & \ \ \ \  \min_{\pi\in\Pi_n} \hat{F}^{n}_1(\pi,\mathcal{Z}^n_t,\Phi^n_t)- \min_{\pi\in\Pi_{n+1}} \hat{F}^{n+1}_1(\pi,\mathcal{Z}^{n+1}_t,\Phi^{n+1}_t)\\
              &\leq  \min_{\pi\in\Pi_n} \Bigg\{ \frac{\gamma}{2}\left|\hat{\sigma}^{n}(\Phi^{n}_s) \pi - \left(\mathcal{Z}^{n}_t + \frac{\hat{\alpha}^{n}(\Phi^{n}_t)}{\gamma}\right)^2 \right| - \hat{\alpha}^{n}(\Phi^{n}_t)'\mathcal{Z}^{n}_t \\
              &\quad - \frac{|\hat{\alpha}^{n}(\Phi^{n+1}_t)|^2}{2\gamma}\Bigg\} -  \min_{\pi\in\Pi_{n+1}} \Bigg\{ \frac{\gamma}{2}\left|\hat{\sigma}^{n+1}(\Phi^{n+1}_s) \pi - \left(\mathcal{Z}^{n+1}_t + \frac{\hat{\alpha}^{n+1}(\Phi^{n+1}_t)}{\gamma}\right)^2 \right| \\
              &\quad - \hat{\alpha}^{n+1}(\Phi^{n+1}_t)'\mathcal{Z}^{n+1}_t - \frac{|\hat{\alpha}^{n+1}(\Phi^{n+1}_t)|^2}{2\gamma}\Bigg\} \\
              &\leq \frac{\gamma}{2}|\mathcal{Z}^{n}_t|^2 - \hat{\alpha}^{n}(\Phi^{n}_t)'\mathcal{Z}^{n}_t -  \frac{|\hat{\alpha}^{n}(\Phi^{n}_t)|^2}{2\gamma} + \hat{\alpha}^{n+1}(\Phi^{n+1}_t)'\mathcal{Z}^{n+1}_t + \frac{|\hat{\alpha}^{n+1}(\Phi^{n+1}_t)|^2}{2\gamma} \\
              &\leq -\frac{|\hat{\alpha}^{n}(\Phi^{n}_t)|^2}{2\gamma}    + \frac{|\hat{\alpha}^{n+1}(\Phi^{n+1}_t)|^2}{2\gamma} + \frac{\gamma (K_{Z^n})^2}{2}  + \hat{\alpha}^{n+1}(\Phi^{n+1}_t)'(\mathcal{Z}^{n+1}_t-\mathcal{Z}^n_t) \\
              &\quad -\left(\hat{\alpha}^n(\Phi^n_t)-\hat{\alpha}^{n+1}(\Phi^{n+1}_t) \right)' \mathcal{Z}^{n}_t \\
              &\leq -\frac{|\hat{\alpha}^{n}(\Phi^{n}_t)|^2}{2\gamma}    + \frac{|\hat{\alpha}^{n+1}(\Phi^{n+1}_t)|^2}{2\gamma} + \frac{\gamma (K_{Z^n})^2}{2} + |\hat{\alpha}^{n+1}(\Phi^{n+1}_t)||\mathcal{Z}^{n+1}_t - \mathcal{Z}^n_t| \\
              &\quad + \left|\hat{\alpha}^n(\Phi^n_t)-\hat{\alpha}^{n+1}(\Phi^{n+1}_t) \right|K_{Z^{n}} \\
              &\leq 0,
        \end{align*}
    where the last inequality follows from \eqref{eq:alpha:ergodic:ass}. Hence, we have 
              \begin{align*}
            \mathcal{Y}^n_t - \mathcal{Y}^{n+1}_t
              &\leq  \mathcal{Y}^n_\tau - \mathcal{Y}^{n+1}_\tau + \int_t^\tau \left(\mathcal{Z}^{n+1}_s - \mathcal{Z}^n_s \right)'  dW_s. 
        \end{align*}
    In particular, take $t=\theta_n$, $\Phi^n_{\theta_n}(\theta_{(n)},l_{(n)}) = \hat{\phi}_n$, so that $\Phi^{n+1}_{\theta_n}((\theta_{(n)},\theta_n),(l_{(n)},l)) = \tilde{\phi}_{n+1}: = \hat{\phi}_n +\varphi^n(\hat{\phi}_n)$, and 
        \begin{align*}
           & \ \ \ \ \bar{y}^n(\hat{\phi}_n) - \bar{y}^{n+1}\left( \tilde{\phi}_{n+1} \right)\\
            &\leq  \mathbb{E}\left[\bar{y}^n\left( \Phi^{n,\theta_n,\hat{\phi}_n}_\tau(\theta_{(n)},l_{(n)}) \right) - \bar{y}^{n+1}\left( \Phi^{n+1,\theta_n,\tilde{\phi}_{n+1} }_\tau((\theta_{(n)},\theta_n),(l_{(n)},l)) \right)  |\mathcal{F}_{\theta_n}  \right].
        \end{align*}

By the uniform-in-$\rho$ linear growth property of $\bar{y}^n$ and $\bar{y}^{n+1}$ \eqref{eq:linear:growth:delta}, the  uniform integrability of $\Phi^n$, $\Phi^{n+1}$ (see Lemma \ref{lem:exp:estimate:factor}), and the  convergence of $\bar{y}^{n,\rho_i}(\cdot)\to \bar{y}^n(\cdot)$ as $\rho_i\to 0$, for any $\varepsilon>0$, there exists $N>0$ such that for any $\tau\geq \theta_n$ and $i\geq N$,
     \begin{align*}
            & \ \ \ \ \bar{y}^{n,\rho_i}(\hat{\phi}_n) - \bar{y}^{n+1,\rho_i}\left(\tilde{\phi}_{n+1}  \right)\\
            &\leq \varepsilon + \mathbb{E}\left[\bar{y}^{n,\rho_i}\left( \Phi^{n,\theta_n,\hat{\phi}_n}_\tau(\theta_{(n)},l_{(n)}) \right) - \bar{y}^{n+1,\rho_i}\left( \Phi^{n+1,\theta_n,\tilde{\phi}_{n+1} }_\tau((\theta_{(n)},\theta_n),(l_{(n)},l)) \right)  |\mathcal{F}_{\theta_n}  \right],
        \end{align*}
whenever $i\geq N$, thanks to the Vitali convergence theorem. In addition, by dominated convergence and the vanishing limit of $Y^{n,\rho},Y^{n+1,\rho}$ at infinity,\footnote{This is an immediate consequence of Gr\"onwall's inequality with the discount rate $\rho$, thanks to the boundedness of the drivers.} we have, by passing to the limit $\tau\to\infty$, 
    \begin{equation*}
        \bar{y}^{n,\rho_i}(\hat{\phi}_n) - \bar{y}^{n+1,\rho_i}\left(\tilde{\phi}_{n+1}  \right) \leq \varepsilon + y^{n+1,\rho_i}(\hat{\phi}_{n+1}) - y^{n,\rho_i}(\hat{\phi}_n).
    \end{equation*}
Hence, whenever $i\geq N$, 
    \begin{align*}
      \varepsilon &\geq   y^{n,\rho_i}(\hat{\phi}_n) - y^{n+1,\rho_i}\left(\tilde{\phi}_{n+1}  \right) \\
      &\geq y^{n,\rho_i}(\hat{\phi}_n) - y^{n+1,\rho_i}(\hat{\phi}_{n+1}) - \left|y^{n+1,\rho_i}(\hat{\phi}_{n+1}) - y^{n+1,\rho_i}(\tilde{\phi}_{n+1}) \right| \\
      &\geq  y^{n,\rho_i}(\hat{\phi}_n) - y^{n+1,\rho_i}(\hat{\phi}_{n+1}) - C(1+|\hat{\phi}_{n+1}|+|\hat{\phi}_n|), 
    \end{align*}
where $C>0$ is independent of $\rho_i$. This contradicts with the assumption that $\delta_n=0$, and therefore we deduce that $\delta_n>0$. The proof is then complete by an iterative argument along with extractions of subsequences of $(\rho_i)_{i=1}^\infty$. } \hfill $\square$

\subsection{Proof of Theorem \ref{th:EBSDE}}
\label{sec:pf:th:EBSDE}
    The existence of solution has been proved in the above discussion. Henceforth, we only consider the uniqueness of solution, which shall be proven recursively.

   For $n=0,\dots,m$, let $(\mathcal{Y}^{n,1},\mathcal{Z}^{n,1},\varrho_1)_{n=0}^m$ and $(\mathcal{Y}^{n,2},\mathcal{Z}^{n,2},\varrho_2)_{n=0}^m$ be two solutions of the system of ergodic BSDEs \eqref{eq:ergodic:Ym}-\eqref{eq:ergodic:Yn}. Define $\delta \mathcal{Y} := \mathcal{Y}^{n,1} - \mathcal{Y}^{n,2}$, $\delta \mathcal{Z} := \mathcal{Z}^{n,1} - \mathcal{Z}^{n,2}$, and  $\delta \varrho := \varrho_1-\varrho_2$. For any  $(\theta_{(m)},l_{(m)})\in \Delta_m\times E^m$ and $t\geq \theta_m$, 
        \begin{align*}
            d\delta \mathcal{Y}^m_t(\theta_{(m)},l_{(m)}) &=\bigg( \delta \varrho - \min_{\pi\in\Pi_m} \hat{f}^m\left(\pi,\mathcal{Z}^{m,1}_t(\theta_{(m)},l_{(m)}),\Phi^m_t(\theta_{(m)},l_{(m)})\right) \\
            &\quad + \min_{\pi\in\Pi_m} \hat{f}^m\left(\pi,\mathcal{Z}^{m,2}_t(\theta_{(m)},l_{(m)}),\Phi^m_t(\theta_{(m)},l_{(m)})\right)\bigg)dt  \\
            &\ +\delta \mathcal{Z}^m_t(\theta_{(m)},l_{(m)})'dW_t
        \end{align*}
    For notational convenience, we shall drop the dependence of $(\theta_{(m)},l_{(m)})$ in the remaining proof. 

    It is straightforward to show that the existence of $C>0$ such that, for any $z_1,z_2,\phi\in \mathbb{R}^d$,
        \begin{equation}
        \label{eq:lip:ergodic}
            \left|\min_{\pi\in\Pi_m} \hat{f}^m\left(\pi,z_1,\phi\right)-\min_{\pi\in\Pi_m} \hat{f}^m\left(\pi,z_2,\phi\right) \right| \leq C(1+|z_1|+|z_2|)|z_1-z_2|. 
        \end{equation}
    Using this and the boundedness of $\mathcal{Z}^{m,i}$, $i=1,2$, for any $T>\theta_m$, we define the measure  $\mathbb{Q}^{\xi^m}$ by
        \begin{equation*}
            \frac{d\mathbb{Q}^{\xi^m}}{d\mathbb{P}}  = \mathcal{E}_{\theta_m,T}\left( \int_{\theta_m}^\cdot \xi^m_s ds \right),
        \end{equation*}
    where 
        \begin{equation*}
        \xi_t^m  := \frac{ \min_{\pi\in\Pi_m} \hat{f}^m\left(\pi,\mathcal{Z}^{m,1}_t,\Phi^m_t\right) -  \min_{\pi\in\Pi_m} \hat{f}^m\left(\pi,\mathcal{Z}^{m,2}_t,\Phi^m_t\right)}{\delta \mathcal{Z}^m_t}  \mathbbm{1}_{ \{\delta \mathcal{Z}^m_t \neq 0 \} }. 
        \end{equation*}
   Therefore, for any $T>\theta_m$, we have 
        \begin{equation*}
            \delta \varrho = \frac{\mathbb{E}^{\mathbb{Q}^{\xi^m}}[ \delta\mathcal{Y}^m_T(\theta_{(m)},l_{(m)}) - \delta\mathcal{Y}^m_{\theta_m}(\theta_{(m)},l_{(m)}) \mid \mathcal{F}_{\theta_m} ]}{T-\theta_m}. 
        \end{equation*}
    
    Let $\bar{{\bf y}}_i(\cdot)$, $i=1,2$, be the Markovian representation of $(\mathcal{Y}^{n,i})_{n=0}^m$, where for $\boldsymbol{\phi}\in \mathbb{R}^d$, $\bar{\bf y}_i(\boldsymbol{\phi}) = (\bar{y}^{n-1}_i(\phi_{n-1}))_{n=1}^{m+1}$. In addition, we may assume that $\bar{y}^n_1(0) = \bar{y}^n_2(0)$ for all $n=0,\dots,m$. Using the ergodicity of $\Phi^m_\cdot(\theta_{(m)},l_{(m)})$, the growth property  \eqref{eq:linear:growth:ergodic}
 and Proposition B.1 of \cite{hu2020systems}, there exists $C>0$ such that 
        \begin{align*}
          &\   \mathbb{E}^{\mathbb{Q}^{\xi^m}}[ \delta\mathcal{Y}^m_T(\theta_{(m)},l_{(m)}) - \delta\mathcal{Y}^m_{\theta_m}(\theta_{(m)},l_{(m)}) \mid \mathcal{F}_{\theta_m} ] \\
          =&\  \mathbb{E}^{\mathbb{Q}^{\xi^m}}\bigg[ \left(\bar{y}^{m}_1\left(\Phi^m_T(\theta_{(m)},l_{(m)}) \right)-\bar{y}^{m}_2\left(\Phi^m_T(\theta_{(m)},l_{(m)}) \right) \right)  \\
          &\quad - \left(\bar{y}^{m}_1\left(\Phi^m_{\theta_m}(\theta_{(m)},l_{(m)}) \right)-\bar{y}^{m}_2\left(\Phi^m_{\theta_m}(\theta_{(m)},l_{(m)}) \right) \right)   \bigg] \\
          \leq&\  C\left(1+ |\Phi^m_{\theta_m}(\theta_{(m)},l_{(m)})| \right). 
        \end{align*}
    Hence, by passing to the limit $T\to\infty$, we have  $\delta\varrho=0 $. 

    To show $\mathcal{Y}^{m,1}=\mathcal{Y}^{m,2}$ and $\mathcal{Z}^{m,1} = \mathcal{Z}^{m,2}$, it suffices to show that $\mathcal{Y}^{m,1}_{\theta_m}(\theta_{(m)},l_{(m)}) = \mathcal{Y}^{m,2}_{\theta_m}(\theta_{(m)},l_{(m)})$. The rest of the proof then follows from Theorem 3.11 in \cite{DEBUSSCHE2011407}. Notice that, for any $T>\theta_m$, \eqref{eq:linear:growth:ergodic} and Proposition B.1 in \cite{hu2020systems} imply the existence of $C,K_\phi>0$ such that 
        \begin{align*}
             \delta \mathcal{Y}^m_{\theta_m} &= \mathbb{E}^{\mathbb{Q}^{\xi^m}}\left[ \bar{y}^{m}_1\left(\Phi^m_T(\theta_{(m)},l_{(m)}) \right) - \bar{y}^{m}_2\left(\Phi^m_T(\theta_{(m)},l_{(m)}) \right) \right] \\
             &= \mathbb{E}^{\mathbb{Q}^{\xi^m}}\left[\left(\bar{y}^{m}_1\left(\Phi^m_T(\theta_{(m)},l_{(m)}) \right)-\bar{y}^{m}_1(0) \right) - \left(\bar{y}^{m}_2\left(\Phi^m_T(\theta_{(m)},l_{(m)}) \right)-\bar{y}^{m}_2(0) \right) \right] \\
             &\leq  C\left( 1 + \left|\Phi^m_{\theta_m}(\theta_{(m)},l_{(m)}) \right|^2 \right) e^{-K_\phi(T-\theta_m)}. 
        \end{align*}
    By passing to the limit $T\to \infty$, we deduce that $\delta \mathcal{Y}^m(\cdot,\cdot) \equiv 0$. This also implies $\bar{y}^{m}_1(\cdot) = \bar{y}^{m}_2(\cdot)$. 

    For $n=0,\dots,m-1$, $(\theta_{(n)},l_{(n)})\in \Delta_n\times E^n$, and $t\geq 
    \theta_n$, it suffices to show that  $\mathcal{Y}^{n,1}_{\theta_n}(\theta_{(n)},l_{(n)}) = \mathcal{Y}^{n,2}_{\theta_n}(\theta_{(n)},l_{(n)})$, since we have already shown $\varrho_1=\varrho_2$. By the induction assumption, we have 
      \begin{align*}
            d\delta \mathcal{Y}^n_t  &=\bigg[ - \min_{\pi\in\Pi_n} \hat{f}^n\left(t,\pi,\mathcal{Y}^{n,1}_t ,\mathcal{Z}^{n,1}_t,\Phi^n_t\right) \\
            &\quad + \min_{\pi\in\Pi_n} \hat{f}^n\left(t,\pi,\mathcal{Y}^{n,2}_t,\mathcal{Z}^{n,2}_t,\Phi^n_t\right)\bigg]dt   +(\delta \mathcal{Z}^n_t)'dW_t, 
        \end{align*}
    where we have again omit writing the dependence of $(\theta_{(n)},l_{(n)})$. Using the induction assumption, and following the derivation of        \eqref{eq:Yn:proof:lipschitz:unique}, there exists $C>0$ such that 
        \begin{align*}
             &\    \left|  \min_{\pi\in\Pi_n} \hat{f}^n\left(t,\pi,\mathcal{Y}^{n,1}_t,\mathcal{Z}^{n,2}_t,\Phi^n_t\right)    - \min_{\pi\in\Pi_n}\hat{f}^n\left(t,\pi,\mathcal{Y}^{n,2}_t,\mathcal{Z}^{n,2}_t,\Phi^n_t \right)    \right|\\
             \leq&\  C|\delta \mathcal{Z}^n_t|(1+|\mathcal{Z}^{n,1}_t|+|\mathcal{Z}^{n,2}_t|).
        \end{align*}
   On the other hand, the map $y\mapsto \hat{f}^n(t,\pi,y,z,\phi)$ is non-increasing. Hence, 
    \begin{equation*}
        \delta\mathcal{Y}^n_t\left[   \min_{\pi\in\Pi_n} \hat{f}^n\left(t,\pi,\mathcal{Y}^{n,1}_t,\mathcal{Z}^{n,2}_t,\Phi^n_t\right)    - \min_{\pi\in\Pi_n}\hat{f}^n\left(t,\pi,\mathcal{Y}^{n,2}_t,\mathcal{Z}^{n,2}_t,\Phi^n_t \right)  \right] \leq 0. 
    \end{equation*}
    Using these, Lemma 3.4 in \cite{confortola:briand:QBSDE} and Proposition B.1 in \cite{hu2020systems}, we have, for any $T>\theta_n$, 
          \begin{align*}
             \delta \mathcal{Y}^n_{\theta_n} 
             &= \mathbb{E}^{\mathbb{Q}^{\xi^n}}\left[\left(\bar{y}^{n}_1\left(\Phi^n_T(\theta_{(n)},l_{(n)}) \right)-\bar{y}^{n}_1(0) \right) - \left(\bar{y}^{n}_2\left(\Phi^n_T(\theta_{(n)},l_{(n)}) \right)-\bar{y}^{n}_2(0) \right) \right]  \\
             &\leq  C\left( 1 +  \left|\Phi^n_{\theta_n}(\theta_{(n)},l_{(n)}) \right|^2 \right) e^{-K_\phi(T-\theta_n)}, 
        \end{align*}
    where the measure $\mathbb{Q}^{\xi^n}$ is defined by 
        \begin{align*}
             \frac{d\mathbb{Q}^{\xi^n}}{d\mathbb{P}}  &:= \mathcal{E}_{\theta_n,T}\left( \int_{\theta_n}^\cdot \xi^n_s ds \right), \\
             \xi^n_t &:= \frac{ \min_{\pi\in\Pi_n} \hat{f}^n\left(t,\pi,\mathcal{Z}^{n,1}_t,\Phi^n_t\right) -  \min_{\pi\in\Pi_n} \hat{f}^n\left(t,\pi,\mathcal{Z}^{n,2}_t,\Phi^n_t\right)}{\delta \mathcal{Z}^n_t}  \mathbbm{1}_{ \{\delta \mathcal{Z}^n_t \neq 0 \} }. 
        \end{align*}
    The result then follows by passing to the limit $T\to\infty$. \hfill $\square$
\end{appendices}








\end{document}